\documentclass[showpacs,prd,twocolumn,floatfix,
nofootinbib,superscriptaddress]{revtex4}

\usepackage{hyperref}
\hypersetup{
    bookmarks=true,         
    unicode=false,          
    pdftoolbar=true,        
    pdfmenubar=true,        
    pdffitwindow=false,     
    pdfstartview={FitH},    
    pdftitle={My title},    
    pdfauthor={Author},     
    pdfsubject={Subject},   
    pdfcreator={Creator},   
    pdfproducer={Producer}, 
    pdfkeywords={keyword1} {key2} {key3}, 
    pdfnewwindow=true,      
    colorlinks=true,       
    linkcolor=blue,          
    citecolor=blue,        
    filecolor=blue,      
    urlcolor=blue           
}

\usepackage{graphicx}
\usepackage{nicefrac}
\usepackage{lipsum}
\usepackage{amsmath,bm}
\usepackage{subfig}
\usepackage{multirow}

\newcommand\T{\rule{0pt}{2.6ex}}

\begin{document}


\title{Rare decay $B \to K \ell^+ \ell^-$ form factors from lattice QCD}

\author{Chris Bouchard}
\thanks{bouchard.18@osu.edu}
\affiliation{Department of Physics,
The Ohio State University, Columbus, Ohio 43210, USA}

\author{{G.\ Peter} Lepage}
\affiliation{Laboratory of Elementary Particle Physics,
Cornell University, Ithaca, New York 14853, USA}

\author{Christopher Monahan}
\affiliation{Physics Department,
College of William and Mary, Williamsburg, Virginia 23187, USA}

\author{Heechang Na} 
\affiliation{Argonne Leadership Computing Facility,
ANL, Argonne, Illinois 60439, USA}

\author{Junko Shigemitsu}
\affiliation{Department of Physics,
The Ohio State University, Columbus, OH 43210, USA}

\collaboration{HPQCD Collaboration}
\noaffiliation
\date{\today}


\begin{abstract}
We calculate, for the first time using unquenched lattice QCD, form factors for the rare decay $B \to K \ell^+ \ell^-$ in and beyond the Standard Model.  
Our lattice QCD calculation utilizes a nonrelativistic QCD formulation for the $b$ valence quarks, the highly improved staggered quark formulation for the light valence quarks, and employs the MILC $2+1$ asqtad ensembles.  
The form factor results, based on the $z$ expansion, are valid over the full kinematic range of $q^2$.
We construct the ratios $f_0/f_+$ and $f_T/f_+$, which are useful in constraining new physics and verifying effective theory form factor symmetry relations.
We also discuss the calculation of Standard Model observables.

\end{abstract}


\maketitle


\section{ Introduction }

The rare decay $B \rightarrow K \ell^+ \ell^-$ involves a $b \rightarrow s$ flavor-changing neutral current transition 
which can proceed only through loop diagrams in the Standard Model, making this a particularly sensitive probe for new physics. 
Experimentalists have already started to collect more and more data on this decay and comparisons with accurate Standard Model predictions have become very important and timely.  
Calculations of the branching fraction ${\cal B}(B \rightarrow K \ell^+ \ell^-)$ require knowledge of several form factors, which in turn depend on having control over long distance QCD phenomena and on being able to evaluate hadronic matrix elements of vector and tensor currents 
between the $B$ and the $K$ meson states.  The only first-principles method for carrying out such nonperturbative QCD calculations is lattice QCD.  
In this article we use lattice QCD to determine the three relevant form factors $f_+$, $f_0$ and $f_T$, which can then be used to obtain branching fraction information both in and beyond the Standard Model.

There is an active effort ~\cite{Becirevic:2012, Bobeth:2011, Beaujean:2012, Altmannsofer:2012, Bobeth:2013} to constrain new physics using experimental results for $B\to K\ell^+\ell^-$, often in combination with other rare $B$ decays.  Form factor information in these works typically comes from light cone sum rule results ({\it cf.} Refs.~\cite{Ball:2005, Khodjamirian:2010, Khodjamirian:2013}), valid at low $q^2$.  Ref.~\cite{Becirevic:2012} calculates all three form factors in lattice QCD using the quenched approximation ({\it i.e.} neglecting virtual quark loops in the sea) and extrapolates to low $q^2$ using the model-dependent Be\v{c}irevi\'{c} Kaidalov parameterization~\cite{Becirevic:2000}.  
In Ref.~\cite{Al-Haydari:2009} the QCDSF collaboration published results for $f_0$ and $f_+$ using the quenched approximation.
There are preliminary unquenched results by Liu {\it et al.}~\cite{Liu:2011} and the Fermilab Lattice and MILC collaborations~\cite{Zhou:2012}.
Measurements related to this decay have been made at the $B$-factories BABAR~\cite{Lees:2012} and Belle~\cite{Wei:2009}, by CDF~\cite{Aaltonen:2011}, and most recently by LHCb~\cite{Aaij:2012, Aaij:2012b}.

This work reports the first unquenched lattice QCD calculation of the form factors for this rare decay.
In~\cite{Bouchard:PRL} we explore the phenomenological implications of these form factors for several Standard Model Observables, comparing to experiment where possible and making predictions elsewhere.
In this article we give details of the lattice calculations leading to the form factors and also provide more information on the relation between 
these form factors and various Standard Model observables.

\section{ Form Factors and Matrix Elements }
\label{sec-FFs_and_MEs}
The phenomenologically relevant quantities are the form factors $f_{0,+,T}$.
However, the fundamental quantities, and therefore the quantities directly accessible on the lattice, are hadronic matrix elements.  
In this section we summarize the relations between the form factors and the hadronic matrix elements.

The vector hadronic matrix element is parameterized by the scalar and vector form factors $f_{0,+}$
\begin{eqnarray}
\langle K | V^\mu | B \rangle &=& f_+(q^2) \left( p_B^\mu +p_K^\mu - \frac{ M_B^2 - M_K^2 }{ q^2 }\,q^\mu \right) \nonumber \\
& &  +\ f_0(q^2)\ \frac{ M_B^2 - M_K^2 }{ q^2 }\, q^\mu,
\end{eqnarray}
where $V^\mu = \bar{s} \gamma^\mu b$ and $q^\mu \equiv p_B^\mu - p_K^\mu$, the four-momentum transferred to the final state leptons.  At intermediate stages of the calculation we recast $f_{0,+}$ in terms of the more convenient form factors $f_{\parallel,\perp}$
\begin{equation}
\langle K | V^\mu | B \rangle = \sqrt{2M_B} \left[ \frac{p_B^\mu}{M_B}\ f_\parallel(q^2) + p_\perp^\mu\  f_\perp(q^2) \right],
\end{equation}
where $p_\perp^\mu \equiv p_K^\mu - (p_K\cdot p_B)\nicefrac{p_B^\mu}{M_B^2}$.  In the $B$ meson rest frame $f_{\parallel,\perp}$ are simply related to the temporal and spatial components of the hadronic vector matrix elements,
\begin{eqnarray}
\langle K | V^0 | B \rangle &=& \sqrt{2M_B}\ f_\parallel(q^2), \label{eq-fpardef} \\
\langle K | V^k | B \rangle &=& \sqrt{2M_B}\ p_K^k\ f_\perp(q^2). \label{eq-fperpdef}
\end{eqnarray}
The scalar and vector form factors are related to $f_{\parallel,\perp}$ by
\begin{eqnarray}
f_0 &=& \frac{\sqrt{2M_B}}{M_B^2-M_K^2} \left[ (M_B-E_K) f_{\parallel} + {\bf p}_K^2 f_{\perp} \right], \label{eq-f0def} \\
f_+ &=& \frac{1}{\sqrt{2M_B}}\left[ f_{\parallel} + (M_B-E_K) f_{\perp} \right], \label{eq-fplusdef}
\end{eqnarray}
where ${\bf p}_K$ is the kaon three-momentum.  The tensor hadronic matrix element is parameterized by the tensor form factor $f_T$
\begin{equation}
\langle K | T^{k0} | B \rangle = \frac{ 2i M_B p^k_K }{ M_B+M_K }\  f_T(q^2),
\label{eq-fTdef}
\end{equation}
where $T^{\mu\nu} = \bar{s}\, \sigma^{\mu\nu} b$ and $\sigma_{\mu\nu}=\frac{1}{2}[\gamma_\mu,\gamma_\nu]$.  We extract the hadronic matrix elements from lattice simulations, then use them to reconstruct the various form factors.

\section{ Generating Lattice Data }
\label{sec-lattice}
In this section we discuss how lattice data are generated in the form of two and three point correlation functions and how effective lattice currents in the three point data are matched to physical currents.

\subsection{Two and Three Point Correlators}
\begin{table*}[t]
\begin{tabular}{cccccccccccc}
\hline\hline	
	\T ens   	& $L^3\times N_t$	& $r_1/a$	& $au_0m_{\rm sea}$	& $u_0$ 	& $N_{\rm conf}$	& $N_{\rm tsrc}$	& $am_l^{\rm val}$	& $am_s^{\rm val}$	& $am_b$	& $aE_{b\bar{b}}^{\rm sim}$	& $T$	\\ [0.5ex]
	\hline
	\T C1 	& $24^3 \times 64$	& 2.647(3)& 0.005/0.05			& 0.8678	& 1200			& 2				& 0.0070			& 0.0489			& 2.650	& 0.28356(15)				& 12 -- 15	\\
	\T C2	& $20^3 \times 64$	& 2.618(3)& 0.01/0.05			& 0.8677	& 1200			& 2				& 0.0123			& 0.0492			& 2.688	& 0.28323(18)				&12 -- 15 	\\
	\T C3	& $20^3 \times 64$	& 2.644(3)& 0.02/0.05			& 0.8688	& 600			& 2				& 0.0246			& 0.0491			& 2.650	& 0.27897(20)				& 12 -- 15	\\ 
	\T F1		& $28^3 \times 96$	& 3.699(3)& 0.0062/0.031			& 0.8782	& 1200			& 4				& 0.00674			& 0.0337			& 1.832	& 0.25653(14)				& 21 -- 24	\\
	\T F2		& $28^3 \times 96$	& 3.712(4)& 0.0124/0.031			& 0.8788	& 600			& 4				& 0.01350			& 0.0336			& 1.826	& 0.25558(28)				& 21 -- 24	\\ [0.5ex]
\hline\hline
\end{tabular}\caption{Left to right:  labels for the three coarse and two fine ensembles used in this analysis; lattice volume; inverse lattice spacing in $r_1$-units; light/strange sea-quark masses; tadpole improvement factor $u_0 = \langle {\rm plaquette}\rangle^{\nicefrac{1}{4}}$; number of configurations; number of time sources; valence light-quark mass; valence strange-quark mass; $b$-quark mass;  spin-averaged $b\bar{b}$ ground state energies used to relate our $B$ meson simulation energies to the physical $M_B$; and the range of temporal separations between the $B$ meson and the kaon.}
\label{tab-ens}
\end{table*}
Ensemble averages are performed using the MILC $2+1$ asqtad gauge configurations~\cite{Bazavov:2010} listed in Table~\ref{tab-ens}.  The valence quarks in our simulation are nonrelativistic QCD (NRQCD)~\cite{Lepage:1992} $b$ quarks, tuned in Ref.~\cite{Na:2012}, and highly improved staggered quark (HISQ)~\cite{Follana:2007} light and strange quarks, propagators for which were generated in Refs.~\cite{Na:2010, Na:2011}.  Valence quark masses for each ensemble used in the simulations are tabulated in Table~\ref{tab-ens}.

\begin{figure}[t]
\vspace{0.0in}
\centering
{\scalebox{0.7}{\includegraphics[angle=0,width=0.7\textwidth]{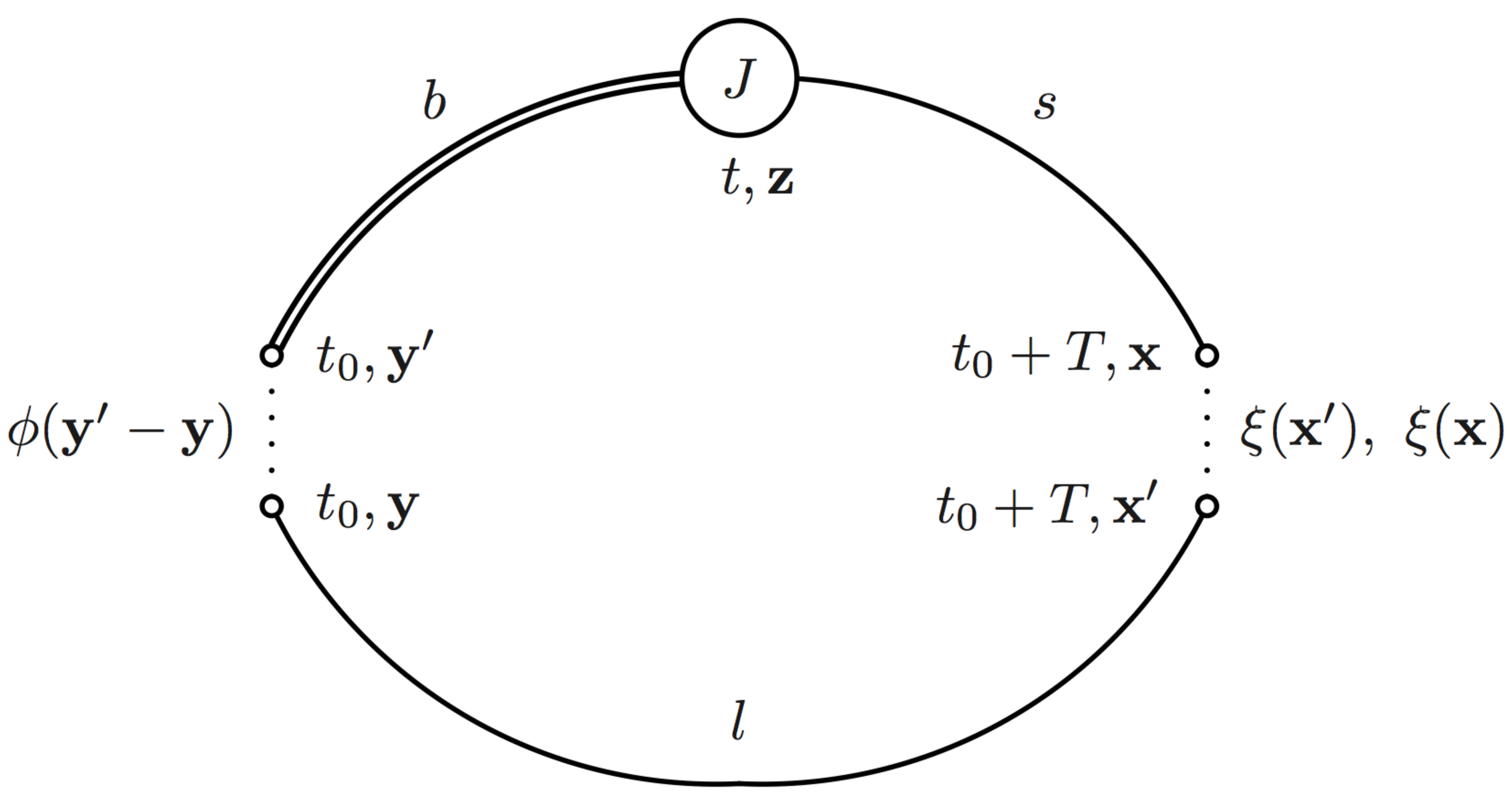}}}
\caption{Setup for three point correlator data generation.}
\vspace{0.0in}
\label{fig-feyndiag}
\end{figure}

Heavy-light $B$ meson bilinears $\Phi_B^\alpha$ are built from NRQCD $b$ and HISQ light quarks (for details see Ref.~\cite{Na:2012}) and light-light kaon bilinears $\Phi_K$ are built from HISQ light and strange quarks (for details see Ref.~\cite{Na:2010}).  The bilinears are used to build two and three point correlation functions
\begin{eqnarray}
C^{\alpha\beta}_B(t_0,t) &=& \frac{1}{L^3} \sum_{{\bf x}, {\bf y}}  \langle \Phi^\beta_B(t,{\bf y})\ \Phi^{\alpha\dagger}_B(t_0,{\bf x}) \rangle, \label{eq-B2pt} \\
C_K(t_0,t;{\bf p}) &=& \frac{1}{L^3} \sum_{{\bf x}, {\bf y}}  e^{i\,{\bf p} \cdot ({\bf x} - {\bf y})}\langle \Phi_K(t,{\bf y})\ \Phi^\dagger_K(t_0,{\bf x}) \rangle, \nonumber \\
& & \label{eq-X2pt} \\
C^\alpha_J(t_0,t,T; {\bf p}) &=& \frac{1}{L^3} \sum_{{\bf x}, {\bf y}, {\bf z}} e^{i\,{\bf p}\cdot ({\bf z} - {\bf x})} \times \nonumber \\
& &\!\! \langle \Phi_K(t_0+T,{\bf x})\ J(t,{\bf z})\ \Phi^{\alpha\dagger}_B(t_0,{\bf y}) \rangle, \label{eq-3pt}
\end{eqnarray}
where the inserted lattice current $J(t,{\bf z})$ is a heavy-light bilinear described in the next section.  The three point correlator setup is depicted in Fig.~\ref{fig-feyndiag}.

Working in the $B$ meson rest frame, a sequential propagator is built from smeared NRQCD $b$ and spectator HISQ quarks.  The $b$ quark smearing function $\phi({\bf y}'-{\bf y})$ is either a delta function or Gaussian, as specified by indices $\alpha,\beta$
\begin{equation}
\phi({\bf y}'-{\bf y}) = \begin{cases} \delta_{{\bf y}'{\bf y}}, & \alpha=l \mbox{(ocal)} \\ \frac{1}{\sqrt{2\pi \sigma^2}} \exp[- \frac{({\bf y}'-{\bf y})^2}{2\sigma^2}], & \alpha=s \mbox{(meared)}, \end{cases}
\end{equation}
with $\sigma=5a$ on the coarse and $\sigma=7a$ on the fine ensembles.  The smearing function is introduced by the replacement $\sum_{\bf y} \to \sum_{{\bf y},{\bf y}'}\phi({\bf y}'-{\bf y})$ in Eqs.~(\ref{eq-B2pt}, \ref{eq-3pt}).  The spectator $l$ quark source includes a U(1) phase $\xi({\bf x}')$.  The daughter $s$ quark, with U(1) phase and momentum insertion at ${\bf x}$, is tied to the sequential quark propagator, with $\sum_{{\bf x}}$ in Eqs.~(\ref{eq-X2pt}, \ref{eq-3pt}) accomplished via random wall sources, {\it ie.} $\sum_{{\bf x}} \to \sum_{{\bf x},{\bf x}'}\xi({\bf x})\xi({\bf x}')$.  

$B$ meson two point correlator data is generated for all four combinations of $b$ quark smearing:  $C_B^{ll}$, $C_B^{ls}$, $C_B^{sl}$, and $C_B^{ss}$.  Kaon two point data are generated for each of the momenta ${\bf p} \in \nicefrac{2\pi}{L}\times \{(0,0,0), (1,0,0), (1,1,0), (1,1,1)\}$.

In three point correlator data the $B$ meson is created at time-slice $t_0$, the kaon is annihilated at $t_0+T$, and a flavor-changing current $J(t,{\bf z})$ is inserted at intermediate times $t_0 \leq t \leq t_0+T$, where $t_0$ is chosen at random to reduce autocorrelations.  Prior to fitting, all data are shifted to a common $t_0=0$. 
The structure of the current $J(t,{\bf z})$ determines whether generated data is for the vector or tensor matrix element and whether it is lowest order in NRQCD or a $\nicefrac{1}{m_b}$ correction.  The details of the currents used in this work are given in the next section.  Three point data are generated over the ranges of $B$ meson and kaon temporal separations $T$ listed in Table~\ref{tab-ens}.

\subsection{Matching Lattice Currents}
\label{sec-match}
We generate data for the lattice effective currents that contribute through $\mathcal{O}(\alpha_s, \nicefrac{\Lambda_{\rm QCD}}{m_b}, \nicefrac{\alpha_s}{am_b})$.  Through this order, the relevant lattice vector ($J = \mathcal{V}_\mu$) and tensor ($J = \mathcal{T}_{\mu\nu}$) currents are
\begin{eqnarray}
\mathcal{V}_\mu^{(0)} &=& \overline{\Psi}_s\, \gamma_\mu\, \Psi_b, \\
\mathcal{V}_\mu^{(1)} &=&-\frac{1}{2 am_b} \overline{\Psi}_s\, \gamma_\mu\, {\boldsymbol \gamma} \cdot {\boldsymbol \nabla}\, \Psi_b, \\
\mathcal{T}_{\mu\nu}^{(0)} &=& \overline{\Psi}_s\, \sigma_{\mu\nu}\, \Psi_b, \\
\mathcal{T}_{\mu\nu}^{(1)} &=&-\frac{1}{2 am_b} \overline{\Psi}_s\, \sigma_{\mu\nu}\, {\boldsymbol \gamma} \cdot {\boldsymbol \nabla}\, \Psi_b .
\end{eqnarray}
For the tensor current we focus on the $\mathcal{T}_{k0}$ component, where heavy-quark symmetry allows us to relate it to the vector current.

The continuum vector current $\langle V_\mu \rangle$ is matched to the lattice vector current by
\begin{equation}
\langle V_\mu\rangle = (1+\alpha_s \rho_0^{(V_\mu)})\langle \mathcal{V}_\mu^{(0)}\rangle + \langle \mathcal{V}_\mu^{(1), {\rm sub}}\rangle ,
\label{eq-Vmatch}
\end{equation}
where
\begin{equation}
\langle \mathcal{V}_\mu^{(1), {\rm sub}}\rangle \equiv \langle \mathcal{V}_\mu^{(1)}\rangle - \alpha_s \zeta_{10}^{V_\mu} \langle \mathcal{V}_\mu^{(0)}\rangle.
\end{equation}
The matching calculation is done to one loop using massless HISQ lattice perturbation theory.  Details of the calculation, and values for the matching coefficients, are given in~\cite{Monahan:2013}.  
In matching the temporal component of the vector current we omit $\mathcal{O}\left( \nicefrac{ \alpha_s\Lambda_{\rm QCD}}{m_b} \right)$ contributions specified in~\cite{Monahan:2013}.  To justify their omission, we generated data for these terms and verified their contributions are sub-percent, consistent with the findings of~\cite{Gulez:2007}.

The continuum tensor current $\langle T_{k0}\rangle$ is matched to the lattice current by
\begin{equation}
\langle T_{k0} \rangle = ( 1 + \alpha_s \rho_0^{(T)} ) \langle \mathcal{T}_{k0}^{(0)} \rangle + \langle \mathcal{T}_{k0}^{(1), {\rm sub}} \rangle ,
\label{eq-Tmatch}
\end{equation}
where
\begin{equation}
\langle \mathcal{T}_{k0}^{(1), {\rm sub}} \rangle = \langle \mathcal{T}_{k0}^{(1)} \rangle - \alpha_s \zeta_{10}^T \langle \mathcal{T}_{k0}^{(0)} \rangle.
\end{equation}
As mentioned above, heavy-quark symmetry of the NRQCD $b$ quark allows the tensor current renormalization to be recast in terms of vector current quantities:  $\mathcal{T}_{k0}^{(0)} =  \mathcal{V}_k^{(0)}$, $\mathcal{T}_{k0}^{(1)} = -  \mathcal{V}_k^{(1)}$, and $\zeta_{10}^T = - \zeta_{10}^{V_k}$.

\section{ Extracting Matrix Elements }
Hadronic matrix elements are extracted from fits to two and three point correlator data using Bayesian fitting techniques \cite{Lepage:2002}.

\subsection{$B$ Meson Two Point Fits}
\begin{figure}[t!]
\vspace{0.0in}
{\scalebox{0.9}{\includegraphics[angle=-90,width=0.5\textwidth]{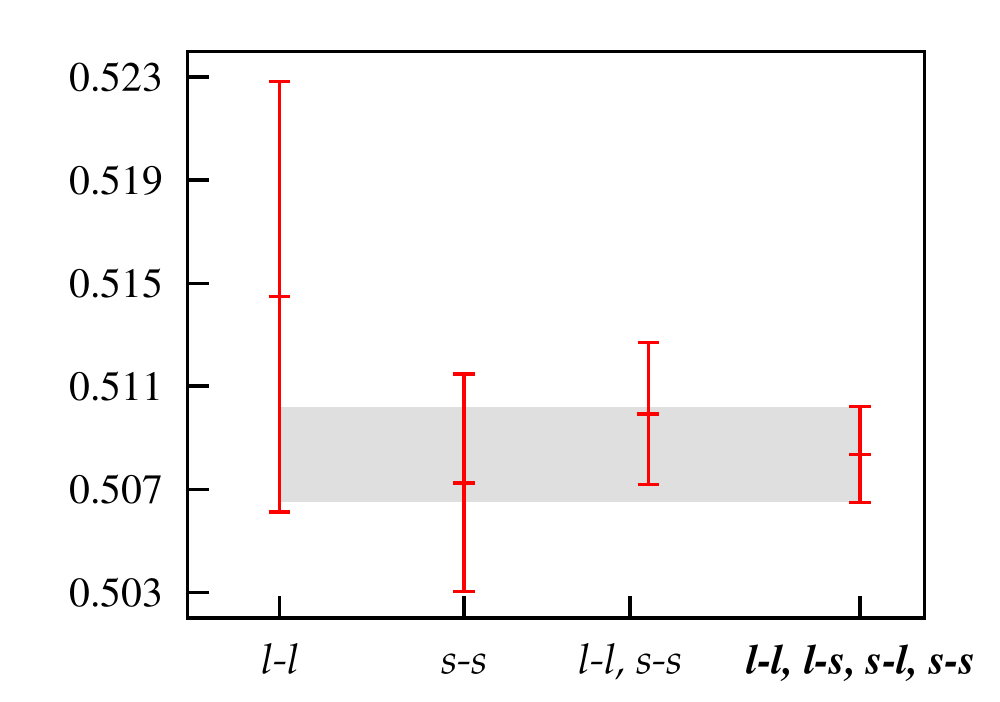}}}
\caption{Ensemble C2 fit results to the ground state energy $aE_B^{{\rm sim}(0)}$ for various source-sink smearing combinations.  There is a factor of roughly four improvement from simultaneous fits to local (``$l$'') and smeared (``$s$'') sources and sinks.  The shaded band shows the best-fit result using all smearing combinations.}
\vspace{0.0in}
\label{fig-Bsmear}
\end{figure}
Two point correlator data for $B$ mesons are fit to the ansatz
\begin{equation}
C^{\alpha \beta}_B(t) = \sum^{2N-1}_{n=0} b^{\alpha (n)} b^{\beta (n) \dagger} (-1)^{nt} e^{-E_B^{{\rm sim}(n)}t},
\end{equation}
where
\begin{equation}
b^{\alpha(n)} = \frac{ a^3\langle \Phi_B^\alpha | B^{(n)} \rangle }{ \sqrt{2 a^3E_B^{(n)}} }.
\end{equation}
We have studied fits to all possible combinations of local and smeared data and found a simultaneous fit to all four combinations yields the smallest errors.  Fig.~\ref{fig-Bsmear} shows the improvement achieved from the simultaneous fit for data on a coarse ensemble and is indicative of the improvement seen on all ensembles.

\begin{figure}[t!]
\vspace{0.0in}
{\scalebox{0.9}{\includegraphics[angle=-90,width=0.5\textwidth]{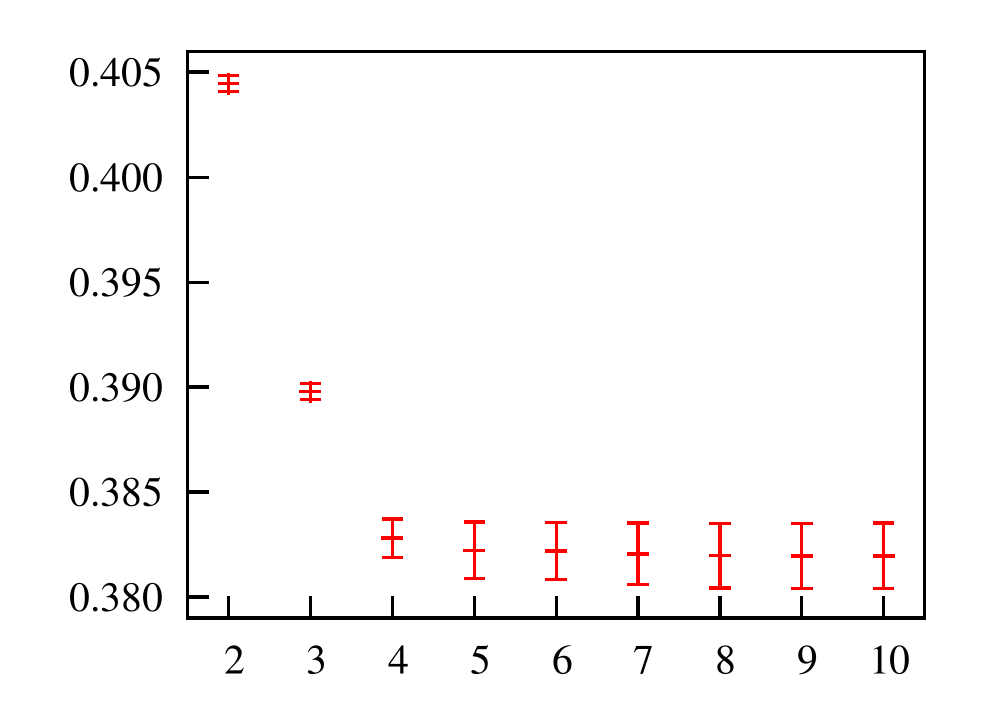}}}
\caption{\label{fig-BfitsvN}Ensemble F1 fit results for the ground state energy $aE_B^{{\rm sim}(0)}$ vs. the number of states $N$ with $\nicefrac{t_{\rm min}}{a}=2$ and $\nicefrac{t_{\rm max}}{a}=25$.}
\end{figure}

To ensure excited state contributions are adequately accounted for, we increase the number of exponentials $N$ in the fit ansatz until both the central values and errors stabilize.
This increase in $N$ is balanced by the practical constraint that additional exponentials add to the complexity of the fit ansatz and increase the time required for convergence.
Fig.~\ref{fig-BfitsvN} plots fit results on fine ensemble F1 as a function of $N$ and suggests that $N \geq 5$ is sufficient.  
Similar behavior is seen on the other ensembles and we report results for $N=8$.

\begin{figure}[t!]
{\scalebox{0.9}{\includegraphics[angle=-90,width=0.5\textwidth]{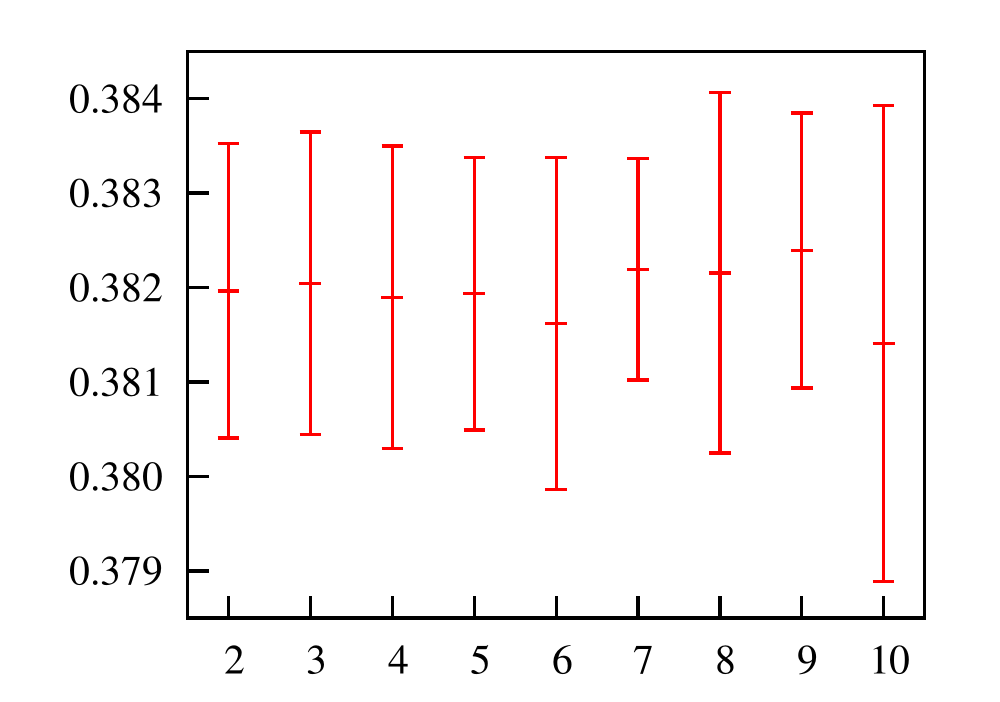}}}
\\
{\scalebox{0.9}{\includegraphics[angle=-90,width=0.5\textwidth]{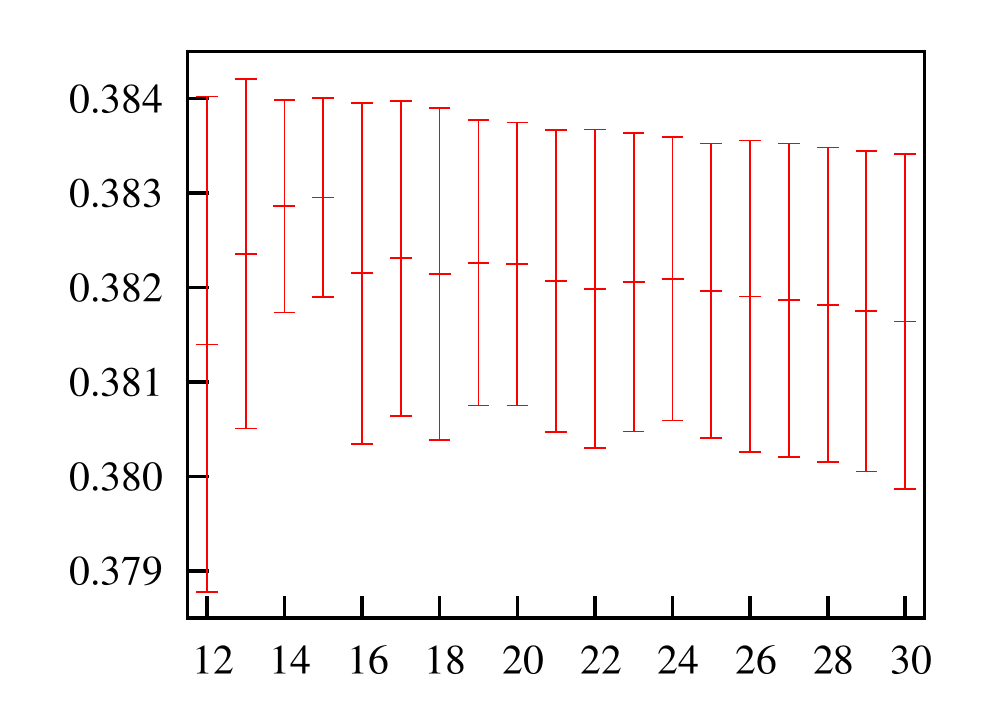}}}
\caption{Ensemble F1 fit results for the ground state energy $aE_B^{{\rm sim}(0)}$ vs. $\nicefrac{t_{\rm min}}{a}$ with $\nicefrac{t_{\rm max}}{a}=25$ ({\it top}) and vs. $\nicefrac{t_{\rm max}}{a}$ with $\nicefrac{t_{\rm min}}{a}=2$ ({\it bottom}).  All fits use $N=8$.}
\label{fig-Bfitsvdata}
\end{figure}
In addition, we have studied the impact on our fit results of varying the timeslices of data included in the fit.  For a given fit, the timeslices of data included are specified by the range $t_{\rm min} \leq t \leq t_{\rm max}$.  
An exponential decrease in signal to noise, especially prevalent in $B$ meson data, leads to greater precision data at smaller values of $t$, while data at larger values of $t$ have a greater relative amount of information on the ground state - the information we are after.  
Data at very large $t$ are redundant, as is evident from Fig.~\ref{fig-Bfitsvdata}, and omitting them speeds up the fit.
In practice, we study fits on each ensemble to determine values of $t_{\rm min}$ and $t_{\rm max}$ for which the fit results are stable.  Fig.~\ref{fig-Bfitsvdata} illustrates the ranges of $t_{\rm min}$ and $t_{\rm max}$ considered on the fine ensembles and demonstrates that fit results are largely insensitive to reasonable changes to the data included in the fit.

We choose our $B$ meson best-ft results from simultaneous fits to all four combinations of local and smeared data.  On the coarse ensembles (C1, C2, and C3) we fit these data on timeslices $2 \leq \nicefrac{t}{a} \leq 19$ and with $N=8$.  On the fine ensembles (F1 and F2) we fit data on timeslices $2 \leq \nicefrac{t}{a} \leq 25$ and with $N=8$.  Prior choices and $B$ meson fit results are given in Appendix~\ref{app-Bpriors}.

On each ensemble the fitted $B$ meson energy $E_B^{\rm sim}$ is related to the physical $B$ meson mass by a shift associated with the NRQCD shift in the $b$ quark rest mass
\begin{equation}
M_{B} = E^{{\rm sim}(0)}_{B} + \frac{1}{2}\left( M_{b\bar{b}}^{\rm{expt}} - E_{b\bar{b}}^{\rm{sim}} \right),
\label{eq-PhysMB}
\end{equation}
where $M_{b\bar{b}}^{\rm{expt}} = 9.450(4)$ GeV~\cite{Gregory:2010} is adjusted from experiment to remove electromagnetic, $\eta_b$ annihilation, and charmed sea effects not present in our simulations, and $E_{b\bar{b}}^{\rm{sim}}$ is the spin-averaged energy of $b\bar{b}$ states calculated on the lattice ensembles used in our simulation.

As a byproduct of this analysis we obtain the leading order contribution to the $B$ meson decay constant, $\Phi = F_B \sqrt{M_B}$.  Though not particularly useful by itself, a comparison with previous results using the same ensembles provides a cross-check of our $B$ meson fit results.  Accounting for a numerical factor of $\sqrt{2}$ from the HISQ inversion, the unsmeared amplitude from this analysis is related to the leading order decay constant by
\begin{equation}
a^{3/2} \Phi^{(0)} = 2 b^{l(0)}.
\end{equation}
We calculated values for $\Phi^{(0)}$ on each ensemble and verified our results are consistent with previous work~\cite{Na:2012}.

\subsection{Kaon Two Point Fits}
For each simulated momentum, two point correlator data for the kaons are fit to
\begin{equation}
C_K(t;{\bf p}) = \sum^{2N-1}_{n=0} \big| d_{\bf p}^{(n)} \big|^2 (-1)^{nt} \left( e^{-E_K^{(n)}t} + e^{-E_K^{(n)}(N_t-t)} \right),
\label{eq-fitK}
\end{equation}
where
\begin{equation}
d_{\bf p}^{(n)} = \frac{ a^3\langle \Phi_K | K_{\bf p}^{(n)} \rangle }{ \sqrt{2 a^3E_K^{(n)}} }.
\end{equation}
As with the $B$ meson, we studied the effect of varying the number of exponentials $N$ used in the fit ansatz and the range of data included in the fits and found qualitatively similar results to those of Figs.~\ref{fig-BfitsvN} and~\ref{fig-Bfitsvdata}.  Coarse ensemble results are reported for fits using data at timeslices $2 \leq \nicefrac{t}{a} \leq 30$, while on the fine ensembles we fit data at timeslices $3 \leq \nicefrac{t}{a} \leq 40$.  For all fits we use $N=8$.  

\begin{figure}[t]
\vspace{0.0in}
\centering
{\scalebox{0.9}{\includegraphics[angle=-90,width=0.5\textwidth]{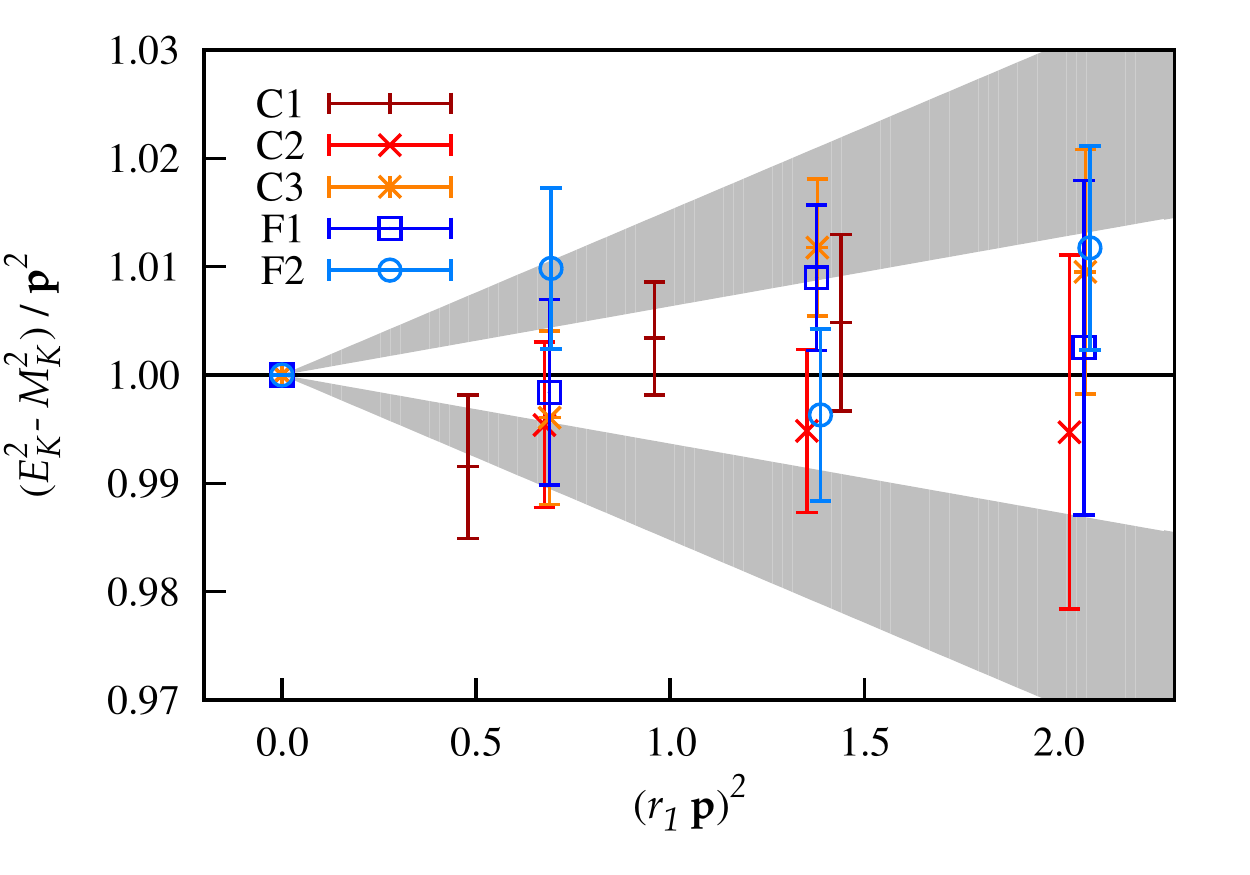}}}
\vspace{0.0in}
\caption{Fit results for $E_K$ and $M_K$, combined with simulated kaon momenta, satisfy the dispersion relation.   Observed discretization errors, over the range of lattice spacings used in the simulations, are represented by the gray bands and correspond to $1\pm \nicefrac{1}{3}\ \alpha_s (a {\bf p})^2$.}
\label{fig-dispreln}
\end{figure}

Kaon mass and energy fit results satisfy the dispersion relation as shown in Fig.~\ref{fig-dispreln}.  We observe an empirical factor of three improvement in the dispersion relation relative to the expected $\mathcal{O}(\alpha_s (a {\bf p})^2 )$.  Prior choices and kaon fit results are given in Appendix~\ref{app-Kpriors}.

As a consistency check, we extract kaon decay constant values on each ensemble using 
\begin{equation}
aF_K = \frac{am_l^{val} + am_s^{val} }{ \big(aM_K\big)^{\nicefrac{3}{2}}} \sqrt{2}\ d^{(0)}_{000}\ ,
\end{equation}
and compare with those obtained using the same ensembles in Ref.~\cite{Na:2010}.

\subsection{Simultaneous Two and Three Point Fits}
For each momentum, three point correlator data for current $J$ and $b$ quark smearing $\alpha$ are fit to
\begin{eqnarray}
C^\alpha_{J({\bf p})}(t,T) &=& \sum_{m,n=0}^{2N-1} d_{\bf p}^{(n)} A^{(n,m)}_{J({\bf p})} b^{\alpha(m) \dagger}  (-1)^{mt+n(T-t)} \nonumber \\
& &\times \  e^{-E_K^{(n)}(T-t)}\ e^{-E_B^{{\rm sim}(m)}t}\ ,
\end{eqnarray}
where the three point amplitude is related to the lattice hadronic matrix element by
\begin{equation}
\frac{4}{\sqrt{2}} A^{(n,m)}_{J({\bf p})} = \frac{a^3 \langle K_{\bf p}^{(n)} | J | B^{(m)} \rangle}{\sqrt{ 2a^3E_K^{(m)} }\sqrt{ 2a^3E_B^{(n)} } }.
\label{eq-lattJ}
\end{equation}
The factor of $\nicefrac{4}{\sqrt{2}}$ accounts for numerical factors introduced in the simulation associated with taste averaging and HISQ inversion.

\begin{figure}[t!]
{\scalebox{0.9}{\includegraphics[angle=-90,width=0.5\textwidth]{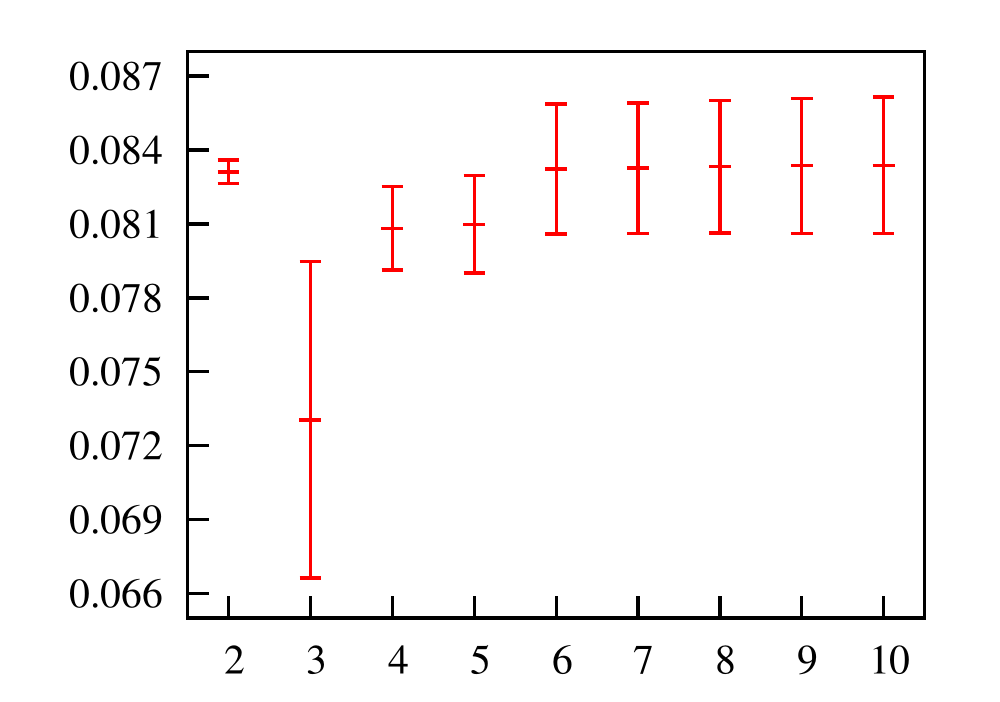}}}
\caption{Ensemble F2 fit results for the ground state amplitude $A^{(0,0)}_{V_k(111)}$ vs. the number of states $N$.}
\label{fig-3ptvN}
\end{figure}

In the fits, we include three point data at timeslices $t_c \leq t \leq T-t_c$, with $t_c=2$, and have verified that fit results are insensitive to small variations, {\it ie.} $t_c = 3,4$.  As with two point correlator fits, the number of exponentials used in the fit ansatz is increased until the central values and errors stabilize, see Fig.~\ref{fig-3ptvN}.  We use $N=8$ in all simultaneous two and three point correlator fits.  Three point data are generated for local and smeared $b$ quarks and both data sets are included in the fits.

The leading order and $\nicefrac{1}{m_b}$ correction three point correlator data can be combined, per the matching prescription of Sec.~\ref{sec-match}, before extracting three point amplitudes via correlator fits.  The resulting hadronic matrix elements are then $\langle V_\mu \rangle$ of Eq.~(\ref{eq-Vmatch}) and $\langle T_{k0} \rangle$ of Eq.~(\ref{eq-Tmatch}).  We have verified that results obtained this way are equivalent to those obtained by separately fitting the leading order and $\nicefrac{1}{m_b}$ correction data, to extract $\langle \mathcal{V}_\mu^{(0,1)}\rangle$ and $\langle \mathcal{T}_{k0}^{(0,1)}\rangle$, and then combining these results per the matching prescription.  We adopt the method of first combining the data as fewer fits are required and correlations between the leading order and $\nicefrac{1}{m_b}$ correction data are automatically accounted for.

\begin{figure}[t]
\vspace{-0.1in}
{\scalebox{1.0}{\includegraphics[angle=-90,width=0.5\textwidth]{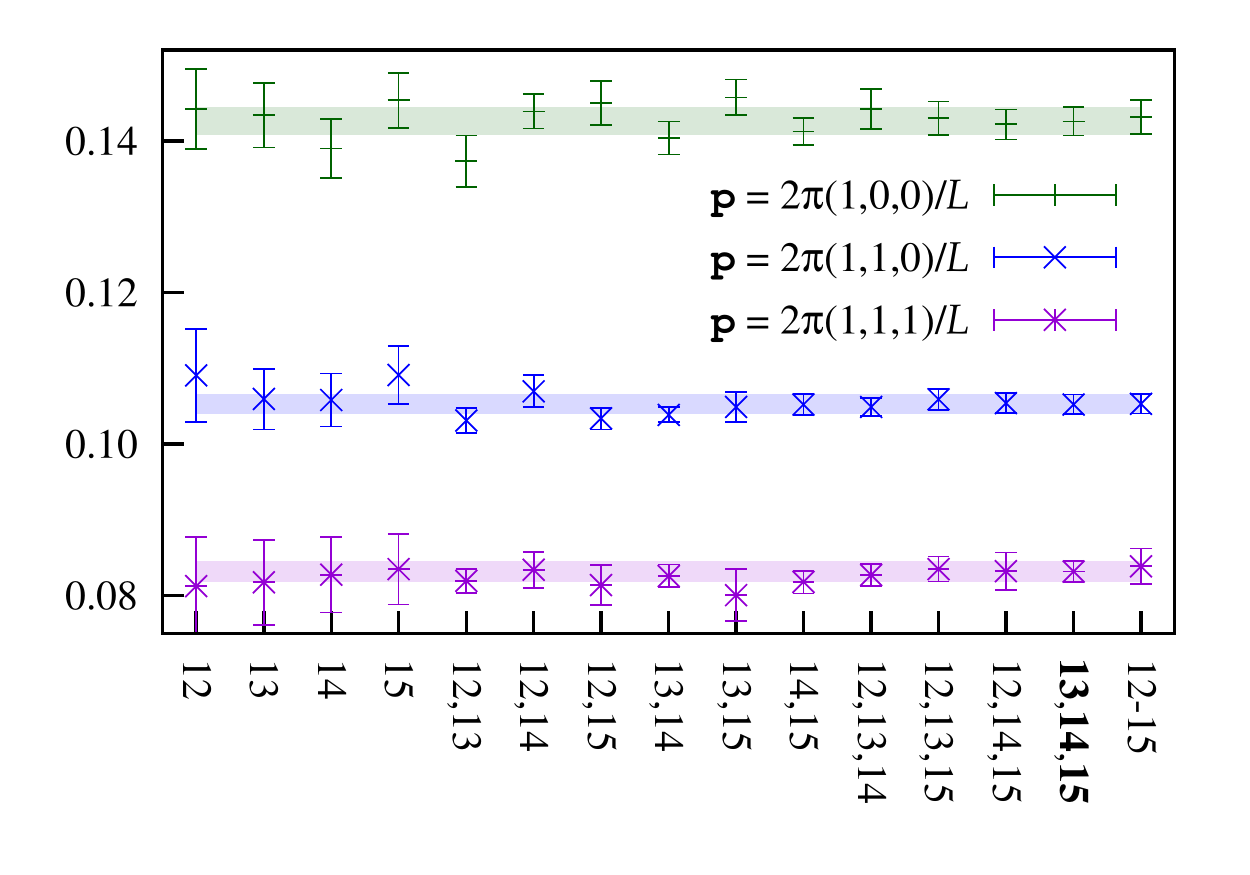}}}
\vspace{-0.2in}
\caption{Ensemble C3 fit results for $a\langle V_k \rangle$ (defined in Sec.~\protect\ref{sec-match}) are shown for combinations of $T$ used in simultaneous fits.   Colored bands correspond to the ``best-fit'' combination $T=13, 14, 15$.}
\label{fig-3ptfits}
\end{figure}

The amplitude $A_J$ is extracted from a simultaneous fit to two and three point data.  
Including three point data at multiple separation times $T$ improves the precision of extracted matrix elements, as shown in Fig.~\ref{fig-3ptfits}.  
By studying fit results for $\langle V_0\rangle$, $\langle V_k \rangle$, and $\langle T_{k0} \rangle$ for various $T$ combinations, we select a best-fit combination of $T$'s to include in the fit for each ensemble.  
Our choices, listed in Table~\ref{tab-Tcombos}, strike a balance between improved fit precision and the complexity of the fits and time required for convergence.
\begin{table}[t]
\begin{tabular}{ccc}
\hline\hline	
	\T ens   	& & $T$'s    \\ [0.5ex]
	\hline
	\T C1 	& & 13, 14, 15	\\
	\T C2	& & 13, 14, 15	\\
	\T C3	& & 13, 14, 15	\\
	\T F1		& & 23, 24	\\
	\T F2		& & 21, 22, 24	\\ [0.5ex]
\hline\hline
\end{tabular}\caption{Best-fit combinations of $T$'s included in simultaneous fits on each ensemble.}
\vspace{-0.1in}
\label{tab-Tcombos}
\end{table}

For each ensemble we performed simultaneous fits including the four smearing combinations of $B$ meson two point data, both smearing combinations of three point data, and the combination of three point data $T$'s listed in Table~\ref{tab-Tcombos}.  
We considered three variations of this fit based on combinations of data for different momenta and currents.  In order of increasing complexity, we performed:
\begin{description}
\item[fit1:]  separate fits to each current and each momentum, 
\item[fit2:]  one fit to each current, including all momenta, and
\item[fit3:]  one fit including all currents and all momenta.
\end{description}
Results for the ground state three point amplitudes for each fit, and priors used, are listed in Table~\ref{tab-3ptfits}.  
The central values vary between the different fits but the differences are not statistically significant.  
Fit results for the more complicated fits, fit2 and fit3, generally have larger errors.
We take fit1 results, given in Table~\ref{tab-f0f+fT}, for use in the subsequent chiral/continuum and kinematic extrapolation.  For each ensemble we extract the full correlation matrix for all form factors and momenta from fit3 and combine it with fit1 errors to build the full covariance matrix.


Results for $\langle V_\mu \rangle$ are converted to $f_{\parallel,\perp}$ and then to $f_{0,+}$ via Eqs.~(\ref{eq-fpardef}, \ref{eq-fperpdef}, \ref{eq-f0def}, \ref{eq-fplusdef}).  Results for $\langle T_{k0} \rangle$ are converted to $f_T$ using Eq.~(\ref{eq-fTdef}).  Results for $f_0$, $f_+$, and $f_T$ are listed in Table~\ref{tab-f0f+fT}.
\setlength{\tabcolsep}{0.06in}
\begin{table}[t]
\begin{tabular}{ccccc}
\hline\hline
	\T ens   	& $f_0(0,0,0)$	& $f_0(1,0,0)$	& $f_0(1,1,0)$	& $f_0(1,1,1)$ \\ [0.5ex]
	\hline
	\T C1 	& 0.8477(74)	& 0.7449(70)	& 0.6878(70)	& 0.6464(98)	\\
	\T C2	& 0.8518(90) 	& 0.7199(70)	& 0.6484(49)	& 0.6027(68)	\\
	\T C3	& 0.8338(65) 	& 0.7159(59)	& 0.6513(38)	& 0.6012(54)	\\
	\T F1		& 0.8396(51) 	& 0.7158(50)	& 0.6502(42)	& 0.582(17)	\\
	\T F2		& 0.8356(46) 	& 0.7159(44)	& 0.6397(52)	& 0.5987(56)	\\ [0.5ex]
\\ [-2.5ex]
	\T ens   	&& $f_+(1,0,0)$& $f_+(1,1,0)$	& $f_+(1,1,1)$ \\ [0.5ex]
	\hline
	\T C1 	&& 1.982(28)	& 1.626(19)	& 1.380(17)	\\
	\T C2	&& 1.827(26) 	& 1.423(13)	& 1.199(19)	\\
	\T C3	&& 1.748(18) 	& 1.416(14)	& 1.197(15)	\\
	\T F1		&& 1.784(27) 	& 1.427(28)	& 1.168(29)	\\
	\T F2		&& 1.805(28) 	& 1.138(19)	& 1.191(29)	\\ [0.5ex]
\\ [-2.5ex]
	\T ens   	&& $f_T(1,0,0)$& $f_T(1,1,0)$	& $f_T(1,1,1)$ \\ [0.5ex]
	\hline
	\T C1 	&& 1.706(25)	& 1.422(19)	& 1.220(24)	\\
	\T C2	&& 1.607(29) 	& 1.236(14)	& 1.053(23)	\\
	\T C3	&& 1.555(22) 	& 1.272(17)	& 1.083(23)	\\
	\T F1		&& 1.615(31) 	& 1.335(65)	& 1.046(31)	\\
	\T F2		&& 1.667(32) 	& 1.267(26)	& 1.090(34)	\\ [0.5ex]
\hline\hline
\end{tabular}\caption{Fit1 results for the scalar, vector, and tensor form factors on each ensemble and for each simulated momentum.}
\label{tab-f0f+fT}
\end{table}

\section{ Chiral/Continuum and Kinematic Extrapolation }
\label{sec-2step}

We perform the chiral/continuum and kinematic extrapolations in separate steps.  We use the results of the chiral/continuum extrapolation to generate a synthetic data set to guide a subsequent kinematic extrapolation.  

\subsubsection{Chiral/Continuum Extrapolation}
\label{sec-ChPT}
\begin{figure*}[t!]
\hspace{-0.07in}
\subfloat[][\label{fig-modzf0fpC123}Fit to data for $f_{0,+}$ on the coarse ensembles.]
{\scalebox{0.98}{\includegraphics[angle=-90,width=0.5\textwidth]{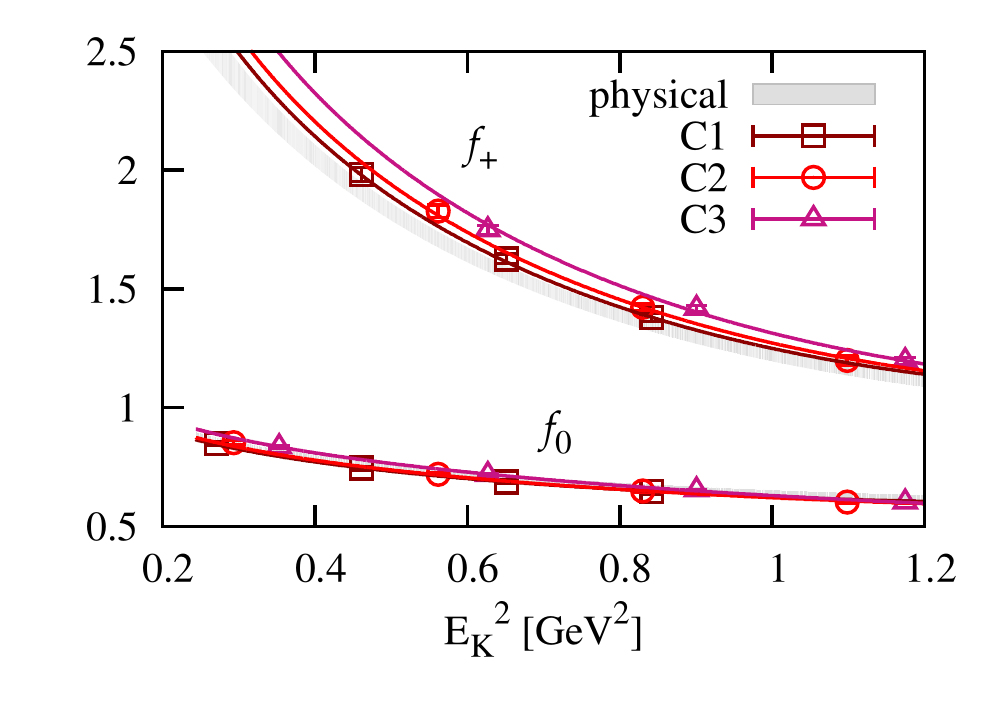}}}
\subfloat[][\label{fig-modzifTC123}Fit to data for $f_T$ on the coarse ensembles.]
{\scalebox{0.98}{\includegraphics[angle=-90,width=0.5\textwidth]{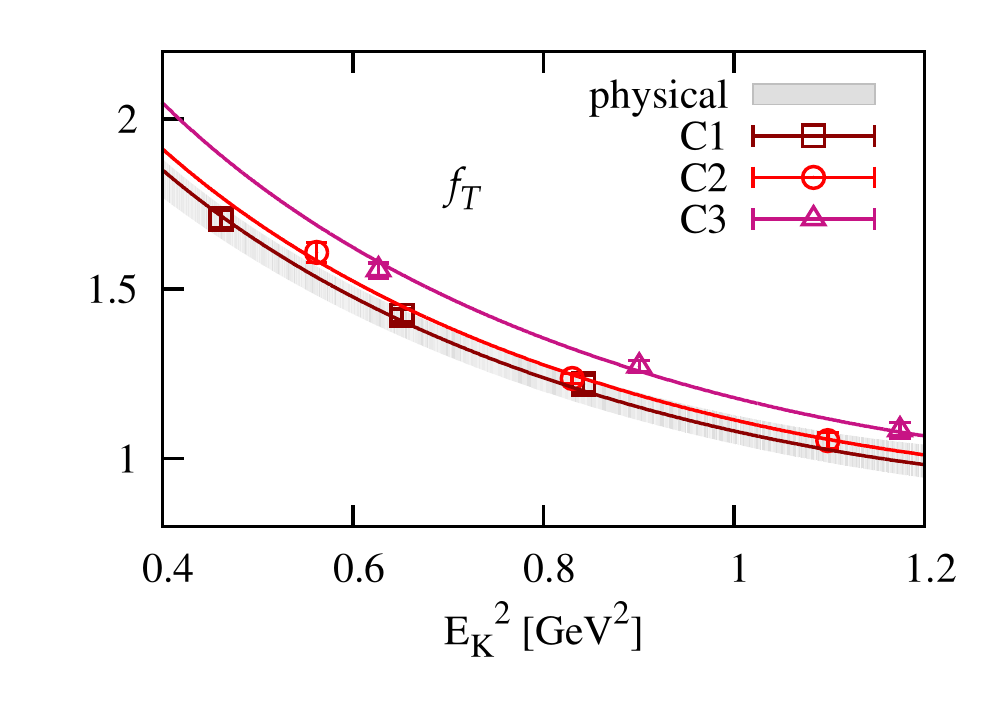}}}
\\
\subfloat[][\label{fig-modzf0fpF12}Fit to data for $f_{0,+}$ on the fine ensembles.]
{\scalebox{0.98}{\includegraphics[angle=-90,width=0.5\textwidth]{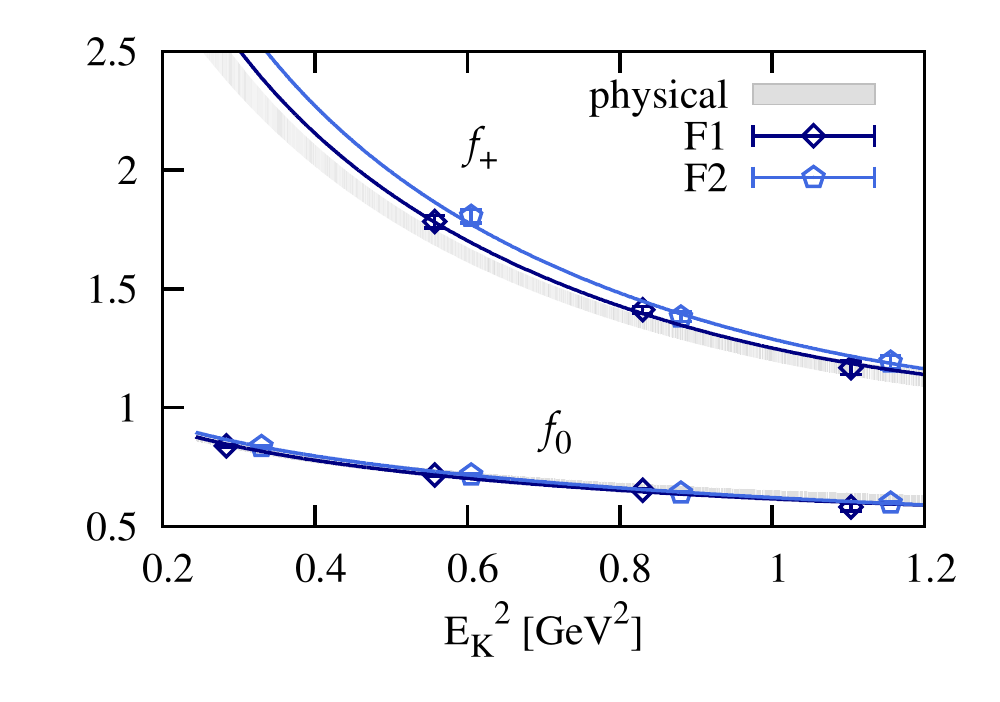}}}
\subfloat[][\label{fig-modzifTF12}Fit to data for $f_T$ on the fine ensembles.]
{\scalebox{0.98}{\includegraphics[angle=-90,width=0.5\textwidth]{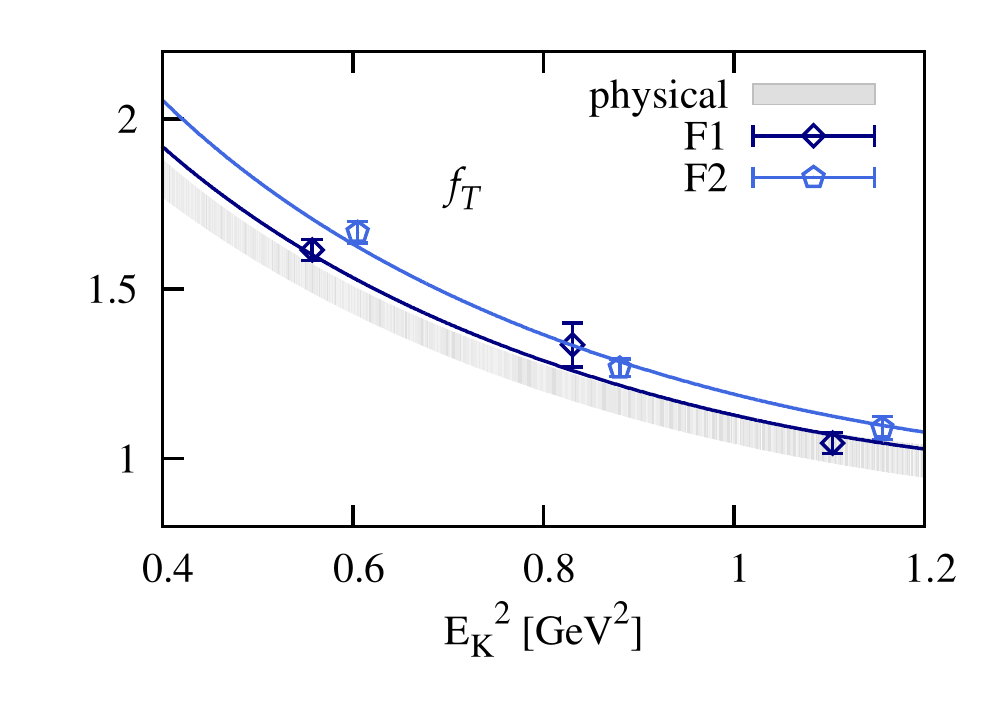}}}
\vspace{0.1in}
\caption{Simultaneous chiral/continuum extrapolation to $f_{0,+,T}$ data.  In each plot, curves indicate the fit to data on each ensemble and the gray band shows the extrapolated physical result.}
\label{fig-chpt}
\end{figure*}

We perform the chiral/continuum extrapolation on the data for $f_0$, $f_+$, and $f_T$ in Table~\ref{tab-f0f+fT}.  For $f_0$ and $f_+$ this is accomplished using fit ans\"atze for $f_{\parallel}$ and $f_{\perp}$ based on the partially quenched staggered chiral perturbation theory of Ref.~\cite{Aubin:2007}, with taste-breaking effects turned off, and Eqs.~(\ref{eq-f0def}, \ref{eq-fplusdef}).
Ref.~\cite{Gulez:2007} studied $B$ semileptonic decays using NRQCD $b$ and asqtad light valence and sea quarks.  There, the chiral/continuum extrapolation was performed using ChPT with and without the staggered taste breaking terms.  Negligible differences were found in the results from these two approaches.  The use of HISQ valence light quarks is known~\cite{MILC:2010} to reduce the already small taste breaking effects seen in~\cite{Gulez:2007}.  We therefore omit the staggered ChPT taste breaking terms in favor of generic, light quark mass-dependent discretization effects as described below.
For $f_T$ we use the fact that, at leading order in $\nicefrac{1}{m_b}$, and for the large values of $q^2$ at which we simulate and extrapolate in this step, $f_T \approx f_+$.  Furthermore, the shape of $f_+$ at large $q^2$ is driven by $f_\perp$.  We therefore use a fit ansatz for $f_T$ that has the same form as that for $f_\perp$ and uses the same chiral logs.
The fit ans\"atze for $f_\parallel$, $f_\perp$, and $f_T$ are
\begin{eqnarray}
f_\parallel &=& \frac{\kappa}{F_\pi}\left( 1 + \delta f_\parallel + A_\parallel + H_\parallel \right) D_\parallel, \label{eq-fpar} \\
f_\perp &=& \frac{\kappa\ g/F_\pi}{E_K + \Delta^* + \delta f_\perp^{(1)}} \left( 1 + \delta f_\perp^{(2)} + A_\perp + H_\perp \right) D_\perp, \label{eq-fperp} \nonumber \\
& & \\
f_T &=& \frac{\kappa_T\ g/F_\pi}{E_K + \Delta^* + \delta f_\perp^{(1)}} \left( 1 + \delta f_\perp^{(2)} + A_T + H_T \right) D_T, \label{eq-fT} \nonumber \\
& &
\end{eqnarray}
where:  $\kappa$ and $\kappa_T$ are leading order low energy constants; $g$ is the $BB^*\pi$ coupling; $\Delta^*$ is the $B_s^*-B$ splitting; the $\delta f$'s represent next-to-leading order (NLO) chiral logs from~\cite{Aubin:2007}; $A$ is a collection of NLO and next-to-next-to-leading order (NNLO) chiral analytic terms; $H$ is a polynomial function of the kaon energy; and $D$ contains discretization effects.  Explicit expressions for the functions $A$, $H$, and $D$ are
\begin{eqnarray}
A &=& a_1 \frac{m_l}{m_c} + a_2 \frac{m_s}{m_c} + a_3 \frac{2 \tilde{m}_l + \tilde{m}_s}{\tilde{m}_c} \nonumber \\
& + & a_4 \left( \frac{m_l}{m_c} \right)^2 + a_5  \left( \frac{m_s}{m_c} \right)^2 + a_6 \left( \frac{2 \tilde{m}_l + \tilde{m}_s}{\tilde{m}_c} \right)^2 \nonumber \\
& + & a_7 \frac{m_l m_s}{m_c^2} + a_8 \frac{m_l(2\tilde{m}_l + \tilde{m}_s)}{m_c \tilde{m}_c} \nonumber \\
& + & a_9 \frac{m_s(2\tilde{m}_l + \tilde{m}_s)}{m_c \tilde{m}_c} + a_{10} \frac{m_l}{m_c} E_K, \label{eq-Adef} \\
H &=& h_1 E_K + h_2 E_K^2 + h_3 E_K^3, \label{eq-Hdef} \\
D &=& 1 + d_1 \left( \nicefrac{a}{r_1} \right)^2 + d_2 \left( \nicefrac{a}{r_1} \right)^4. \label{eq-Ddef}
\end{eqnarray}
Here $A$, $H$, $D$, and the coefficients $a_i$, $h_i$, and $d_i$, have implicit indices $\parallel$, $\perp$, or $T$ specifying the relevant form factor.
We use powers of bare HISQ ($m_{l,s}/m_c$) and asqtad ($\tilde{m}_{l,s}/\tilde{m}_c$) mass ratios for the analytic terms in $A$, with values for the HISQ and asqtad charm quark masses taken from~\cite{Na:2012}.  These quark mass ratios are well defined in the physical limit.  
The NLO valence mass coefficients for $f_\parallel$ and $f_\perp$ are related by $a_1^{\parallel} + a_1^{\perp} = a_2^{\parallel} + a_2^{\perp}$~\cite{Aubin:2007}.  
To obtain acceptable $\chi^2$ and stable fit results, we found it necessary to include NNLO terms proportional to $m_l E_K$ and $E_K^3$.  We note these terms were found to be necessary in chiral/continuum extrapolations for $B\to \pi$ semileptonic data in~\cite{Bailey:2009}.  
\begin{figure*}[t!]
\hspace{-0.07in}  
\subfloat[][\label{fig-modzf0fpC123}Fit to data for $f_{0,+}$ on the coarse ensembles.]
{\scalebox{0.98}{\includegraphics[angle=-90,width=0.5\textwidth]{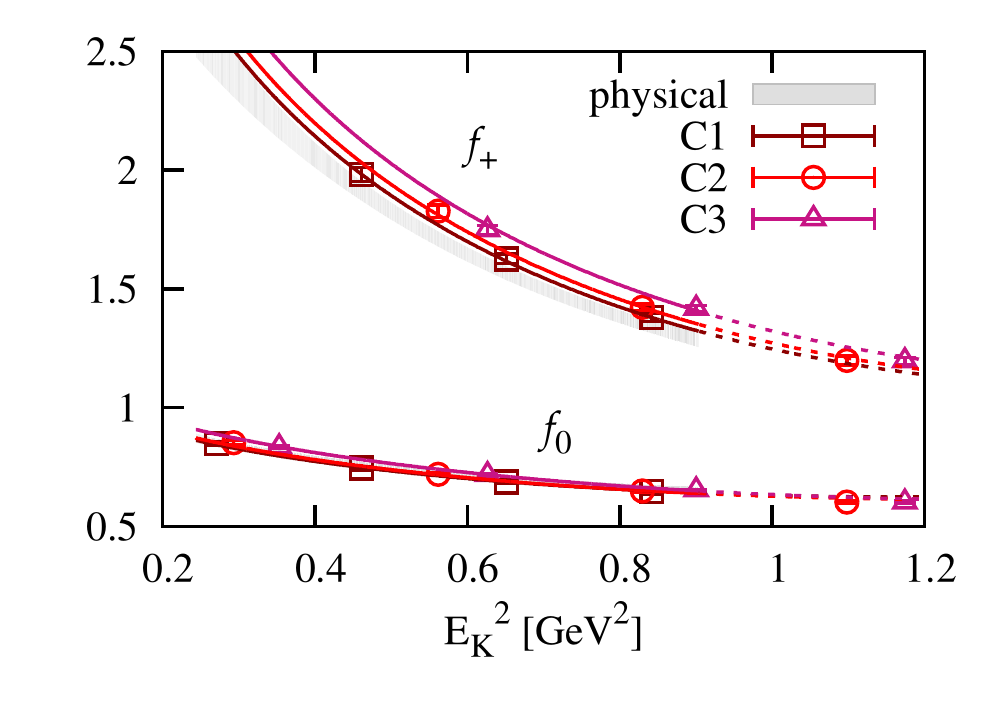}}}
\subfloat[][\label{fig-modzifTC123}Fit to data for $f_T$ on the coarse ensembles.]
{\scalebox{0.98}{\includegraphics[angle=-90,width=0.5\textwidth]{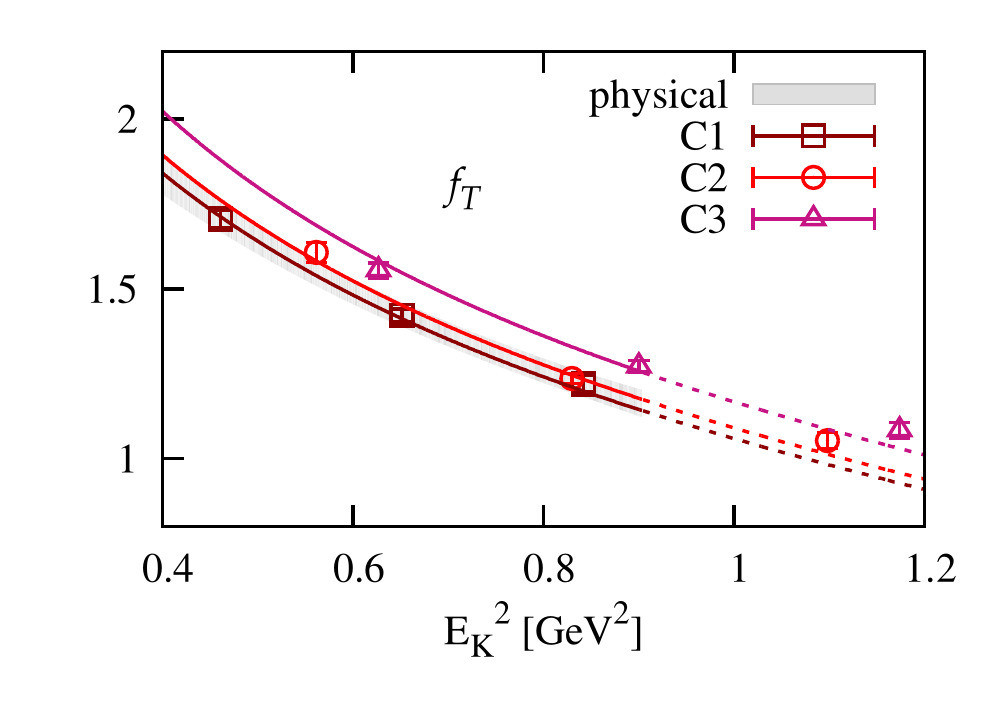}}}
\\
\subfloat[][\label{fig-modzf0fpF12}Fit to data for $f_{0,+}$ on the fine ensembles.]
{\scalebox{0.98}{\includegraphics[angle=-90,width=0.5\textwidth]{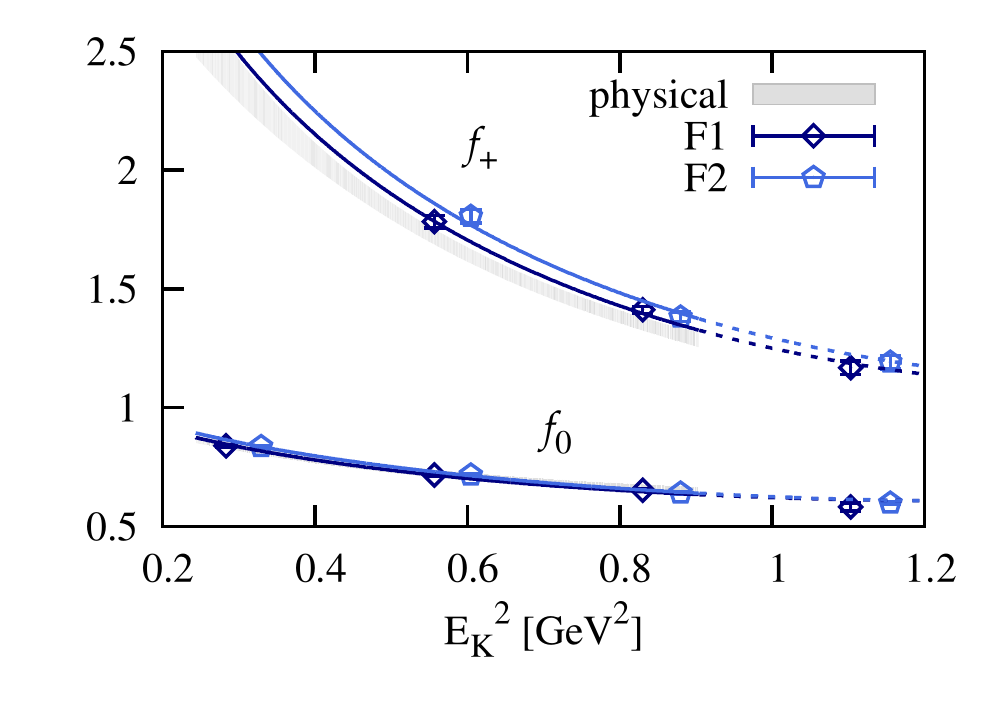}}}
\subfloat[][\label{fig-modzifTF12}Fit to data for $f_T$ on the fine ensembles.]
{\scalebox{0.98}{\includegraphics[angle=-90,width=0.5\textwidth]{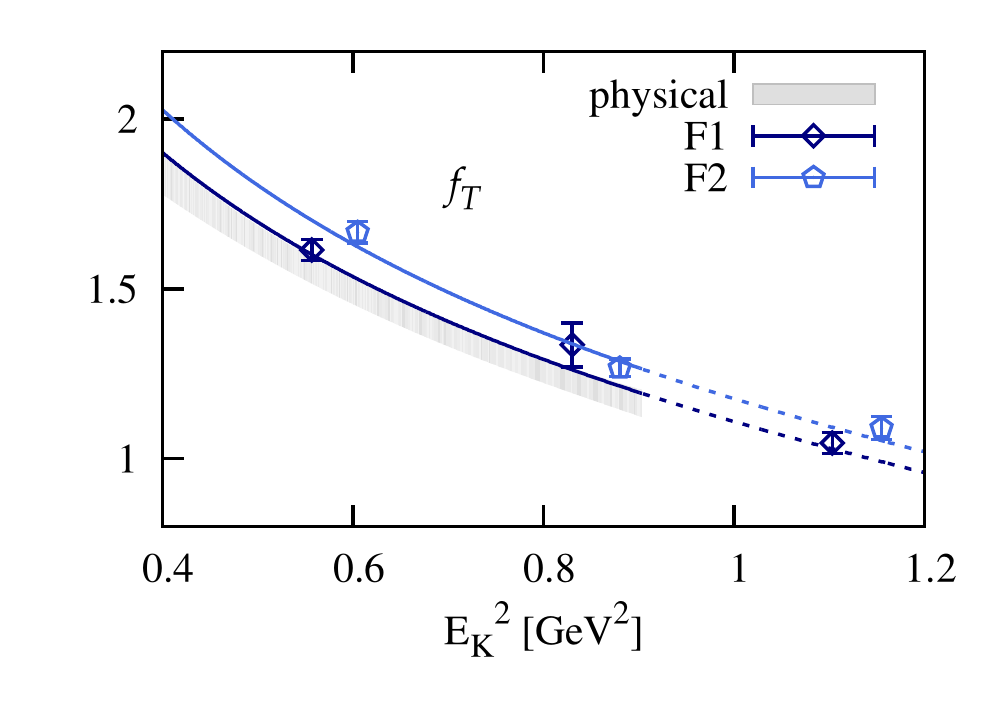}}}
\vspace{0.1in}
\caption{Simultaneous chiral/continuum extrapolation to $f_{0,+,T}$ data with $E_K < 1\ {\rm GeV}$.  Ensemble fit curves are extended beyond the fitted region (dashed lines) to demonstrate the level of agreement with the $\nicefrac{2\pi}{L}(1,1,1)$ data.  The error bands (not shown) for the ensemble curves are comparable in width to the physical band.}
\label{fig-chpt_sans111}
\end{figure*}

As in~\cite{Gregory:2010} we account for heavy-quark discretization effects by making the coefficients $d_i$ mild functions of $am_b$,
\begin{eqnarray}
  d_1 &\to& d_1( 1+ f_1\, \delta x_b + f_2\, \delta x_b^2\, ), \nonumber \\
  d_2 &\to& d_2( 1+ f_3\, \delta x_b + f_4\, \delta x_b^2\, ),
  \label{eq-hvy_disc}
\end{eqnarray}
where $\delta x_b = am_b - 2.26$, chosen so that as $am_b$ varies over the coarse and fine ensembles, $-0.4~\lesssim~\delta x_b~\lesssim~0.4$.  
Similarly, we allow for light-quark mass-dependent discretization effects by making the $d_i$ mild functions of $m_l$,
\begin{eqnarray}
  d_1 &\to& d_1\Big( 1+ g_1 \frac{m_l}{m_c} + g_2 \Big( \frac{m_l}{m_c}\Big)^2\ \Big), \nonumber \\
  d_2 &\to& d_2\Big( 1+ g_3 \frac{m_l}{m_c} + g_4 \Big(\frac{m_l}{m_c}\Big)^2\ \Big).
  \label{eq-lt_disc}
\end{eqnarray}
As a result of these modifications, $d_1$ in Eq.~(\ref{eq-Ddef}) is multiplied by $( 1+ f_1 \delta x_b + f_2 \delta x_b^2 )( 1+ g_1 \frac{m_l}{m_c} + g_2 (\frac{m_l}{m_c})^2 )$, and similarly for $d_2$.
For the chiral logs we follow~\cite{Aubin:2007} but turn off taste-breaking effects.  To the extent these effects are present in our data, they should be accommodated by the light-quark mass-dependent discretization terms introduced in Eq.~(\ref{eq-lt_disc}).  As discussed below, we find negligible contribution from these effects in our fits.  Finite volume effects are included per~\cite{Aubin:2007}.  Meson masses entering the chiral logs are computed using the leading order relation between constituent quark masses $m_x, m_y$ and the corresponding meson mass $M_{xy}$
\begin{equation}
M_{xy}^2 = B_0(m_x + m_y),
\label{eq-B0}
\end{equation}
where $B_0$ is a low energy constant.
Results of a simultaneous fit to data for $f_0$, $f_+$, and $f_T$, in which the $\chi^2/{\rm dof}$ is 35.1/50, are shown in Fig.~\ref{fig-chpt}.  Values for priors and fit results are collected in Tables~\ref{tab-chiptfit_I} and~\ref{tab-chiptfit_II}.  The stability of these fit results with respect to changes in the fit ans\"atze are discussed at the end of the next section.

Chiral perturbation theory is not well-defined for kaon energies above the chiral scale, $\Lambda_\chi$.  Taking the chiral scale to be $\Lambda_\chi = 4\pi F_\pi$ suggests simulation data with kaon energies above $\sim\!1\ {\rm GeV}$ may not be described by chiral perturbation theory.  The data in question are $\nicefrac{2\pi}{L}(1,1,1)$ data for ensembles C2, C3, F1, and F2.  Ensemble C1 has a larger spatial extent so the physical momentum corresponding to $\nicefrac{2\pi}{L}(1,1,1)$ is roughly equivalent to the physical momenta for ensembles C2 and C3 with $\nicefrac{2\pi}{L}(1,1,0)$.  We first note that, as shown in Fig.~\ref{fig-chpt}, our chiral/continuum fit ans\"atez do a good job of fitting all simulation data, including data at ``large" momenta.  Fig~\ref{fig-chpt_sans111} shows the results of a chiral/continuum extrapolation omitting the $\nicefrac{2\pi}{L}(1,1,1)$ data, except for that on C1.  Comparison of the physical bands in Figs.~\ref{fig-chpt} and~\ref{fig-chpt_sans111} shows consistent fit results.  In Fig.~\ref{fig-chpt_sans111} we extend the ensemble chiral/continuum fit curves to energies beyond the fitted region to show the level of agreement between the fit curves and the omitted data.  These comparisons demonstrate that our chiral/continuum fit ans\"atze adequately describe all simulation data.  Kinematic extrapolation of synthetic data sets generated with and without the $\nicefrac{2\pi}{L}(1,1,1)$ data produces similar extrapolated bands over the full kinematic range of $q^2$.  At $q^2=0$, where agreement is the worst, the results remain consistent at the level of $\sim\! 1 \sigma$.

\subsubsection{Kinematic Extrapolation }
\label{sec-stdzexp}
\begin{figure}[t!]
{\scalebox{1}{\includegraphics[angle=-90,width=0.5\textwidth]{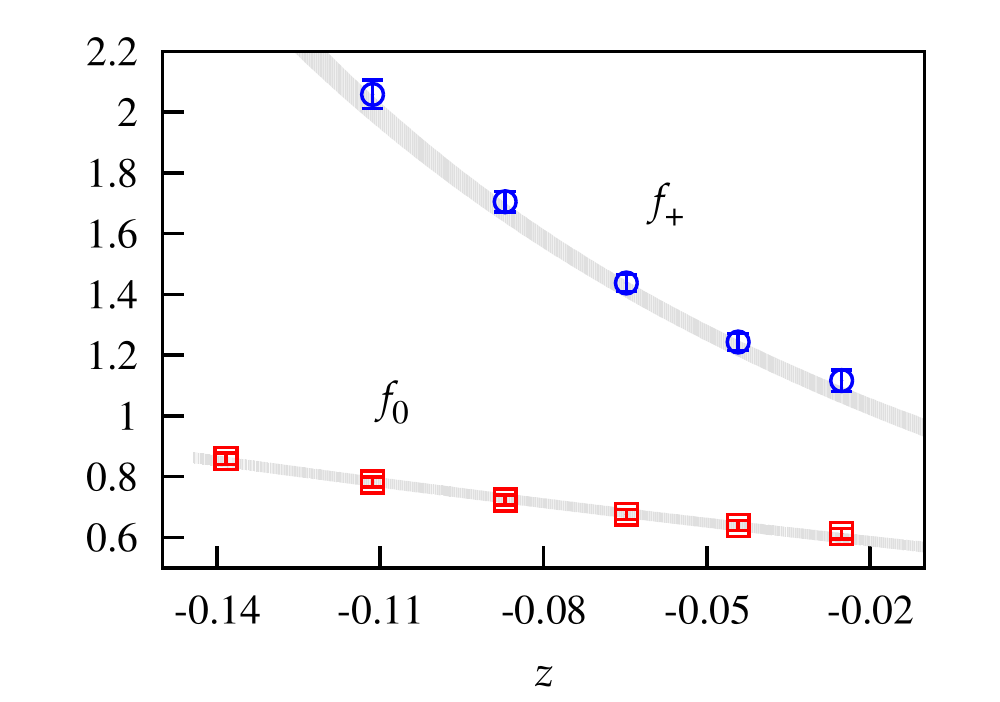}}}
\\
%
{\scalebox{1}{\includegraphics[angle=-90,width=0.5\textwidth]{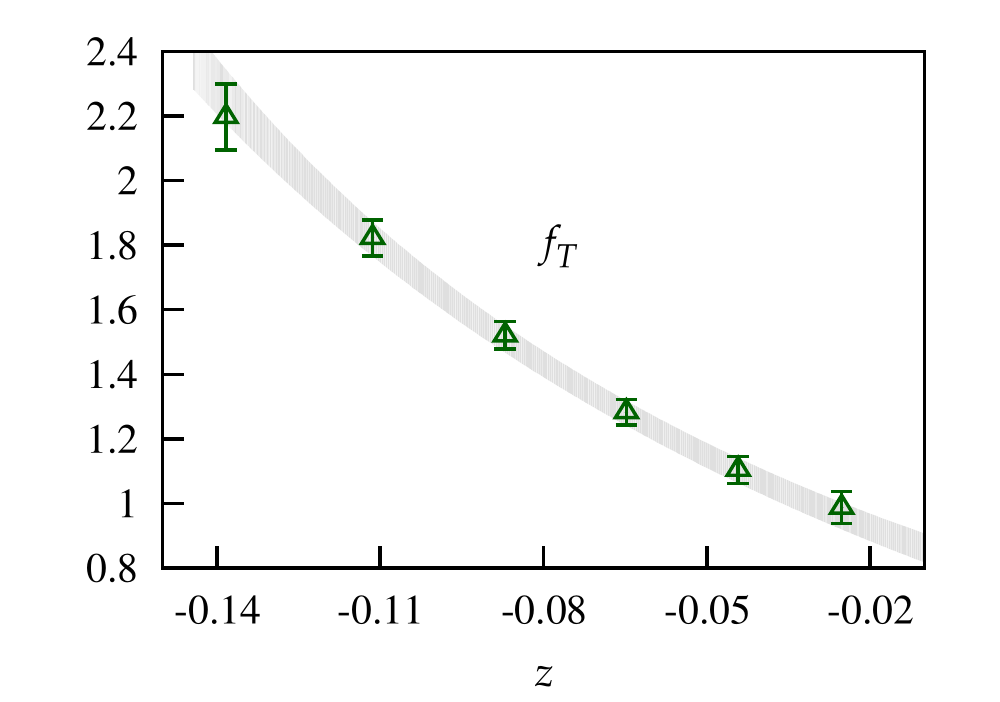}}}
\caption{Synthetic data points from the chiral/continuum extrapolation relative to the standard $z$ expansion fit results.}
\label{fig-zfit}
\end{figure}
Using the fit results from the previous section, including the $\nicefrac{2\pi}{L}(1,1,1)$ data, we generate a synthetic data set of physical values for $f_0$, $f_+$, and $f_T$, restricted to the range of $q^2$ for which the simulations are performed.  These data are then extrapolated to the full kinematic range of $q^2$ using the $z$ expansion~\cite{Boyd:1996, Arnesen:2005, Bourrely:2010}.  
We varied the number of data points in the synthetic data set between 6 and 12 and found the fit results to be largely insensitive to changes in this range, provided correlations among the synthetic data points are accounted for.  We use 6 synthetic data points per form factor in our final fit result.  The synthetic data points are highly correlated and it is necessary to introduce a sizable singular value decomposition cut (in the range $10^{-3}$ to $10^{-2}$) when inverting the covariance matrix.  The synthetic data points are shown in Fig.~\ref{fig-zfit}.

The $z$ expansion maps the kinematic variable $q^2$ onto $z(q^2,t_0)$,
\begin{equation}
z(q^2,t_0) = \frac{\sqrt{t_+ - q^2} - \sqrt{t_+ - t_0}}{\sqrt{t_+ - q^2} + \sqrt{t_+ - t_0}},
\label{eq-defz}
\end{equation}
where $t_\pm = (M_B \pm M_K)^2$ and $t_0$ is a free parameter whose choice determines the values $z$ assumes over the kinematic range of $q^2$.  The size of the interval ($z_{\rm min},  z_{\rm max})$ is largely independent of $t_0$ and the optimum choice of $t_0$ results in $z_{\min} \approx -z_{\rm max}$~\cite{Boyd:1996}.  We considered several possible values and found $t_0 \approx 14.8\ {\rm GeV}^2$ results in $-0.15 < z < 0.15$.  This is similar to the recommendations of Becher and Hill~\cite{Becher:2006}, $t_0 = t_+ - ( t_+^2 - t_+t_-)^{1/2} \approx 14.65\ {\rm GeV}^2$, and Bourrely, Caprini, and Lellouch (BCL)~\cite{Bourrely:2010}, $t_0=(M_B + M_K)(\sqrt{M_B} - \sqrt{M_K})^2 \approx 14.69\ {\rm GeV}^2$.  We use the BCL parameterization for the $z$ expansion below and also choose their recommended value of $t_0$.

Based on the BCL~\cite{Bourrely:2010} parameterization for $f_+$, we fit the form factors to
\begin{eqnarray}
f_0(q^2) &=& \sum_{k=0}^{K} a^0_k\,  z(q^2)^k , \label{eq-zf0} \\
f_i(q^2) &=& \frac{1}{P_i(q^2)} \sum_{k=0}^{K-1} a^i_k \Big[ z(q^2)^k -(-1)^{k-K}\frac{k}{K} z(q^2)^K \Big] , \nonumber \label{eq-zfpT} \\
\end{eqnarray}
where $i=+,T$.
The expected scaling behavior of $f_+$ at large $q^2$ leads to a constraint involving the $k=K$ term in the sum.  For the reasons discussed in Sec.~\ref{sec-ChPT} we expect $f_T$ to display $q^2$-scaling behavior similar to that of $f_+$ and we therefore impose the same $K$ constraint when fitting the tensor data.  We performed fits for $f_+$ and $f_T$ with and without this constraint and found negligible difference.
We also impose the kinematic constraint $f_0(0)=f_+(0)$ by adding an additional data point at $q^2=0$, equal to $0\pm\epsilon$, and defining the fit function at this data point to be $f_+(0)-f_0(0)$.  The value of $\epsilon$ is chosen small, though non-zero to avoid singularities when inverting the covariance matrix.  In practice, values ranging from $10^{-3}$ to $10^{-14}$ give indistinguishable fit results.

In Eq.~(\ref{eq-zfpT}) the Blaschke factor $P_i(q^2)$ accounts for poles above the physical range of $q^2$ but below the $BK$ production threshold, {\it ie.} $t_- < q^2 < t_+$.  The resonances responsible for the poles must have quantum numbers consistent with the flavor-changing current being considered.  
Within this energy range there are no $J^P = 0^+$ bound states, and therefore no Blaschke factor for $f_0$, the only vector $J^P=1^-$ state is the $B^*$, and there are no known tensor $J^P = 2^-$ states.  However, the tensor current is equivalent to the vector current at leading order in $\nicefrac{1}{m_b}$.  We therefore take the pole for the tensor form factor to lie at the same location as the vector pole, but with a width 100 times larger.  
The Blaschke factors are then
\begin{eqnarray}
P_+(q^2) &=& 1 - \nicefrac{ q^2 }{ (M_+^{\rm pole})^2 } , \label{eq-P+} \\
P_T(q^2) &=& 1 - \nicefrac{ q^2 }{ (M_T^{\rm pole})^2 } , \label{eq-PT}
\end{eqnarray}
and we parameterize the pole masses in terms of splittings above the $B$ meson mass
\begin{eqnarray}
M_+^{\rm pole} &=& M_B + \Delta^*_+ , \label{eq-f+polemass}  \\
M_T^{\rm pole} &=& M_B + \Delta^*_T. \label{eq-fTpolemass}
\end{eqnarray}

We apply Eqs.~(\ref{eq-zf0}, \ref{eq-zfpT}) separately to $f_{0,+,T}$ in a simultaneous, correlated fit to the synthetic data sets generated from the chiral/continuum extrapolation.  We take as our result the fit using $K=3$. The fit has a $\chi^2/{\rm dof}$ of 8.58/11.  

\begin{figure}[t!]
\vspace{0.0in}
\hspace{0.0in}
{\scalebox{0.7}{\includegraphics[angle=-90,width=0.75\textwidth]{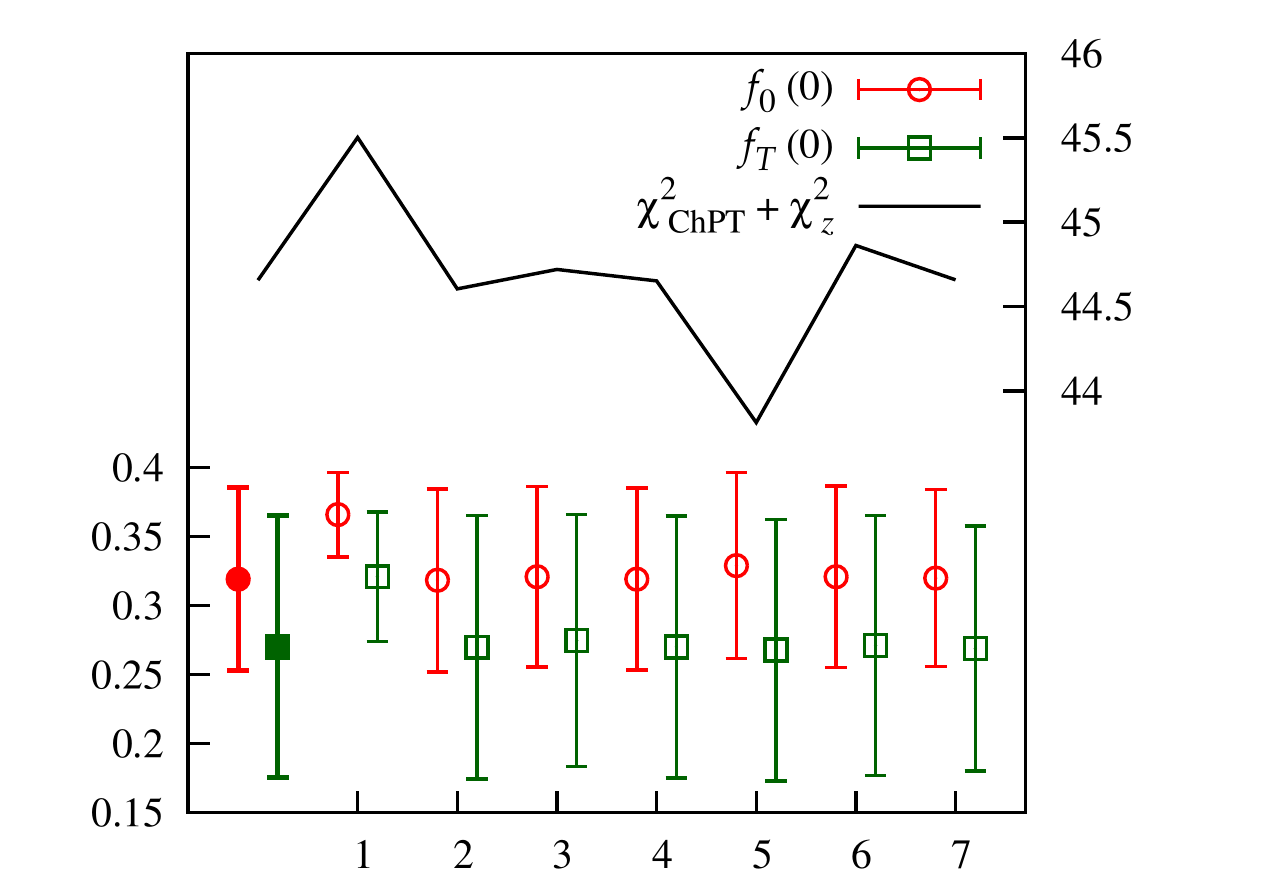}}}
\caption{Two-step extrapolation results for $f_{0,T}(0)$, the left-most points, are stable under various modifications to the fit ansatz.  Fit quality is represented by the sum of $\chi^2$'s from the chiral/continuum and $z$ expansion fits, with a total of 61 degrees of freedom.  The $x$-axis label corresponds to the modifications listed in the text.}
\vspace{0.0in}
\label{fig-ChPTz_stab}
\end{figure}
\begin{figure}[t!]
{\scalebox{0.7}{\includegraphics[angle=-90,width=0.71\textwidth]{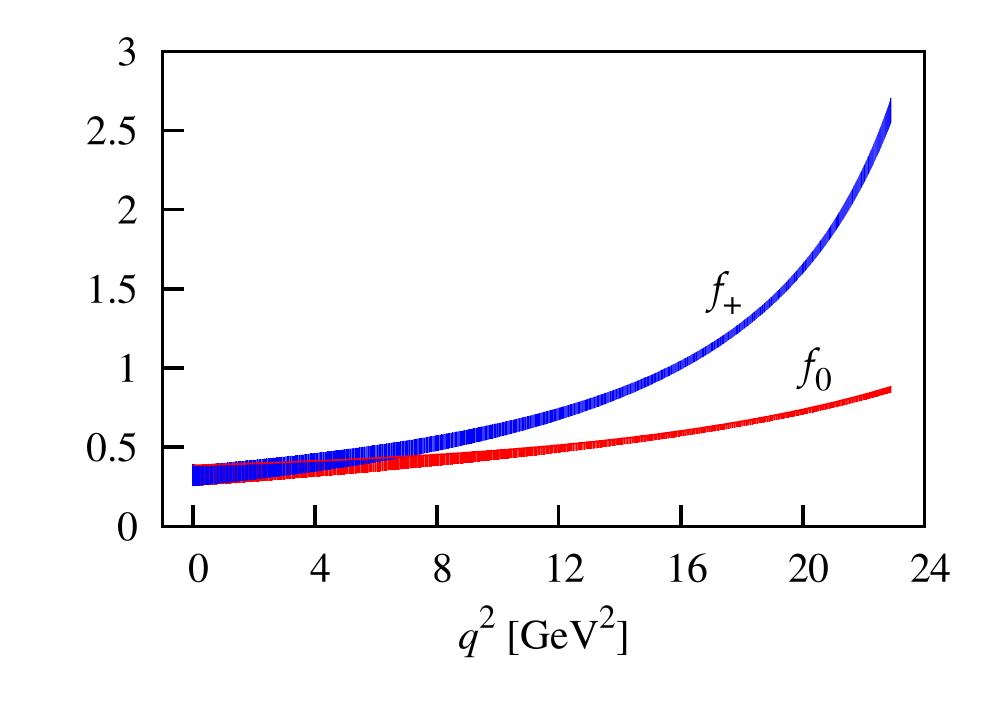}}}
\\
{\scalebox{0.7}{\includegraphics[angle=-90,width=0.71\textwidth]{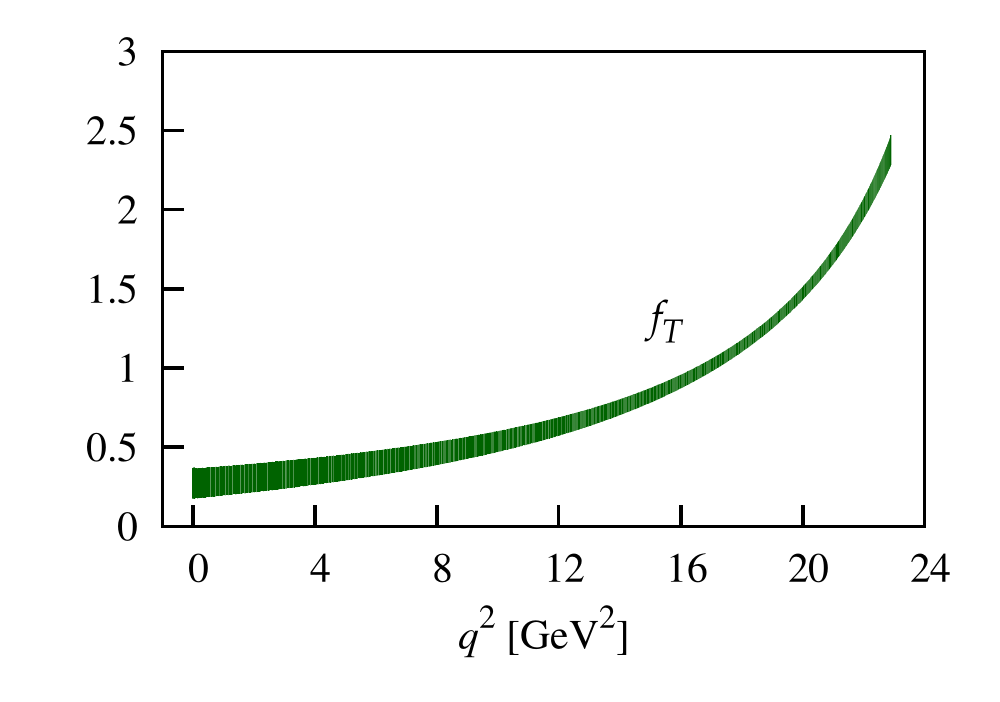}}}
\caption{Form factors from the two-step chiral/continuum and kinematic extrapolation over the full kinematic range of $q^2$.  The error bands include only fit errors and do not account for the additional systematic errors addressed in Sec.~\ref{sec-syserr}.}
\label{fig-ChPTzextrap}
\end{figure}
We study the stability of the fit results to numerous modifications to the fit ans\"atze.  
For purposes of comparison, we study the effects of these modifications on the values of $f_{0,T}(q^2=0)$.  This point is furthest from the region of $q^2$ for which simulation data exists and exhibits maximum sensitivity to changes in the fit ans\"atze.  The kinematic constraint ensures the fit results for $f_{0,+}(0)$ are equivalent.  The modifications studied are:
\begin{enumerate}
  \item[1.] Perform the $z$ expansion through $\mathcal{O}(z^2)$.
  \item[2.] Perform the $z$ expansion through $\mathcal{O}(z^4)$.
  \item[3.] Drop the heavy-quark mass-dependent discretization effects from the $d_i$ in the chiral/continuum fit ans\"atze.
  \item[4.] Drop the light-quark mass-dependent discretization effects from the $d_i$ in the chiral/continuum fit ans\"atze.
  \item[5.] Drop the finite volume effects from the chiral/continuum fit ans\"atze.
  \item[6.] Drop the NNLO analytic sea and strange quark mass terms, {\it ie.} those associated with the coefficients $a_5, ..., a_9$ in Eq.~(\ref{eq-Adef}).
  \item[7.] Add a constraint based on the heavy quark, large recoil (small $q^2$) symmetry relationship among the form factors~\cite{Beneke:2002, Hill:2006}
\begin{equation}
f_T(q^2) = \frac{M_B(M_B+M_K)}{q^2}\Big[ f_+(q^2) - f_0(q^2) \Big],
\label{eq-statreln}
\end{equation}
which, in the $q^2\to 0$ limit, gives
\begin{equation}
f_T(0) = M_B(M_B + M_K) \left. \left(\frac{\partial f_+}{\partial q^2} - \frac{\partial f_0}{\partial q^2}\right)\right|_{q^2=0} .
\label{eq-statlim}
\end{equation}
To impose this constraint we add to the fit an additional data point at $q^2=0$, given by $0\pm \nicefrac{\Lambda_{\rm QCD}}{m_b}$, and define the fit function to be $\text{l.h.s.} - \text{r.h.s. of Eq.~(\ref{eq-statlim})}$, where the slopes of the form factors at $q^2=0$ are evaluated numerically.  The error for this data point accommodates higher order effects from the $\nicefrac{1}{m_b}$ expansion and we conservatively take $\nicefrac{\Lambda_{\rm QCD}}{m_b} \sim 0.2$.
\end{enumerate}  

Fig.~\ref{fig-ChPTz_stab} shows the results of these tests and demonstrates that our fit results are stable against reasonable modifications to the fit ans\"atze.
Tests 1 and 2, together with the final fit results, show that by $\mathcal{O}(z^3)$ the $z$ expansion fit results and $\chi^2$ have stabilized.  The fit errors have also saturated by $\mathcal{O}(z^3)$ and account for the error associated with truncating the $z$ expansion.  
In tests 3 and 4 we drop heavy- and light-quark mass-dependent discretization effects, as introduced in Eqs.~(\ref{eq-hvy_disc},~\ref{eq-lt_disc}), and find negligible change in our fit results.  
We drop finite volume effects from the chiral/continuum extrapolation in test 5 and find a slightly lower $\chi^2$ but little impact on fit results.  
Test 6 demonstrates the insensitivity of our fit results to NNLO sea and strange quark chiral analytic terms.
In test 7 we study the consistency of our fit results with the expected symmetry relation among the form factors, valid for heavy quarks and large recoil, and find excellent agreement.  In addition to these tests we have verified the consistency of our results for $f_0(0)$ and $f_+(0)$ with and without the kinematic constraint $f_0(0) = f_+(0)$.  Without the constraint, $f_0(0)$ and $f_+(0)$ central values shift by $\sim \nicefrac{\sigma}{2} $, the errors increase by $\sim\! 40\%$, and the constraint remains satisfied within errors.

In Fig.~\ref{fig-ChPTzextrap} we plot the extrapolated form factors over the full kinematic range of $q^2$.  The error bands represent errors associated with the fit and do not include additional systematic errors discussed in Sec.~\ref{sec-syserr} below.

Lattice calculations of $B$ semileptonic decay form factors suffer from the need to perform a significant extrapolation in $q^2$ -- simulation data in this analysis have largest momenta of $\nicefrac{2\pi}{L}(1,1,1)$, which corresponds to a smallest simulated $q^2$ of $\sim\! 17\ {\rm GeV}^2$.  An obvious way to improve lattice form factor calculations is to include data at larger momenta, thereby reducing the kinematic extrapolation.  Though the ability of chiral perturbation theory to describe our simulation data at the ``large" lattice momenta $\nicefrac{2\pi}{L}(1,1,1)$ was demonstrated in Sec.~\ref{sec-ChPT}, its applicability to larger momenta is doubtful.  Hard pion chiral perturbation theory~\cite{Bijnens:2010} may provide a way to handle larger lattice momenta and we intend to study its effectiveness in future $B_{(s)}$ semileptonic decay analyses.  Another approach that may prove useful in this regard is a simultaneous chiral/continuum and kinematic extrapolation via the modified $z$ expansion, introduced in~\cite{Na:2010, Na:2011}.  We apply this method to our $B\to K\ell^+\ell^-$ data in Appendix ~\ref{app-modz} and verify results consistent with Fig.~\ref{fig-ChPTzextrap} are obtained.



\section{ Form Factor Results }
\label{sec-results}
In this section we summarize final results, with complete error budgets, for the form factors obtained in Sec.~\ref{sec-2step}.

\begin{figure*}[t!]
\hspace{-0.07in}  
\subfloat[][\label{fig-f0ChPTerr}Chiral/continuum extrapolation \% errors for $f_0$.]
{\scalebox{0.98}{\includegraphics[angle=0,width=0.5\textwidth]{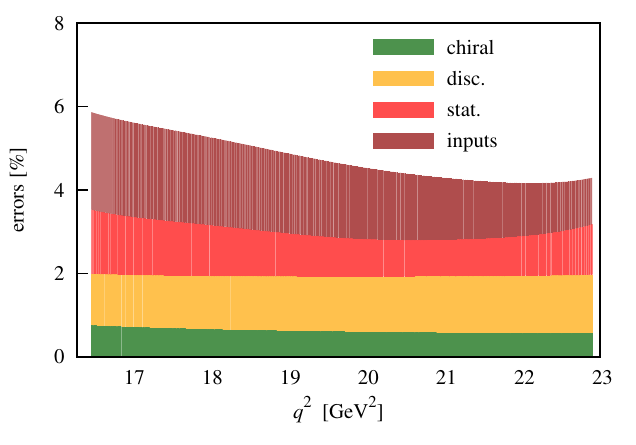}}}
\subfloat[][\label{fig-f0stdzerr}Kinematic extrapolation \% errors for $f_0$.]
{\scalebox{0.98}{\includegraphics[angle=0,width=0.5\textwidth]{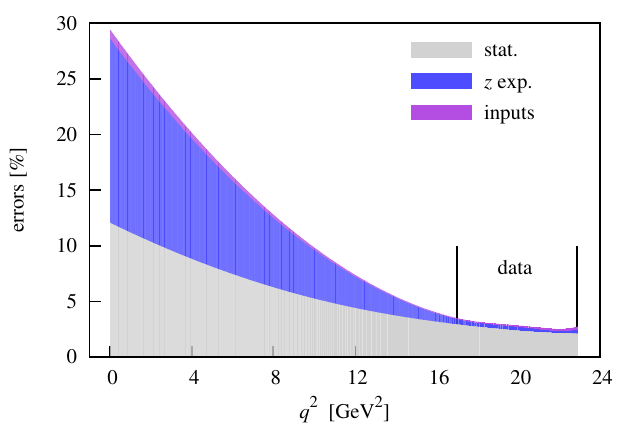}}}
\\
\subfloat[][\label{fig-fpChPTerr}Chiral/continuum extrapolation \% errors for $f_+$.]
{\scalebox{0.98}{\includegraphics[angle=0,width=0.5\textwidth]{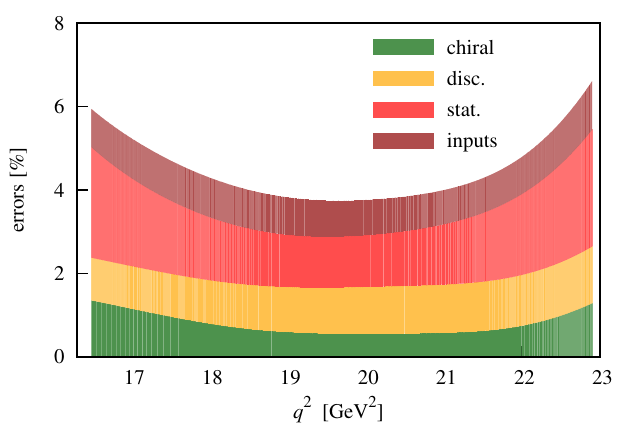}}}
\subfloat[][\label{fig-fpstdzerr}Kinematic extrapolation \% errors for $f_+$.]
{\scalebox{0.98}{\includegraphics[angle=0,width=0.5\textwidth]{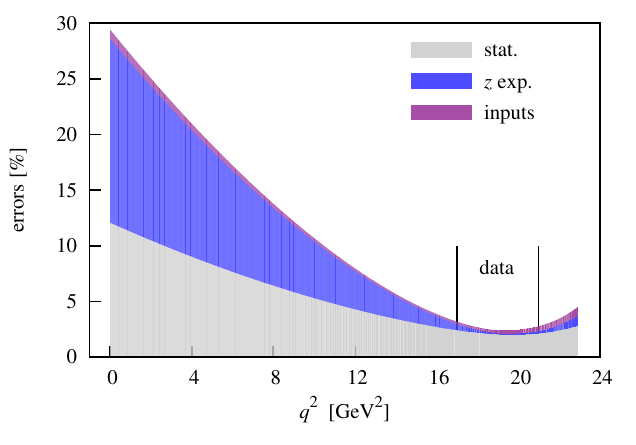}}}
\\
\subfloat[][\label{fig-fTChPTerr}Chiral/continuum extrapolation \% errors for $f_T$.]
{\scalebox{0.98}{\includegraphics[angle=0,width=0.5\textwidth]{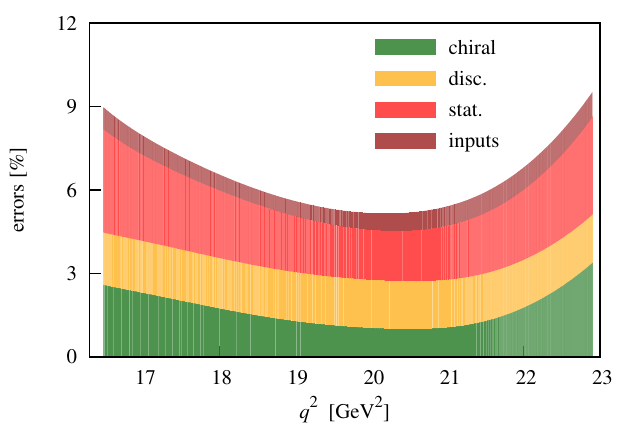}}}
\subfloat[][\label{fig-fTstdzerr}Kinematic extrapolation \% errors for $f_T$.]
{\scalebox{0.98}{\includegraphics[angle=0,width=0.5\textwidth]{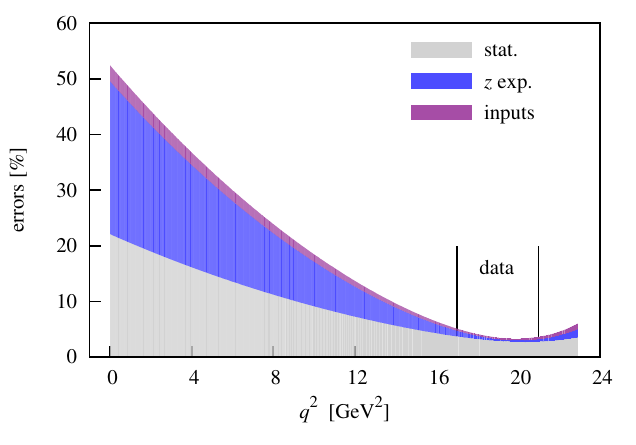}}}
\vspace{0.1in}
\caption{Breakdown, by percentage, of form factor fit errors from the two-step chiral/continuum and kinematic extrapolations plotted in relevant regions of $q^2$.  Plots on the left show fit errors from the chiral/continuum extrapolation and represent the error budget for the synthetic data used in the kinematic extrapolation.  Plots on the right show the fit errors for the kinematic extrapolation.  The total \% fit error for the form factors is the sum in quadrature of the three components of the kinematic extrapolation fit error.}
\label{fig-2step_errors}
\end{figure*}

An advantage of constrained curve fitting is the ability to incorporate certain errors directly in the fit.  For example, the uncertainty associated with input parameters is built into the fit by making these quantities fit parameters with widths set by their uncertainty.  Part of the resulting fit error is then due to these uncertainties.  
Using the method outlined in Ref.~\cite{Davies:2008}, we extract the components of the fit errors and plot them, as a percentage of the central value, in Fig.~\ref{fig-2step_errors}.  
The three plots on the left (Figs.~\ref{fig-f0ChPTerr},~\ref{fig-fpChPTerr}, and~\ref{fig-fTChPTerr}) show errors for the chiral/continuum extrapolation and the three plots on the right (Figs.~\ref{fig-f0stdzerr},~\ref{fig-fpstdzerr}, and~\ref{fig-fTstdzerr}) show errors for the subsequent kinematic extrapolation.  
In each plot the sum in quadrature of the listed errors yields the total fit error in percent, corresponding to the fit bands in Fig.~\ref{fig-ChPTzextrap}.  Listed errors are groupings of the errors associated with related fit parameters and are described in detail below.

\subsection{ Chiral/Continuum Extrapolation Fit Errors }
\label{sec-Errors_ChPT}
Here we discuss the components of the fit errors from the chiral/continuum extrapolation of Sec.~\ref{sec-ChPT}.  The components of the fit errors are plotted in Figs.~\ref{fig-f0ChPTerr},~\ref{fig-fpChPTerr}, and~\ref{fig-fTChPTerr} over the region of $q^2$ for which simulation data exists and for which the chiral/continuum extrapolation is performed.  The components of the fit errors are:

\begin{itemize}

\item{\bf chiral:}  The chiral extrapolation error is the sum in quadrature of the errors due to uncertainty in the $a_i$ of Eq.~(\ref{eq-Adef}), the $h_i$ of Eq.~(\ref{eq-Hdef}), and the appropriate leading order low energy constant $\kappa_{(T)}$.  The resulting error describes fit uncertainty due to the extrapolation in light-quark mass, slight interpolation in strange-quark mass due to mistuning, small mass differences due to the mixed action used in the simulation, and the $E_K$ dependence introduced in $H$ and via the NNLO term with coefficient $a_{10}$.
For $f_{+,T}$ our simulation data is restricted to $17\ {\rm GeV}^2 \lesssim q^2 \lesssim 21\ {\rm GeV}^2$.  Extrapolation beyond the region of simulated $q^2$ leads to an increase in their error at large $q^2$.

\item{\bf disc.:}  The chiral/continuum extrapolation includes discretization effects via the $d_i$ terms of Eq.~(\ref{eq-Ddef}).  These terms are modified to incorporate potential heavy-quark mass-dependent discretization effects via the $f_i$ terms introduced in Eq.~(\ref{eq-hvy_disc}) and further modified via Eq.~(\ref{eq-lt_disc}) to include possible light-quark mass-dependent discretization effects from the $g_i$ terms.
In the reported discretization error we combine these effects by adding in quadrature the form factor errors due to uncertainty in the $d_i$, $f_i$, and $g_i$.

\item{\bf stat.:}  The statistical error is the error associated with the uncertainty in the data being fit, {\it ie.} the errors from form factor data of Table~\ref{tab-f0f+fT}.  

\item{\bf inputs:}  The input error is the sum in quadrature of the errors associated with the ``Group I" priors of Table~\ref{tab-chiptfit_I}.

\end{itemize}

\subsection{ Kinematic Extrapolation Fit Errors }
\label{sec-Errors_z}
Here we discuss the components of the fit errors from the kinematic extrapolation, performed via $z$ expansion in Sec.~\ref{sec-stdzexp}.  The components are plotted in Figs.~\ref{fig-f0stdzerr},~\ref{fig-fpstdzerr}, and~\ref{fig-fTstdzerr} over the full kinematic range of $q^2$.  The region of $q^2$ for which simulation data exist is indicated in the plots.  The components of the fit errors are:

\begin{itemize}

\item{\bf stat.:}  This is the statistical error associated with the synthetic data generated from the chiral/continuum extrapolation.  It is composed of components whose individual contributions are shown in Figs.~\ref{fig-f0ChPTerr}, ~\ref{fig-fpChPTerr}, and~\ref{fig-fTChPTerr}, and described above.

\item{\bf $z$ exp.:}  This is the error in the form factors due to uncertainty in the coefficients $a_k$ of the $z$ expansion in Eqs.~(\ref{eq-zf0},~\ref{eq-zfpT}).  A comparison in Fig.~\ref{fig-ChPTz_stab} of our final fit results with those from tests 1 and 2 shows that by $\mathcal{O}(z^3)$ our fit errors have saturated.  Therefore, these errors also include the error associated with truncating the $z$ expansion at $\mathcal{O}(z^3)$.
The error associated with the $z$ expansion grows as we extrapolate beyond the region of $q^2$ for which simulation data exist.

\item{\bf inputs:}  The input error is the sum in quadrature of the errors from the parameters labled ``Group I'' priors in Table~\ref{tab-zpriors}.

\end{itemize}

\subsection{Additional Systematic Errors}
\label{sec-syserr}
In addition to the errors accounted for directly in the fit, several other systematic errors must be addressed:

\begin{itemize}

\item {\bf matching:}   With the matching coefficients calculated in~\cite{Monahan:2013}, we find the $\mathcal{O}(\alpha_s, \nicefrac{\Lambda_{\rm QCD}}{m_b}, \nicefrac{\alpha_s}{am_b})$ contributions to $\langle V_0\rangle$ to be $\sim\!4\%$, of which $\sim\!3.5\%$ comes from the one loop $\mathcal{O}(\alpha_s)$ correction and $<\!1\%$ from the NRQCD matching via $\langle J_0^{(1),{\rm sub}}\rangle$.  For $\langle V_k\rangle$ we find the leading order corrections to be $\sim\!2\%$ with $\sim\!1\%$ coming from the $\mathcal{O}(\alpha_s)$ correction and $<\!1\%$ from the NRQCD matching.  We estimate higher order corrections based on observed leading order effects and conservatively use the larger 4\%.  We consider the following options for estimating the size of higher order terms.  

(i.) It can be argued that higher order corrections should be suppressed by a factor of $\alpha_s$ relative to the observed 4\% leading order corrections, resulting in an estimated matching error of $\mathcal{O}(\alpha_s \times 0.04)$, or $\sim\!1$\%.  
We can alternatively characterize the higher order corrections by the size of the higher order matching coefficients.  If we were to multiply the r.h.s. of Eq.~(\ref{eq-Vmatch}) by $1~+~\rho_{\rm 2\, loop} \alpha_s^2$ and assume the matching coefficient for the $\mathcal{O}(\alpha_s^2)$ correction is approximately the same size as the coefficient of the $\mathcal{O}(\alpha_s)$ correction, $\sim\! 0.1$, then the $\mathcal{O}(\alpha_s^2)$ correction would be $\sim\!1\%$.

(ii.) More conservatively, we could argue that higher order corrections should be no larger than the observed leading order corrections and therefore estimate the matching error at 4\%.  This is equivalent to taking the $\mathcal{O}(\alpha_s^2)$ matching coefficient to be  four times larger than the $\mathcal{O}(\alpha_s)$ matching coefficient $\rho_0^{(V_0)}$ (13 times larger than $\rho_0^{(V_k)}$).

(iii.) Alternatively, and following the approach taken in~\cite{Gulez:2007}, we could argue that higher order corrections are $\mathcal{O}(\alpha_s^2)$, giving an estimated matching error of 9\%.  This is equivalent to assuming an $\mathcal{O}(\alpha_s^2)$ matching coefficient that is 10 times larger than $\rho_0^{(V_0)}$ (29 times larger than $\rho_0^{(V_k)}$).
We note that, absent knowledge of the $\mathcal{O}(\alpha_s)$ matching coefficients, this approach suggests a leading order contribution of $\mathcal{O}(\alpha_s)\approx 30\%$, nearly an order of magnitude larger than what is observed.

Given that approach (iii.) significantly overestimates the leading order contribution and approach (i.) does not allow for the possibility of a moderate increase in the $\mathcal{O}(\alpha_s^2)$ matching coefficients relative to the $\mathcal{O}(\alpha_s)$ matching coefficient, we choose approach (ii.) and estimate the matching error to be 4\%.

\item {\bf electromagnetic and isospin breaking effects:}  
Our lattice simulation uses degenerate light quarks and omits electromagnetic effects.  The hadronic matrix elements calculated on the lattice are therefore isospin symmetric.  
By adjusting the meson masses used in the subsequent chiral/continuum and kinematic extrapolations, we estimate the ``kinematic" effects of omitting electromagnetic and isospin symmetry breaking in our simulation to be $\lesssim 1\%$. 
It is more difficult to determine the size of the full effects though, in general, electromagnetic and isospin effects are expected to be sub-percent.  
We assume the error in our form factor calculation due to these effects is negligible when added in quadrature to the errors discussed above.

\item{\bf charm sea quarks:}  Our simulations include up, down, and strange sea quarks and we assume omitted charm sea quark effects are negligible.  This has been the case for processes in which it has been possible and appropriate to perturbatively estimate effects of charm quarks in the sea~\cite{Davies:2010}.

\end{itemize}

\begin{figure}[t!]
{\scalebox{1}{\includegraphics[angle=-90,width=0.5\textwidth]{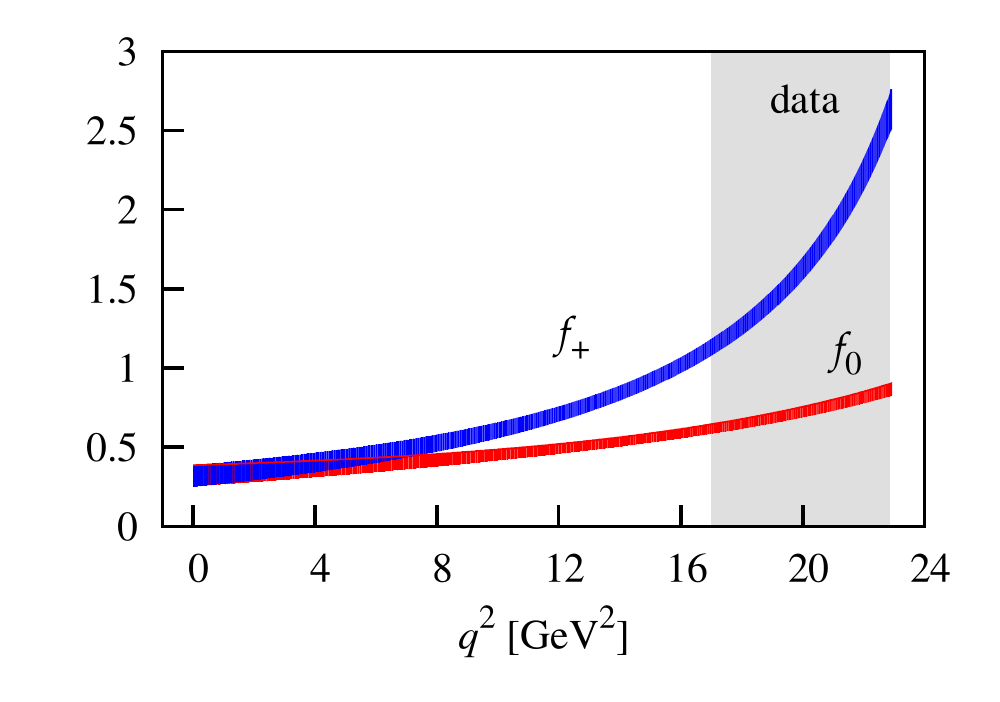}}}
\\
%
{\scalebox{1}{\includegraphics[angle=-90,width=0.5\textwidth]{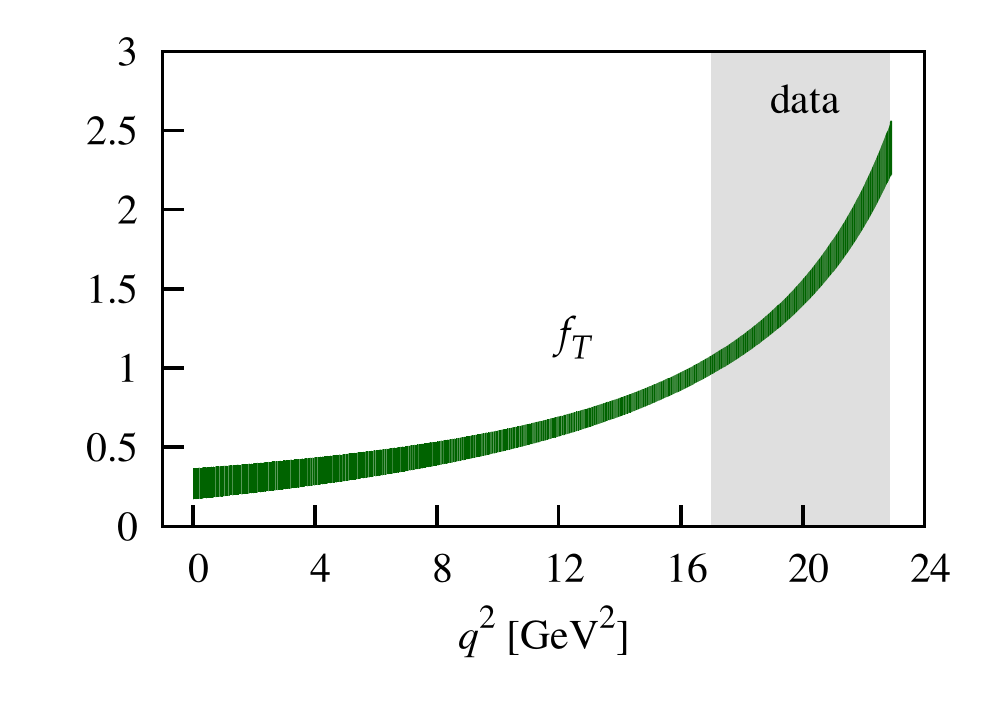}}}
\caption{Final results for the form factors, including all sources of error.  The shaded gray band indicates the region of $q^2$ for which simulation data exists.}
\label{fig-finalFFs}
\end{figure}

Of the additional systematic errors, only the matching error is non-negligible.  We propagate the 4\% matching error on the matrix elements to an error on the form factors.  
We obtain the total error on our form factors by adding in quadrature, at each value of $q^2$, the propagated 4\% matching error to the error obtained from the covariance matrix of Table~\ref{tab-Zcovsimple}.  Our form factors, with full error bands, are shown in Fig.~\ref{fig-finalFFs}, and their values at $q^2=0$ are given in Table~\ref{tab-FFcompare} where they're compared with results from a quenched lattice calculation~\cite{Becirevic:2012} and light cone sum rules~\cite{Khodjamirian:2010}.
\begin{table}[t]
\begin{tabular}{ccc}
\hline\hline	
	\T  							& $f_0(0) = f_+(0)$		& $f_T(0)$		    	\\ [0.5ex]
	\hline
	\T this work					& $0.319 \pm 0.066$	& $ 0.270 \pm 0.095$ 	\\ [0.5ex]
	\T Be\v{c}irevi\'{c} {\it et al.}~\cite{Becirevic:2012}		& $0.33 \pm 0.04$		& $0.31 \pm 0.04$		\\ [0.5ex]
	\T Khodjamirian {\it et al.}~\cite{Khodjamirian:2010}	& $0.34^{+0.05}_{-0.02}$	& $0.39^{+0.05}_{-0.03}$	\\ [0.7ex]
\hline\hline
\end{tabular}\caption{Comparison of form factor results at $q^2=0$.}
\label{tab-FFcompare}
\end{table}

\begin{figure}[t!]
\hspace{-0.2in}
{\scalebox{1}{\includegraphics[angle=0,width=0.5\textwidth]{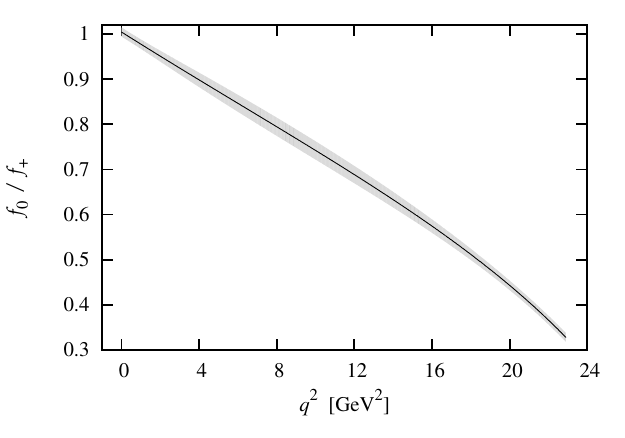}}}
\\
\hspace{-0.2in}
{\scalebox{1}{\includegraphics[angle=0,width=0.5\textwidth]{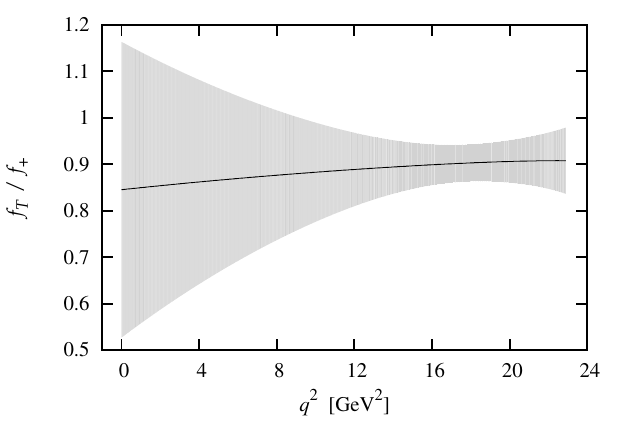}}}
\vspace{-0.2in}
\caption{Ratios of form factors with (gray) error band calculated using correlations between the form factors.}
\label{fig-FFratios}
\end{figure}

In the works of Bobeth {\it et al.}~\cite{Bobeth:2007, Bobeth:2013} the ratios of form factors $f_0/f_+$ and $f_T/f_+$ play an important role in constraining new physics.  The ratio $f_0/f_+$ controls $m_\ell$-suppressed Standard Model contributions, important for future $B\to K \tau^+\tau^-$ studies and beyond the Standard Model scalar and pseudoscalar contributions.  This ratio also provides a test of large $q^2$ operator product expansion and Isgur-Wise relations ({\it c.f.} Ref.~\cite{Bobeth:2013}) and low $q^2$ symmetry relations based on QCD factorization ({\it c.f.} Ref.~\cite{Bobeth:2007}).
The ratio $f_T/f_+$ controls beyond the Standard Model tensor contributions and provides a check of the low $q^2$ symmetry relations.
These ratios are shown in Fig.~\ref{fig-FFratios}.

\subsection{ Reconstructing the Form Factors }

To reconstruct the form factors, one should use the fit result central values for the coefficients $a_k^{0,+,T}$ of the $z$ expansion in Table~\ref{tab-zpriors}, together with Eqs.~(\ref{eq-zf0},~\ref{eq-zfpT}); values for $z$ obtained from Eq.~(\ref{eq-defz}) and surrounding text; and Blaschke factors $P_i$ defined in Eqs.~(\ref{eq-P+}, \ref{eq-PT}, \ref{eq-f+polemass}, \ref{eq-fTpolemass}), with values for the pole mass splittings $\Delta^*_i$ from Table~\ref{tab-zpriors}.  To obtain correct results for the errors, the full covariance matrix of Table~\ref{tab-Zcovsimple} must be used and the resulting errors increased by 4\% to account for the additional systematic errors discussed in Sec.~\ref{sec-syserr}.

Note the dominant additional systematic error comes from matching, and is therefore an error associated with the hadronic matrix elements of Eqs.~(\ref{eq-fpardef}, \ref{eq-fperpdef}, \ref{eq-fTdef}).  We have verified that propagating a 4\% error, applied to the matrix elements, to the form factors $f_{0,+,T}$ is equivalent to applying the error directly to the form factors.

\section{ Phenomenology }

In Ref.~\cite{Bouchard:PRL}, we use the form factor results to calculate several Standard Model observables that either allow comparison with experiment or make predictions.  Here we provide, for completeness, the necessary relations between the various observables and the form factors.

As discussed in Sec.~\ref{sec-syserr}, our form factor results are, within errors, equivalent for $B^0 \to K^0 \ell^+\ell^-$, $\bar{B}^0 \to \bar{K}^0 \ell^+\ell^-$, and $B^\pm \to K^\pm \ell^+\ell^-$.  The observables we calculate from the form factors introduce additional dependence on $M_B$, $M_K$, and $\tau_B$.  Here, and in~\cite{Bouchard:PRL}, we calculate isospin averaged values for all reported observables.
Additional input parameters are required and values used are provided in Table~\ref{tab-SMparams}.  Errors associated with each input parameter are propagated to the error reported for each observable~\cite{gdev}.
\begin{table}[t]
\begin{tabular}{ccc}
\hline\hline	
	\T parameter  				& value							& Ref.    \\ [0.5ex]
	\hline
	\T $m_c$					& 1.275(25) GeV					& \cite{PDG:2012}	\\
	\T $m_b$					& 4.18(3) GeV						& \cite{PDG:2012}	\\
	\T $M_{B^0}$				& 5.27958(17) GeV					& \cite{PDG:2012}	\\
	\T $\tau_{B^0}$				& 1.519(7) ps						& \cite{PDG:2012}	\\
	\T $M_{K^0}$				& 0.497614(24) GeV					& \cite{PDG:2012}	\\
	\T $M_{B^\pm}$			& 5.27925(17) GeV					& \cite{PDG:2012}	\\
	\T $\tau_{B^\pm}$			& 1.641(8) ps						& \cite{PDG:2012}	\\
	\T $M_{K^\pm}$			& 0.497677(16) GeV					& \cite{PDG:2012}	\\
	\T $1/\alpha_{\rm EW}$ 		& 128.957(20)						& \cite{Jegerlehner:2008}	\\
	\T $|V_{tb}V_{ts}^*|$			& 0.0405(8)						& \cite{CKM:2013}	\\
	\T $C_1(m_b)$				& -0.257(5)  						& \cite{Altmannshofer:2008,ASe} 	\\ 
	\T $C_2(m_b)$				& 1.009(20)  						& \cite{Altmannshofer:2008,ASe} 	\\ 
	\T $C_3(m_b)$				& -0.0050(1)  						& \cite{Altmannshofer:2008,ASe} 	\\ 
	\T $C_4(m_b)$				& -0.078(2)  						& \cite{Altmannshofer:2008,ASe} 	\\ 
	\T $C_5(m_b)$				& 0.000(0)  						& \cite{Altmannshofer:2008,ASe} 	\\ 
	\T $C_6(m_b)$				& 0.001(0)  						& \cite{Altmannshofer:2008,ASe} 	\\ 
	\T $C_7^{\rm eff}(m_b)$		& -0.304(6)  						& \cite{Altmannshofer:2008,ASe} 	\\ 
	\T $C_9^{\rm eff}(m_b)$		& $4.211(84) + Y(q^2)$  				& \cite{Altmannshofer:2008,ASe} 	\\ 
	\T $C_{10}^{\rm eff}(m_b)$	& -4.103(82)  						& \cite{Altmannshofer:2008,ASe} 	\\  [0.5ex]
\hline\hline
\end{tabular}\caption{Input parameters used to calculate Standard Model observables.  Parameters not listed here are unambiguously specified in the PDG~\cite{PDG:2012}.}
\label{tab-SMparams}
\end{table}
We begin with the differential decay rate.  Following the notation of Ref.~\cite{Becirevic:2012} and restricting ourselves to the Standard Model we write
\begin{equation}
\frac{d\Gamma_\ell}{dq^2} = 2a_\ell + \frac{2}{3}c_\ell ,
\end{equation}
where $a_\ell$ and $c_\ell$ are given by
\begin{eqnarray}
a_\ell &=& \mathcal{C} \Big[ q^2 |F_P|^2 + \frac{\lambda}{4}(|F_A|^2 + |F_V|^2) + 4m_\ell^2 M_B^2 |F_A|^2 \nonumber \\
&+& 2m_\ell(M_B^2 - M_K^2 + q^2){\rm Re}(F_PF_A^*) \Big], \\
c_\ell &=& -\frac{\mathcal{C} \lambda \beta_\ell^2}{4}(|F_A|^2 + |F_V|^2),
\end{eqnarray}
with
\begin{eqnarray}
\mathcal{C} &=& \frac{G_F^2 \alpha_{EW}^2 |V_{tb} V_{ts}^*|^2}{2^9\pi^5M_B^3}\beta_\ell \sqrt{\lambda}, \\
\lambda &=& q^4 + M_B^4 + M_K^4 - 2(M_B^2M_K^2 + M_B^2q^2 + M_K^2q^2), \nonumber \\
& & \\
\beta_\ell &=& \sqrt{1-4m_\ell^2/q^2}.
\end{eqnarray}
The Standard Model expressions for $F_{P,V,A}$ are
\begin{eqnarray}
F_P &=& -m_\ell C_{10}^{\rm eff} \Big[ f_+ - \frac{M_B^2-M_K^2}{q^2}(f_0-f_+) \Big], \\
F_V &=& C_9^{\rm eff} f_+ + \frac{2m_b}{M_B+M_K}C_7^{\rm eff}f_T, \\
F_A &=& C_{10}^{\rm eff}f_+.
\end{eqnarray}
We take values for the Wilson coefficients from Ref.~\cite{Altmannshofer:2008} with estimated errors of 2\%~\cite{ASe}.  The Wilson coefficient $C_9^{\rm eff}$ is a function of $q^2$ through
\begin{eqnarray}
Y(q^2) &=& \frac{4}{3}C_3 + \frac{64}{9}C_5 + \frac{64}{27}C_6 \nonumber \\
 &-& \frac{1}{2}h(q^2,0)\left( C_3 + \frac{4}{3}C_4 + 16C_5 + \frac{64}{3}C_6 \right)  \nonumber \\
&+&h(q^2,m_c)\left( \frac{4}{3}C_1 + C_2 + 6C_3 + 60C_5 \right) \\
&-& \frac{1}{2}h(q^2,m_b)\left( 7C_3 + \frac{4}{3}C_4 + 76C_5 + \frac{64}{3}C_6 \right), \nonumber
\end{eqnarray}
where
\begin{eqnarray}
h(q^2,m) &=& -\frac{4}{9}\left( \ln \frac{m^2}{q^2} - \frac{2}{3} - x \right) -  \frac{4}{9}(2+x) \nonumber \\
&\times&\hspace{-0.04in}
	\begin{cases} 
		\sqrt{x-1} \arctan \frac{1}{\sqrt{x-1}} \hspace{-0.1in}&\!\!, x>1 \vspace{0.02in}\\
		\sqrt{1-x} \left(\ln \frac{1+\sqrt{1-x}}{\sqrt{x}} - \frac{i\pi}{2}\right)  \hspace{-0.1in}&\!\!, x\leq1,
	\end{cases}
\end{eqnarray}
and $x=4m^2/q^2$.  To compare with experiment, calculated decay rates are converted to branching fractions using the $B$ meson's mean lifetime, $\mathcal{B}_\ell = \Gamma_\ell \tau_B$.  In~\cite{Bouchard:PRL} we calculate:  Standard Model differential branching fractions; branching fractions integrated over experimentally motivated $q^2$ bins; and ratios of branching fractions.  Where possible, we compare with experimental results from BABAR~\cite{Lees:2012}, Belle~\cite{Wei:2009}, CDF~\cite{Aaltonen:2011}, and LHCb~\cite{Aaij:2012, Aaij:2012b}.

\begin{figure}[t!]
{\scalebox{0.653}{\includegraphics[angle=-90,width=0.80\textwidth]{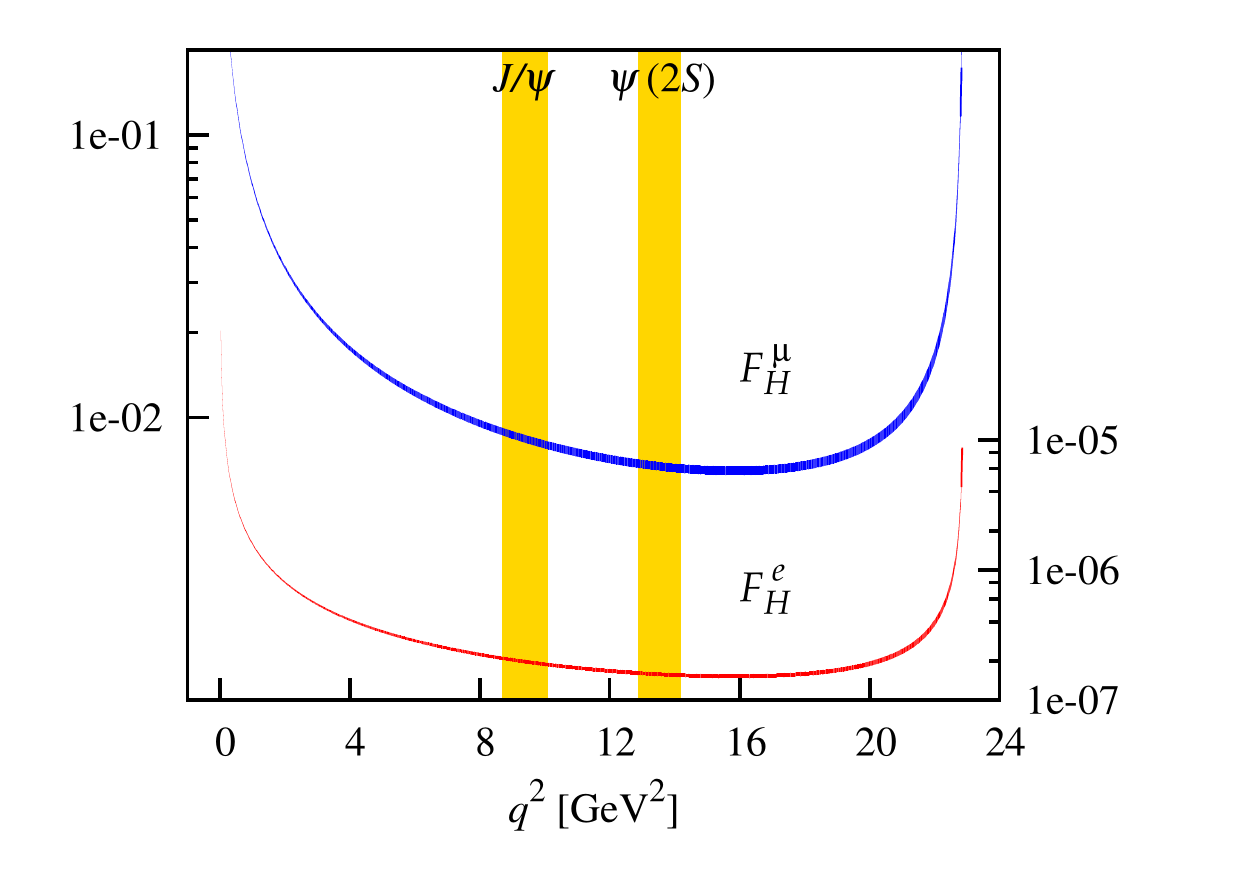}}}
%
\\
{\scalebox{0.653}{\includegraphics[angle=-90,width=0.75\textwidth]{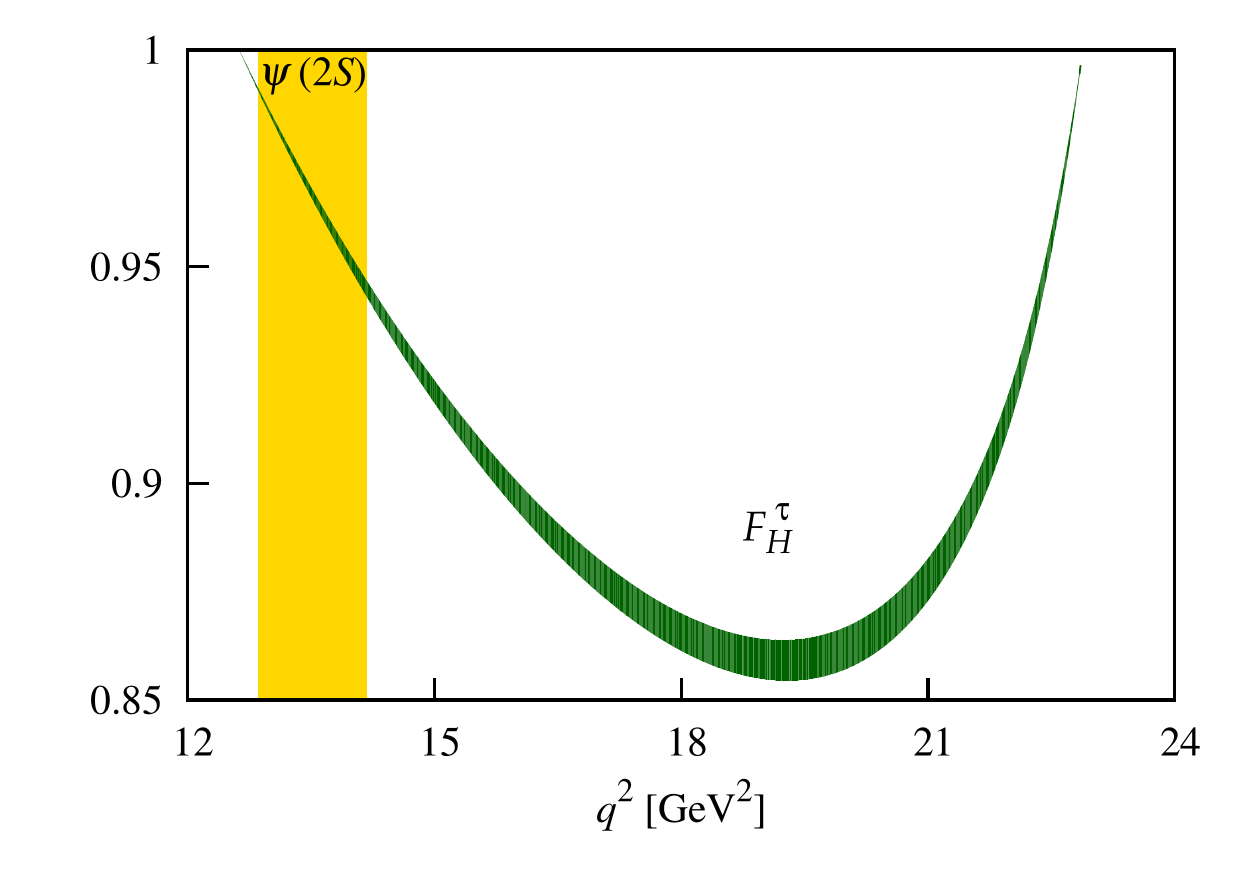}}}
\caption{The flat term in the angular distribution of the differential decay rate for ({\it top}) a light dilepton final state and ({\it bottom}) a ditau final state.}
\label{fig-FH}
\end{figure}

The angular distribution of the differential decay rate is given by
\begin{equation}
\frac{1}{\Gamma_\ell}\frac{d \Gamma_\ell}{d \cos\theta_\ell} = \frac{1}{2}F_H^\ell + A_{FB}^\ell \cos\theta_\ell + \frac{3}{4}(1-F_H^\ell)(1-\cos^2\theta_\ell),
\end{equation}
where $\theta_\ell$ is the angle between the $B$ and $\ell^-$ as measured in the dilepton rest frame.  The ``flat term" $F_H^\ell$, introduced by Bobeth {\it et al.} in Ref.~\cite{Bobeth:2007}, is suppressed by $m_\ell^2$ in the Standard Model and potentially sensitive to new physics~\cite{Bobeth:2013, Becirevic:2012}.  The ``forward-backward asymmetry" $A_{FB}^\ell$ is zero in the Standard Model (up to negligible QED contributions~\cite{Demir:2000, Bobeth:2007}) so is also a sensitive probe of new physics.  The flat term in the angular distribution is given by~\cite{Bobeth:2007}
\begin{equation}
F_H^\ell(q^2_{\rm low}, q^2_{\rm high}) = \frac{ \int_{q^2_{\rm low}}^{q^2_{\rm high}} dq^2\ (a_\ell + c_\ell) }{ \int_{q^2_{\rm low}}^{q^2_{\rm high}} dq^2\ (a_\ell + \frac{1}{3}c_\ell) }.
\end{equation}
This observable is constructed as a ratio to reduce uncertainties.  In~\cite{Bouchard:PRL} we evaluate $F_H^{e,\mu,\tau}$ in experimentally motivated $q^2$ bins.
Taking $a_\ell$ and $c_\ell$ as average values for a bin centered at $q^2=\nicefrac{1}{2}(q^2_{\rm low}+q^2_{\rm high})$, the $q^2$ dependence of $F_H^\ell$ is 
given by
\begin{equation}
F_H^\ell(q^2) = \frac{a_\ell + c_\ell}{a_\ell+\frac{1}{3}c_\ell}.
\end{equation}
Standard Model predictions are shown in Fig.~\ref{fig-FH} for each of the dilepton final states.

\section{ Summary and Outlook }
Using NRQCD $b$ and HISQ light valence quarks with the MILC $2+1$ dynamical asqtad configurations, we report on the first unquenched lattice QCD calculation of the form factors for the rare decay $B \to K \ell^+\ell^-$.  We extrapolate our form factor results over the full kinematic range of $q^2$ using the model-independent $z$ expansion.

Using our form factor results we
determine ratios of form factors, useful both in constraining new physics and verifying effective field theory relations; 
discuss the calculation of Standard Model differential branching fractions; 
and calculate the flat term in the angular distribution of the differential decay rate.
In~\cite{Bouchard:PRL} we present a detailed study of the phenomenological implications of these form factors on several Standard Model observables, including comparison with experiments and previous calculations.

\section*{ Acknowledgements}
This research was supported by the DOE and NSF.
We thank the MILC collaboration for making their asqtad $N_f=2+1$ gauge field configurations available.  
Computations were carried out at the Ohio Supercomputer Center and on facilities of the USQCD collaboration funded by the Office of Science of the U.S. DOE.

\appendix

\section{ Modified $z$ Expansion}
\label{app-modz}

In recent $D\to\ K(\pi)$ semileptonic decay analyses we developed the modified $z$ expansion~\cite{Na:2010, Na:2011} in which the chiral/continuum and kinematic extrapolations are performed in a single step.  
These works, and our more recent $D\to K$ analysis~\cite{Koponen:2013}, demonstrate the utility of the modified $z$ expansion in semileptonic $D$ decays.
The kinematic extrapolations required for semileptonic $D$ decays are, however, mild compared to those needed for semileptonic $B$ decays.  
In addition to the two-step chiral/continuum and kinematic extrapolation of Sec.~\ref{sec-2step}, we perform the extrapolations simultaneously via the modified $z$ expansion.
This allows us to test the modified $z$ expansion for semileptonic decays requiring sizable kinematic extrapolation and provides a consistency check of our final results.
We modify the BCL parameterized $z$ expansion~\cite{Bourrely:2010} and fit the form factor data to
\begin{eqnarray}
f_0(q^2) &=& B_0 \sum_{k=0}^{K} a^0_k D^0_k\,  z(q^2)^k, \label{eq-modzf0} \\
f_i(q^2) &=& \frac{B_i}{P_i(q^2)} \sum_{k=0}^{K-1} a^i_k D^{i}_k \Big[ z(q^2)^k -(-1)^{k-K}\frac{k}{K} z(q^2)^K \Big], \nonumber \label{eq-modzfpT} \\
\end{eqnarray}
where $i=+,T$ and
\begin{eqnarray}
B &=& 1 + b_1(aE_K)^2 + b_2(aE_K)^4, \label{eq-B} \\
D_k &=& 1 + c_1^{(k)} x_l + c_2^{(k)} x_l ( \log{x_l} + \delta ) + c_3^{(k)} \delta x_s \nonumber \\
 &+& d_1^{(k)}(\nicefrac{a}{r_1})^2 +  d_2^{(k)}(\nicefrac{a}{r_1})^4 \nonumber \\
 &+& e^{(k)} \left( \tfrac{1}{2}\delta M_\pi^2 + \delta M_K^2 \right), \label {eq-Dk} \\
 x_l &=& \frac{( M_\pi^{\rm HISQ} )^2}{(4\pi F_\pi)^2}, \\
 \delta x_s &=& \frac{( M_{\eta_s}^{\rm HISQ} )^2 - M_{\eta_s^{\rm phys}}^2}{(4\pi F_\pi)^2}, \\
 \delta M_{\pi,K}^2 &=& \frac{ (M_{\pi,K}^{\rm asqtad} )^2 - ( M_{\pi,K}^{\rm HISQ} )^2}{(4\pi F_\pi)^2}.
\end{eqnarray}
Indices specifying the form factor ($0,+,T$) are implicitly assumed in Eqs.~(\ref{eq-B},~\ref{eq-Dk}) above.  

In the modified $z$ expansion $P$ and $z$ are calculated separately for each ensemble using simulation masses and momenta.  
We include the function $B$ to account for momentum-dependent discretization effects.  
The function $D_k$ contains the NLO chiral analytic terms with coefficients $c_i$, $e$, and $d_1$ and the NNLO $d_2$ term.  
The $c_1$ and $c_2$ terms extrapolate in light quark mass and accommodate finite volume effects via a shift in the chiral log~\cite{Bernard:2002}.  We calculate the shift $\delta$ for each ensemble using
\begin{equation}
\delta = \frac{4}{M_\pi L} \sum_{{\bf r} \neq {\bf 0}} \frac{K_1(rM_\pi L)}{r},
\end{equation}
where ${\bf r}$ is a three-vector whose integer components run over all lattice sites ($r=|{\bf r}|$) and $K_1$ is the order one modified Bessel function of the second kind.  To take the infinite volume limit, we set $\delta=0$.
The $c_3$ term absorbs strange quark mass mistuning by comparing the $\eta_s$ meson mass obtained from simulation strange quark masses~\cite{Na:2010} to the ``physical" $\eta_s$ mass from~\cite{Davies:2009}.  The $e$ term absorbs slight differences between the valence and sea quark masses due to our mixed (HISQ and asqtad) action.  
The $d_i$ terms account for discretization effects. 
\begin{figure*}[t!]
\hspace{-0.07in}  
\subfloat[][\label{fig-modzf0fpC123}Fit to data for $f_{0,+}$ on the coarse ensembles.]
{\scalebox{0.98}{\includegraphics[angle=-90,width=0.5\textwidth]{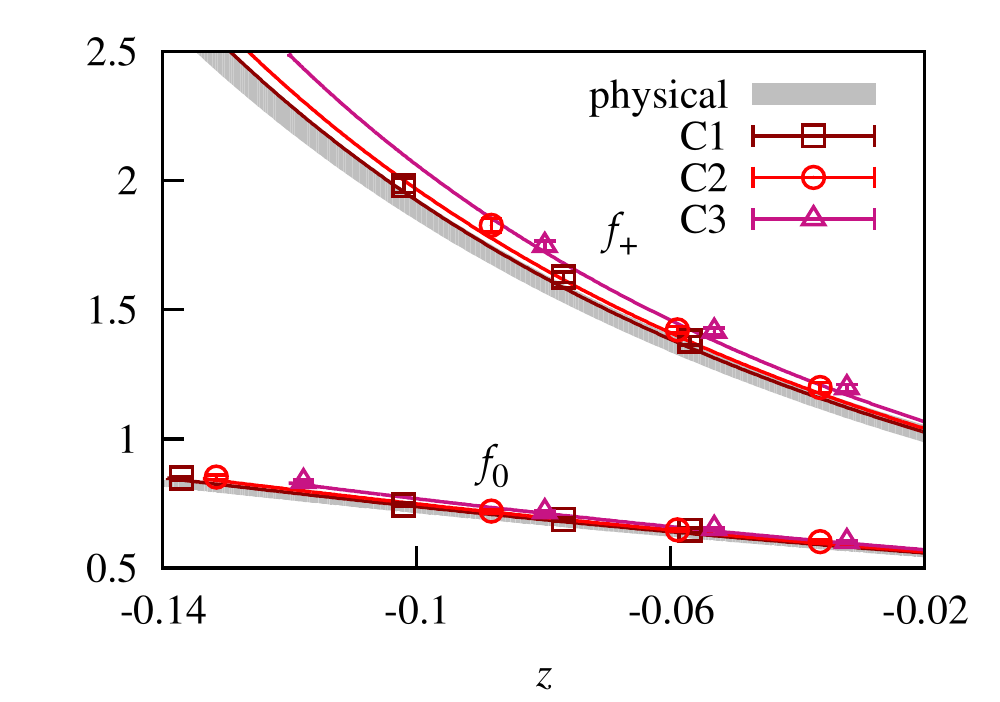}}}
\subfloat[][\label{fig-modzifTC123}Fit to data for $f_T$ on the coarse ensembles.]
{\scalebox{0.98}{\includegraphics[angle=-90,width=0.5\textwidth]{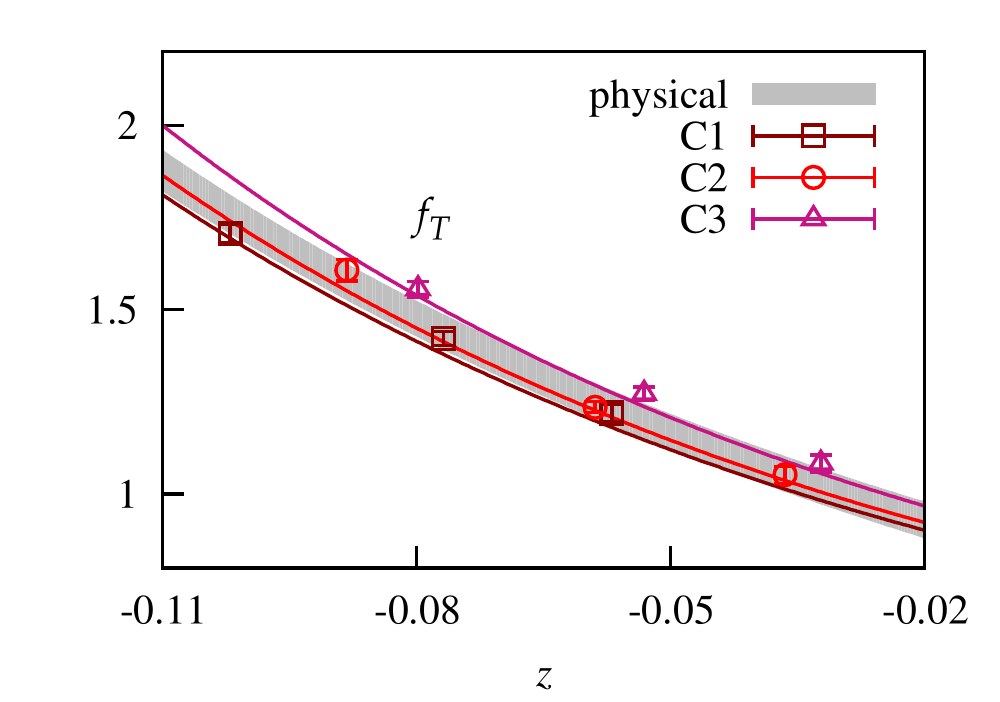}}}
\\
\subfloat[][\label{fig-modzf0fpF12}Fit to data for $f_{0,+}$ on the fine ensembles.]
{\scalebox{0.98}{\includegraphics[angle=-90,width=0.5\textwidth]{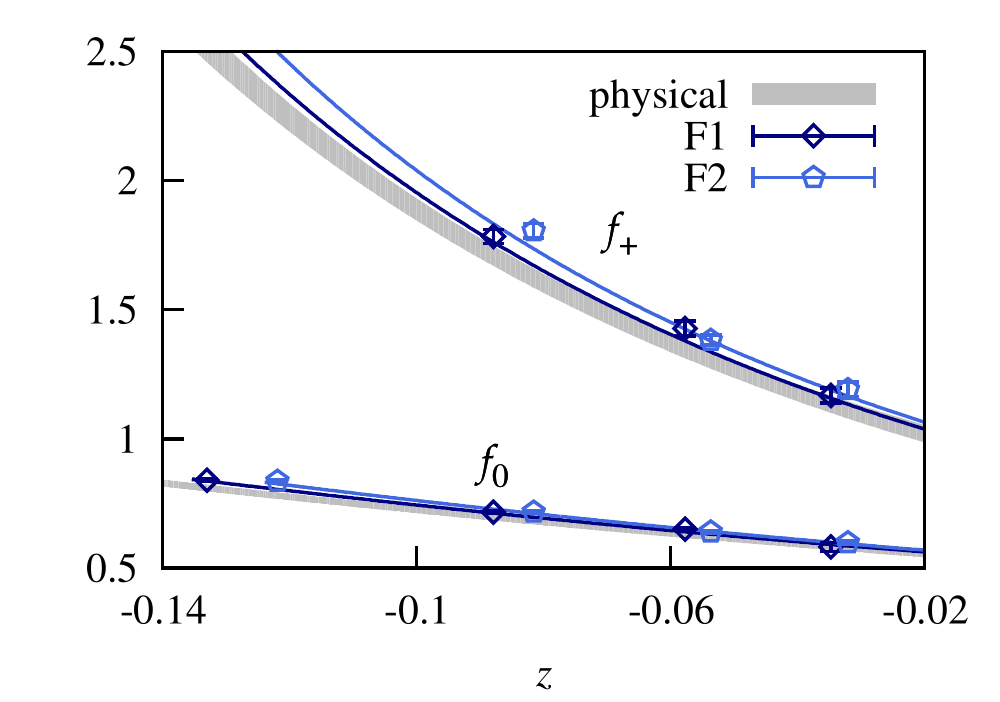}}}
\subfloat[][\label{fig-modzifTF12}Fit to data for $f_T$ on the fine ensembles.]
{\scalebox{0.98}{\includegraphics[angle=-90,width=0.5\textwidth]{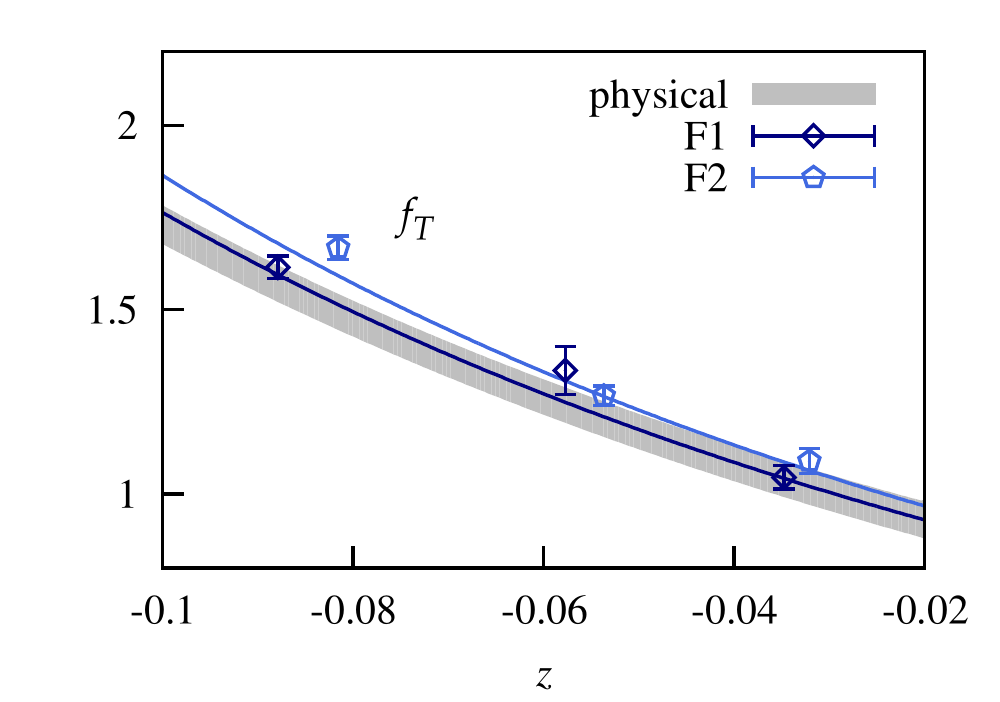}}}
\caption{Results of the simultaneous, modified $z$ expansion fit to the data for $f_0$, $f_+$, and $f_T$.  In each plot, curves indicate the fit to data on each ensemble and the gray band shows the extrapolated physical result.}
\label{fig-modzfits}
\end{figure*}
As in Eq.~(\ref{eq-hvy_disc}) we account for heavy-quark mass-dependent discretization effects by making the $d_i^{(k)}$ mild functions of $am_b$
\begin{eqnarray}
  d_1^{(k)} &\to& d_1^{(k)}( 1+ f_1^{(k)}\, \delta x_b + f_2^{(k)}\, \delta x_b^2\, ), \nonumber \\
  d_2^{(k)} &\to& d_2^{(k)}( 1+ f_3^{(k)}\, \delta x_b + f_4^{(k)}\, \delta x_b^2\, ),
  \label{eq-hvy_disc_modz}
\end{eqnarray}
with $\delta x_b$ as defined in Sec. \ref{sec-ChPT}.  Light-quark mass-dependent discretization effects are similarly accounted for 
\begin{eqnarray}
  d_1^{(k)} &\to& d_1^{(k)}( 1+ g_1^{(k)} x_l + g_2^{(k)} x_l^2 ), \nonumber \\
  d_2^{(k)} &\to& d_2^{(k)}( 1+ g_3^{(k)} x_l + g_4^{(k)} x_l^2 ).
\end{eqnarray}
As a result of these two modifications, $d_1^{(k)}$ in Eq.~(\ref{eq-Dk}) is multiplied by $( 1+ f_1^{(k)} \delta x_b + f_2^{(k)} \delta x_b^2 )( 1+ g_1^{(k)} x_l + g_2^{(k)} x_l^2 )$, and similarly for $d_2^{(k)}$.


We impose the kinematic constraint $f_0(0)=f_+(0)$ ensemble by ensemble using the method outlined in Sec.~\ref{sec-stdzexp}.  
The selection of priors for the simultaneous, modified $z$ expansion is discussed in Sec.~\ref{app-modzpriors}, where prior values and fit results are listed in Tables~\ref{tab-modzpriorsI} and~\ref{tab-modzpriorsII}.  


Fit results for each ensemble, along with the physical extrapolated band, are shown in Fig.~\ref{fig-modzfits}.  A comparison of these plots to those of the chiral/continuum extrapolation in Fig.~\ref{fig-chpt} demonstrates the consistency of the two approaches in the region of $q^2$ for which we simulate.  
The $\chi^2/{\rm dof}$ for this fit is $35.7/60$.

\begin{figure*}[t!]
{\scalebox{0.98}{\includegraphics[angle=-90,width=0.5\textwidth]{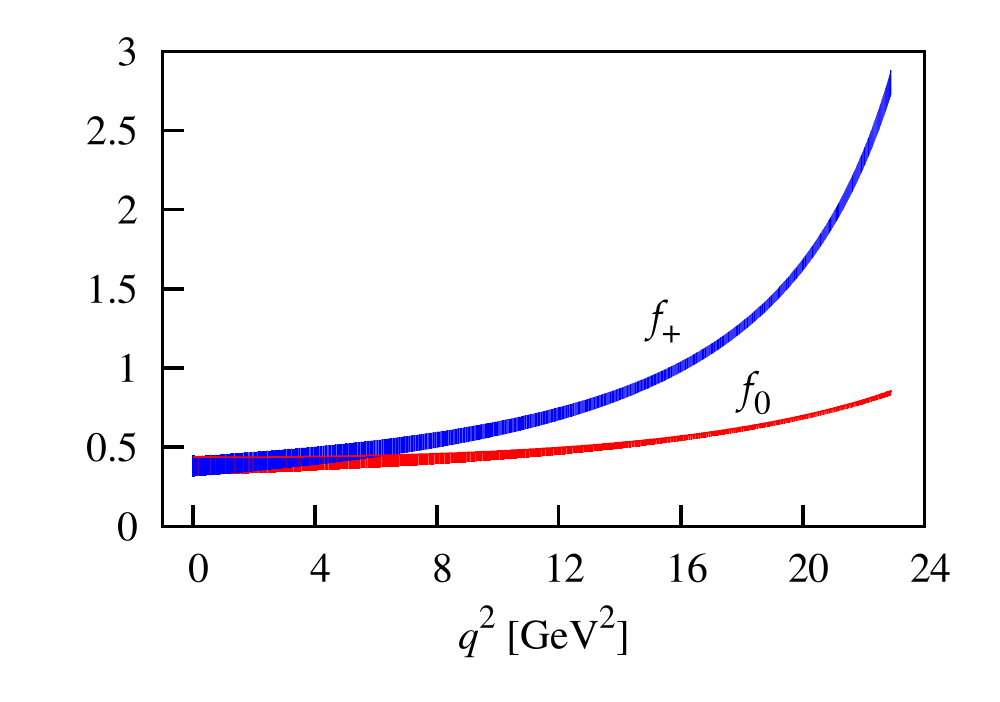}}}
%
{\scalebox{0.98}{\includegraphics[angle=-90,width=0.5\textwidth]{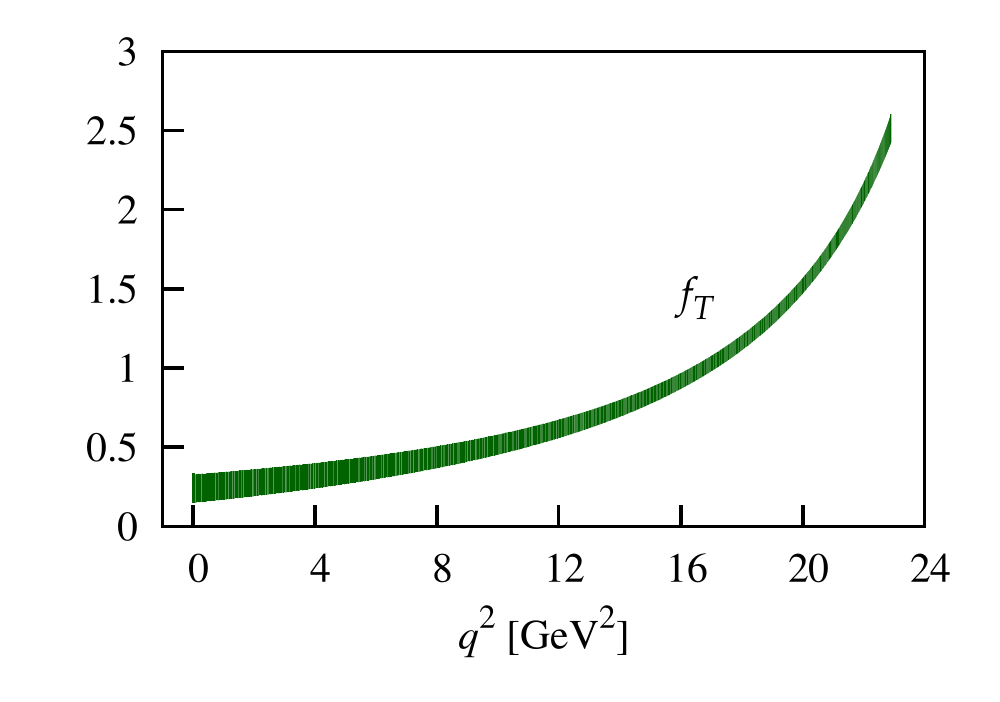}}}
\vspace{-0.15in}
\caption{Modified $z$ expansion curves for the form factors over the full kinematic range of $q^2$.  The width of the plotted bands shows the error associated with the fit and omits additional systematic errors discussed in Sec.~\ref{sec-syserr}.}
\label{fig-modzfull}
\end{figure*}

Note that no synthetic data points are needed in the modified $z$ expansion approach and the results are relevant over the full range of $q^2$.
Extrapolated physical results from the modified $z$ expansion are shown over the full kinematic range of $q^2$ in Fig.~\ref{fig-modzfull}.
 
\begin{figure}[t]
{\scalebox{0.98}{\includegraphics[angle=-90,width=0.5\textwidth]{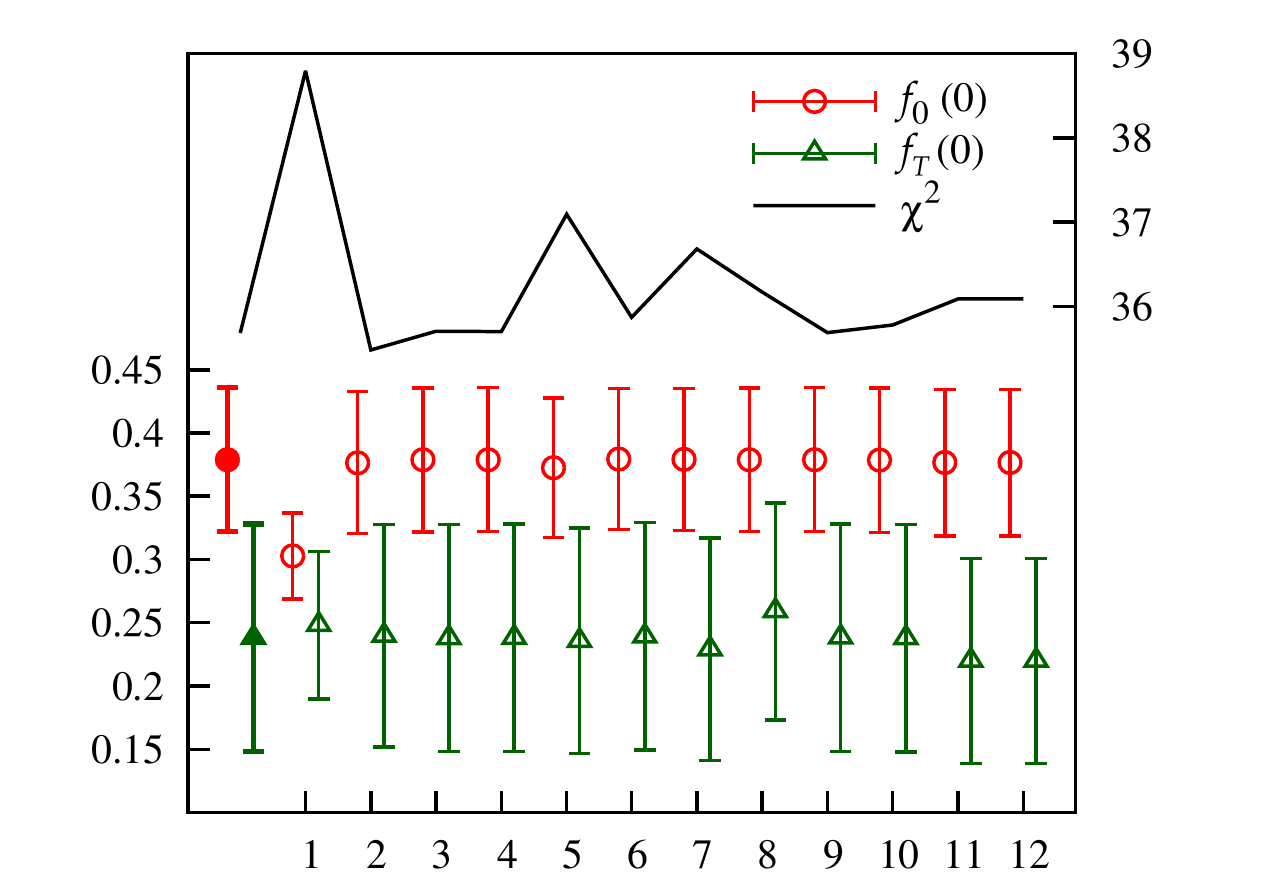}}}
\caption{Modified $z$ expansionion results for $f_{0,T}(0)$, the left-most points, are stable under various modifications to the fit ansatz.  The $x$-axis label corresponds to the modifications listed in the text.}
\label{fig-modzstab}
\end{figure}
We study the stability of these fit results under variations of the fit ans\"atze.  
As in Sec.~\ref{sec-stdzexp} we consider how fit variations impact $\chi^2$ and the values of extrapolated fit results at  $q^2=0$.   
We take as our standard fit ans\"atze Eqs.~(\ref{eq-modzf0},~\ref{eq-modzfpT}) with $K=3$ and $D_k$ as in Eq.~(\ref{eq-Dk}) with the modifications of Eqs.~(\ref{eq-hvy_disc},~\ref{eq-lt_disc}).  We test the stability of modified $z$ expansion fits with respect to the following modifications:
\begin{enumerate}
  \item[1.]  Perform $z$ expansion through $\mathcal{O}(z^2)$.  
  \item[2.]  Perform $z$ expansion through $\mathcal{O}(z^4)$.
  \item[3.]  Drop the NLO analytic sea-quark mass term $e^{(k)} \left( \tfrac{1}{2}\delta M_\pi^2 + \delta M_K^2 \right)$ from $D_k$.
  \item[4.]  Drop the NLO analytic strange valence-quark mass term $c_3^{(k)} \delta x_s$ from $D_k$.
  \item[5.]  Drop the NLO analytic light valence-quark mass term $c_1^{(k)}x_l$ from $D_k$.
  \item[6.]  Drop the discretization term $d_2^{(k)} (\nicefrac{a}{r_1})^{4}$ from $D_k$.
  \item[7.]  Drop the momentum dependent discretization term $b_2(aE_K)^4$ from Eq.~(\ref{eq-B}).
  \item[8.]  Drop the $am_b$-dependent discretization effects from the $d_i^{(k)}$.
  \item[9. ]  Drop the light-quark mass-dependent discretization effects from the $d_i^{(k)}$.
  \item[10.]  Drop the finite volume effects.
  \item[11.]  Add all possible NNLO analytic terms:
  \begin{eqnarray}
  D_k &=&  \text{Eq.}~(\ref{eq-Dk}) +  h^{(k)} x_l^2 + i^{(k)} \left( \tfrac{1}{2}\delta M_\pi^2 + \delta M_K^2 \right)^2    \nonumber \\
 &+&  j^{(k)} \delta x_s^2 + k^{(k)} x_l \delta x_s +  l^{(k)} x_l \left( \tfrac{1}{2}\delta M_\pi^2 + \delta M_K^2 \right)  \nonumber \\
 &+& m^{(k)} \delta x_s \left( \tfrac{1}{2}\delta M_\pi^2 + \delta M_K^2 \right) + n^{(k)} \delta x_s (\nicefrac{a}{r_1})^2 \nonumber \\
 &+& o^{(k)} x_l (\nicefrac{a}{r_1})^2 + p^{(k)}\!\! \left( \tfrac{1}{2}\delta M_\pi^2 + \delta M_K^2 \right)\!\! (\nicefrac{a}{r_1})^2.
  \end{eqnarray}
  \item[12.]  Add the static limit constraint as in test 8 of Sec.~\ref{sec-stdzexp}.
\end{enumerate}

Fig.~\ref{fig-modzstab} shows the results of these tests for $f_0(0)$ and $f_T(0)$ -- the kinematic constraint ensures $f_+(0)$ is equivalent to $f_0(0)$.
Tests 1 and 2 show that by $\mathcal{O}(z^3)$ the fit central values have stabilized and the errors have saturated.  Adding additional terms of higher order in $z$ has no effect on the fit.  This suggests our fit errors include an adequate estimate of the error associated with truncating the $z$ expansion.  Tests 3 and 4 have no noticeable effect on the fit and indicate effects due to mixed action mass differences and strange quark mass mistunings are negligible.  Test 5 results in a slight increase in $\chi^2$.  To ensure errors associated with truncating the perturbative chiral expansion are accounted for in our fit, we include this term.  Tests 6, 7, and 8 show consistent fit central values.  We include the tested terms to ensure the error associated with these discretization effects is accommodated in our fit result.  Test 9 shows no noticeable change in fit central values or errors and indicates that, to the extent they are present in our data, light-quark mass-dependent discretization effects such as taste violations are adequately accounted for by other discretization terms in the fit ansatz.  Test 10 shows finite volume effects to be negligible.  Test 11 includes all NNLO chiral analytic terms and shows negligible change in fit central values, errors, or $\chi^2$ relative to our final result.  This indicates our final result adequately accounts for fit errors associated with truncating the chiral expansion.  Test 12 adds the fit constraint derived from Hill's~\cite{Hill:2006} static limit relation in Eq.~(\ref{eq-statreln}) and demonstrates excellent consistency of the fit results with the symmetry relation.
In addition to these tests we have verified the consistency of fit results for $f_0(0)$ and $f_+(0)$ with and without the kinematic constraint $f_0(0)=f_+(0)$.  With the constraint removed, fit result central values for $f_0(0)$ and $f_+(0)$ shift by $\sim\! 0.1\sigma$ and fit errors increase by $\sim\! 25\%$.  The constraint remains satisfied within errors.

\begin{figure}[h!]
{\scalebox{1}{\includegraphics[angle=-90,width=0.5\textwidth]{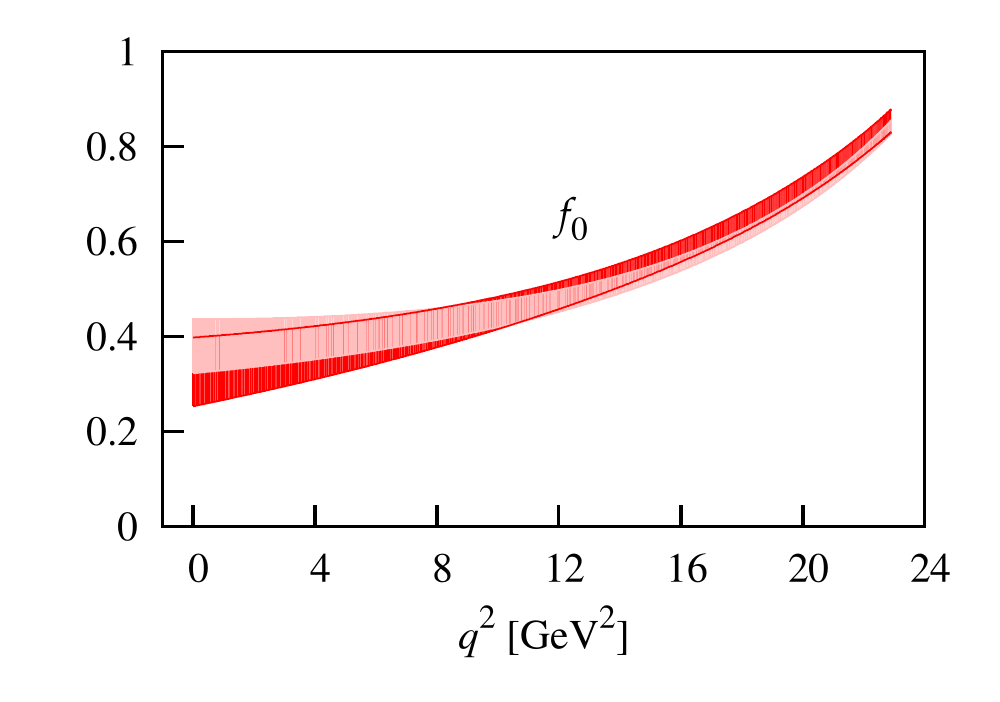}}}
\\
{\scalebox{1}{\includegraphics[angle=-90,width=0.5\textwidth]{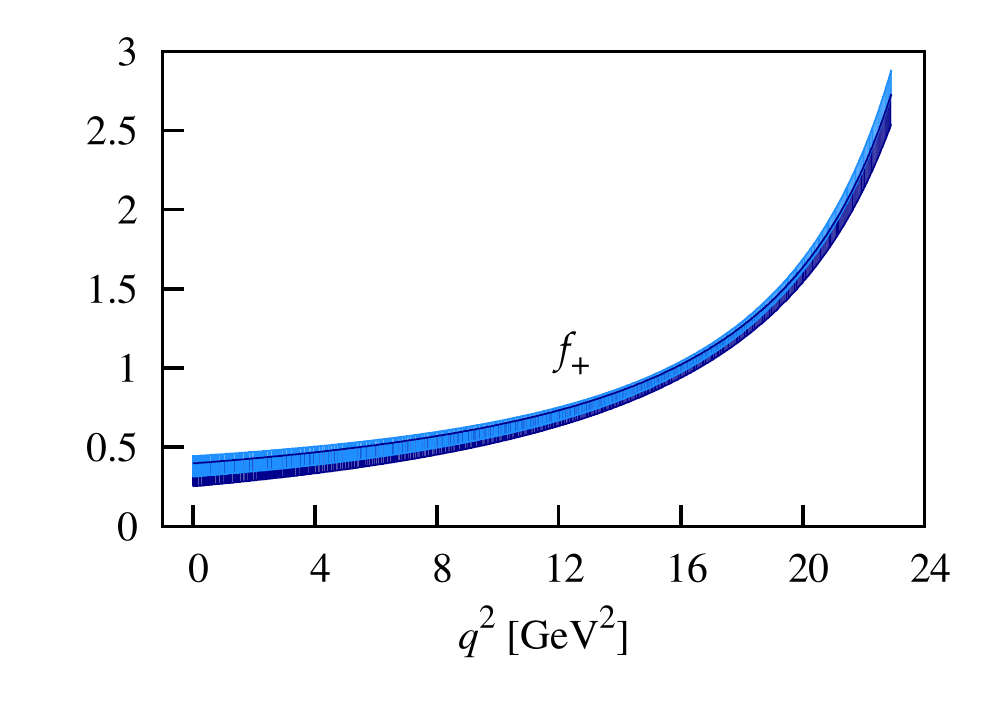}}}
\\
%
{\scalebox{1}{\includegraphics[angle=-90,width=0.5\textwidth]{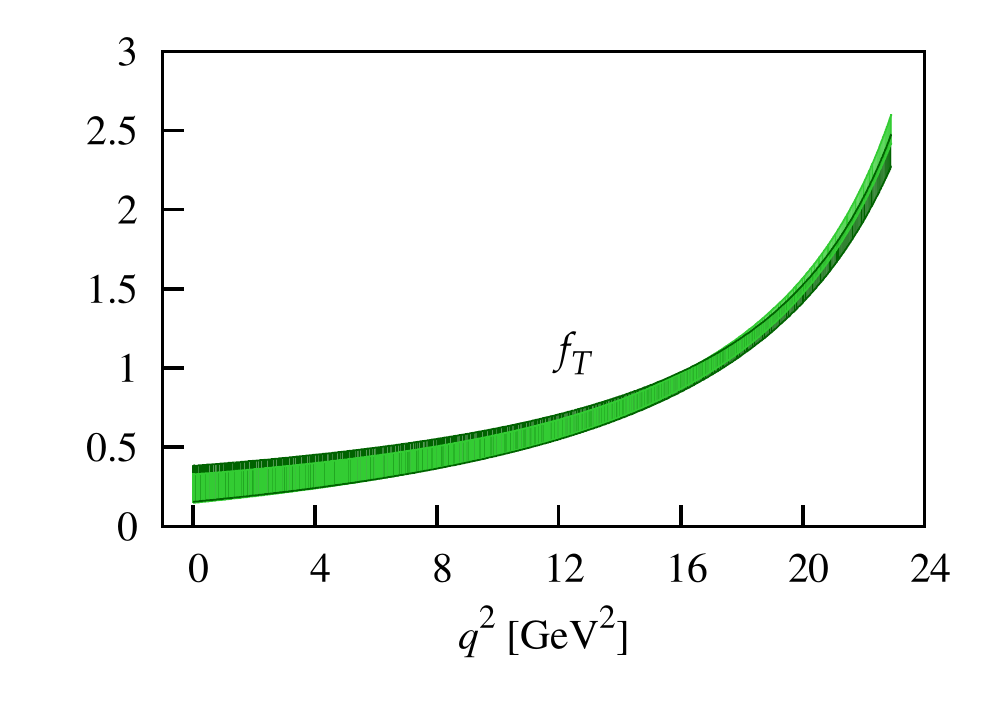}}}
\caption{Comparison of form factors obtained from the two step chiral/continuum and kinematic extrapolation (darker shade) and those obtained from the modified $z$ expansion (lighter shade).}
\label{fig-compare}
\end{figure}

Results of the two-step chiral/continuum and kinematic extrapolation presented in Sec.~\ref{sec-2step} are compared to those of the modified $z$ expansion in Fig.~\ref{fig-compare}.  The modified $z$ expansion is shown to be consistent with the two-step analysis in semileptonic decays requiring significant extrapolation in $q^2$.

\section{ Prior Selection and Fit Results }
\label{app-priors}
In this section of the appendix we discuss the selection of priors for each of the fits performed in this work.  We also provide comparison of fit results with the priors.

\setlength{\tabcolsep}{0.06in}

\subsection{$B$ Meson Correlators}
\label{app-Bpriors}
\begin{figure}[t!]
\subfloat[][\label{fig-Meff} $M_B^{\rm eff}(t)$ vs. $\nicefrac{t}{a}$.]
{\scalebox{1}{\includegraphics[angle=-90,width=0.5\textwidth]{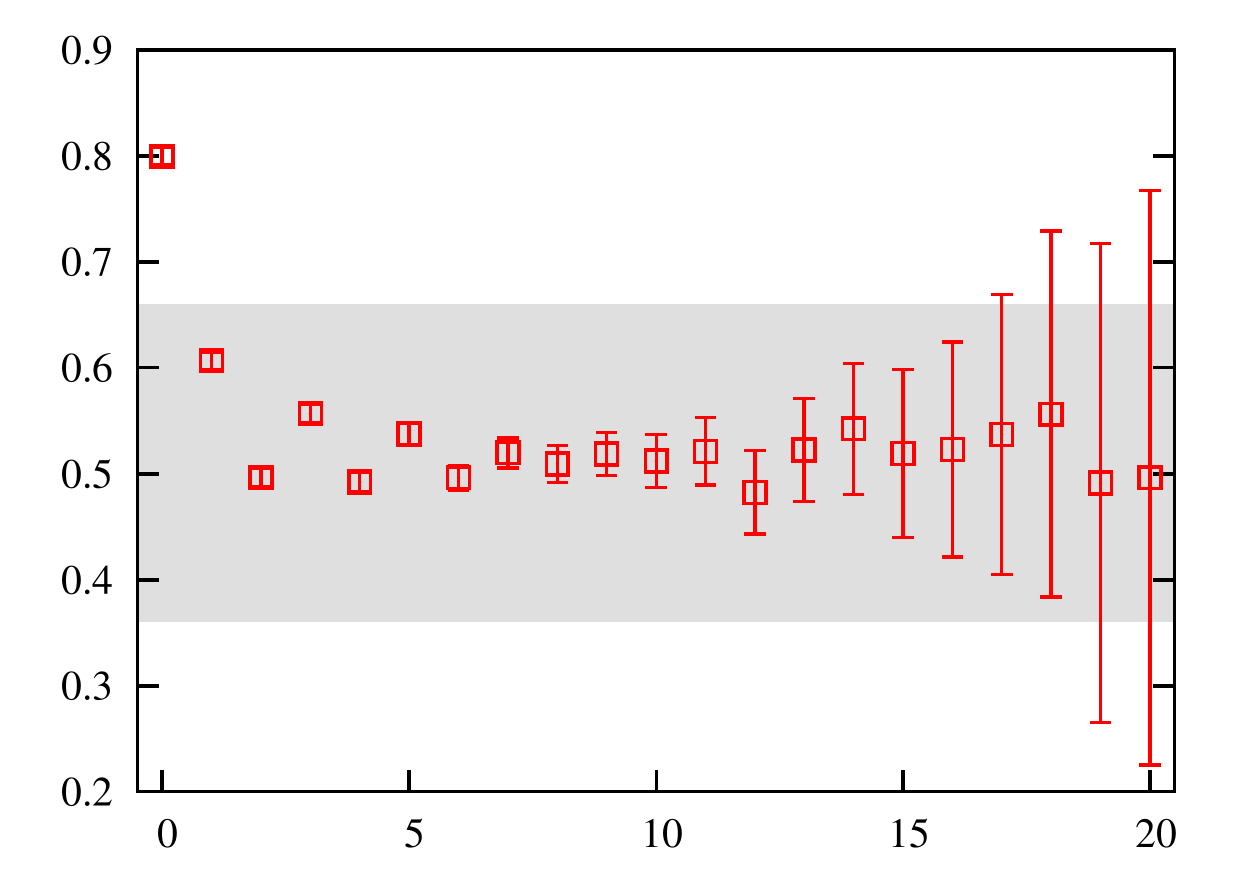}}}
\\
\subfloat[][\label{fig-Zeff}$C_B^{\,ls, {\rm eff}}(t)$ vs. $\nicefrac{t}{a}$.]
{\scalebox{1}{\includegraphics[angle=-90,width=0.5\textwidth]{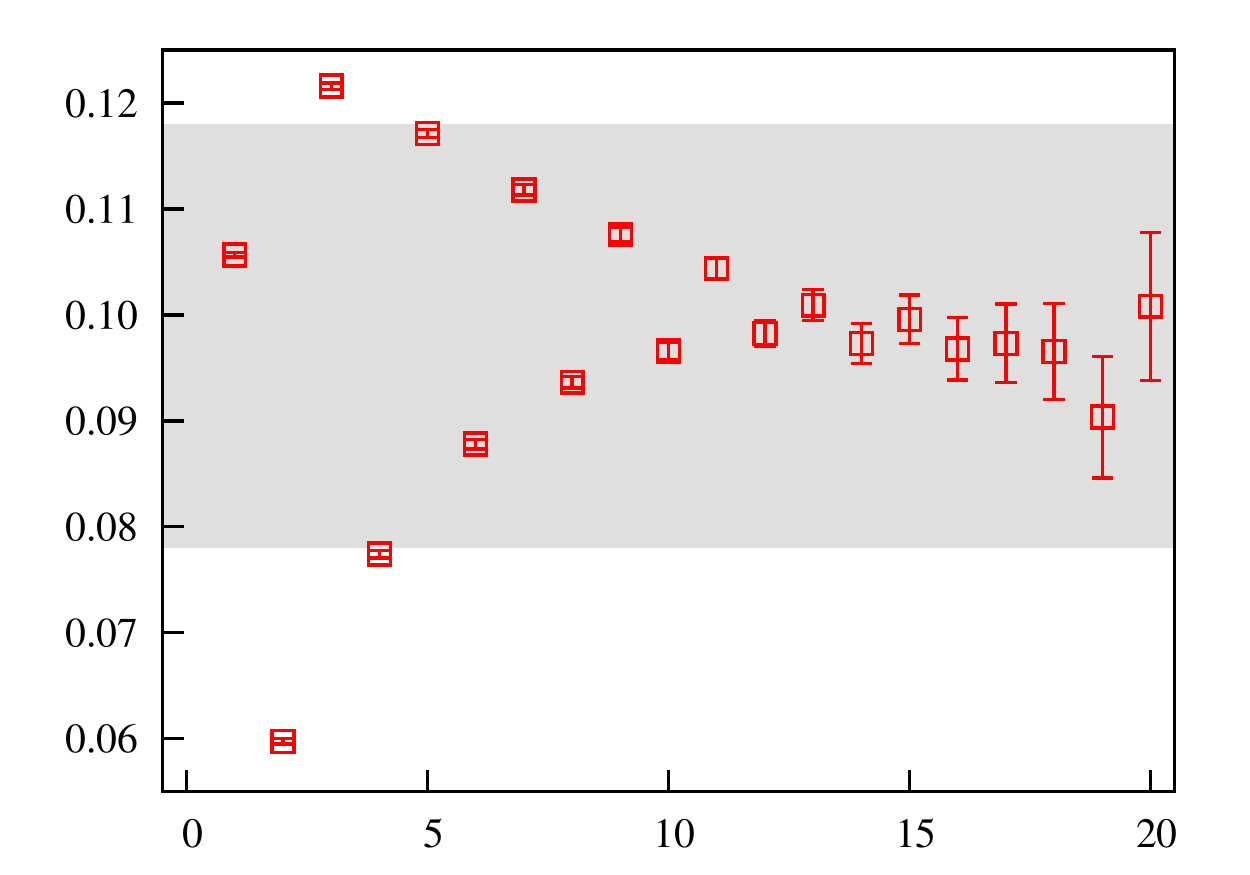}}}
\caption{$B$ meson effective mass ({\it top}) and amplitude ({\it bottom}) plots for ensemble C3.  Shaded bands indicate the choice of priors for $aE_B^{{\rm sim}(0)}$, $b^{l(0)}$, and $b^{s(0)}$.}
\label{fig-BMZeff}
\end{figure}
\begin{table}[t]
\begin{tabular}{clccc}
\hline\hline 
\\ [-3ex]
	\T ens   	&			&$aE_B^{{\rm sim}(0)}$		& $b^{l(0)}$		& $b^{s(0)}$ 	\\ [0.5ex]
	\hline
	\T C1 	& prior		& 0.49(21)		& 0.114(26)		& 0.78(16)	\\
			& fit			& 0.4997(14)		& 0.12011(99)		& 0.8194(31)	\\
			& & & & \\
	\T C2	& prior		& 0.51(15)		& 0.130(20)		& 0.84(18)	\\
			& fit			& 0.5084(19)		& 0.1242(13)		& 0.8191(52)	\\
			& & & & \\
	\T C3	& prior		& 0.51(15)		& 0.130(20)		& 0.78(13) 	\\
			& fit			& 0.5118(23)		& 0.1269(20)		& 0.8379(73)	\\
			& & & & \\
	\T F1		& prior		& 0.38(13)		& 0.071(13)		& 0.73(11) 	\\
			& fit			& 0.3820(16)		& 0.0733(12)		& 0.7705(65)	\\
			& & & & \\
	\T F2		& prior		& 0.39(9)			& 0.079(13)		& 0.80(11) 	\\
			& fit			& 0.3863(16)		& 0.0752(14)		& 0.7828(78)	\\ [0.5ex]
\hline\hline
\end{tabular}\caption{$B$ meson ground state priors and fit results.}
\label{tab-Bfits}
\end{table}
Priors for the ground state masses and amplitudes are obtained from the long time behavior of the effective mass 
\begin{eqnarray}
M^{\rm eff}_B(t) &=& \frac{1}{2} \log\Bigg[ \frac{ C_B^{\alpha\beta}(t) }{ C_B^{\alpha\beta}(t+2) } \Bigg], \\
{\rm prior}[aE_B^{{\rm sim}(0)}] &\sim&M^{\rm eff}_B({\rm long}\ t),
\end{eqnarray}
and amplitude
\begin{eqnarray}
C_B^{\alpha\beta, \rm eff}(t) &=& C_B^{\alpha\beta}(t)\ e^{M^{\rm eff}_B t}, \\
{\rm prior}[b^{\alpha(0)}\ b^{\beta(0)}] &=& C_B^{\alpha\beta, \rm eff}({\rm long}\ t).
\end{eqnarray}
Representative effective mass and amplitude plots are shown in Fig.~\ref{fig-BMZeff}.
Values for ground state priors, and fit results, are given in Table~\ref{tab-Bfits}.  Excited state mass and amplitude priors are based on the PDG~\cite{PDG:2012} and set according to
\begin{eqnarray}
b^{\alpha (n>0)} &=& 0.1 \pm 1.0, \nonumber \\
E^{{\rm sim}(1)}_B - E^{{\rm sim}(0)}_B &=& (400\pm 200) \text{ MeV}, \nonumber \\
E^{{\rm sim}(n+2)}_B - E^{{\rm sim}(n)}_B &=& (600\pm 600) \text{ MeV}.
\end{eqnarray}
Positive parity oscillating states are represented by $E_B^{{\rm sim}(\rm odd)}$ and negative parity even states by $E_B^{{\rm sim}(\rm even)}$.  Positive energy splittings are guaranteed by parameterizing them as $\delta=\log(\Delta E)$ and fitting $\delta$, then reconstructing the towers of states using $\exp(\delta)$.

\subsection{Kaon Correlators}
\label{app-Kpriors}
\begin{figure}[t!]
\subfloat[][\label{fig-Meff} $M_{K(110)}^{\rm eff}(t)$ vs. $\nicefrac{t}{a}$.]
{\scalebox{1}{\includegraphics[angle=-90,width=0.5\textwidth]{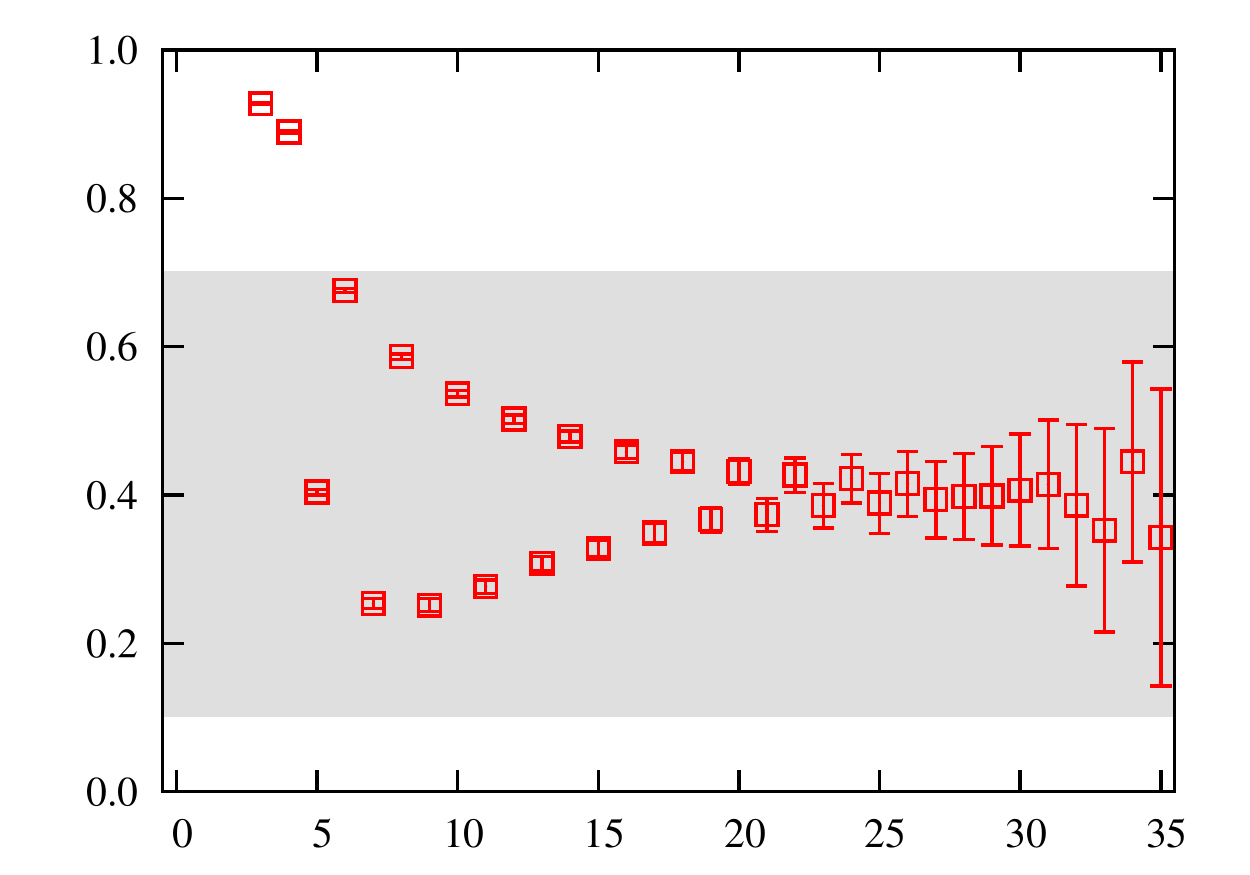}}}
\\
\subfloat[][\label{fig-Zeff}$C_{K(110)}^{\rm eff}(t)$ vs. $\nicefrac{t}{a}$.]
{\scalebox{1}{\includegraphics[angle=-90,width=0.5\textwidth]{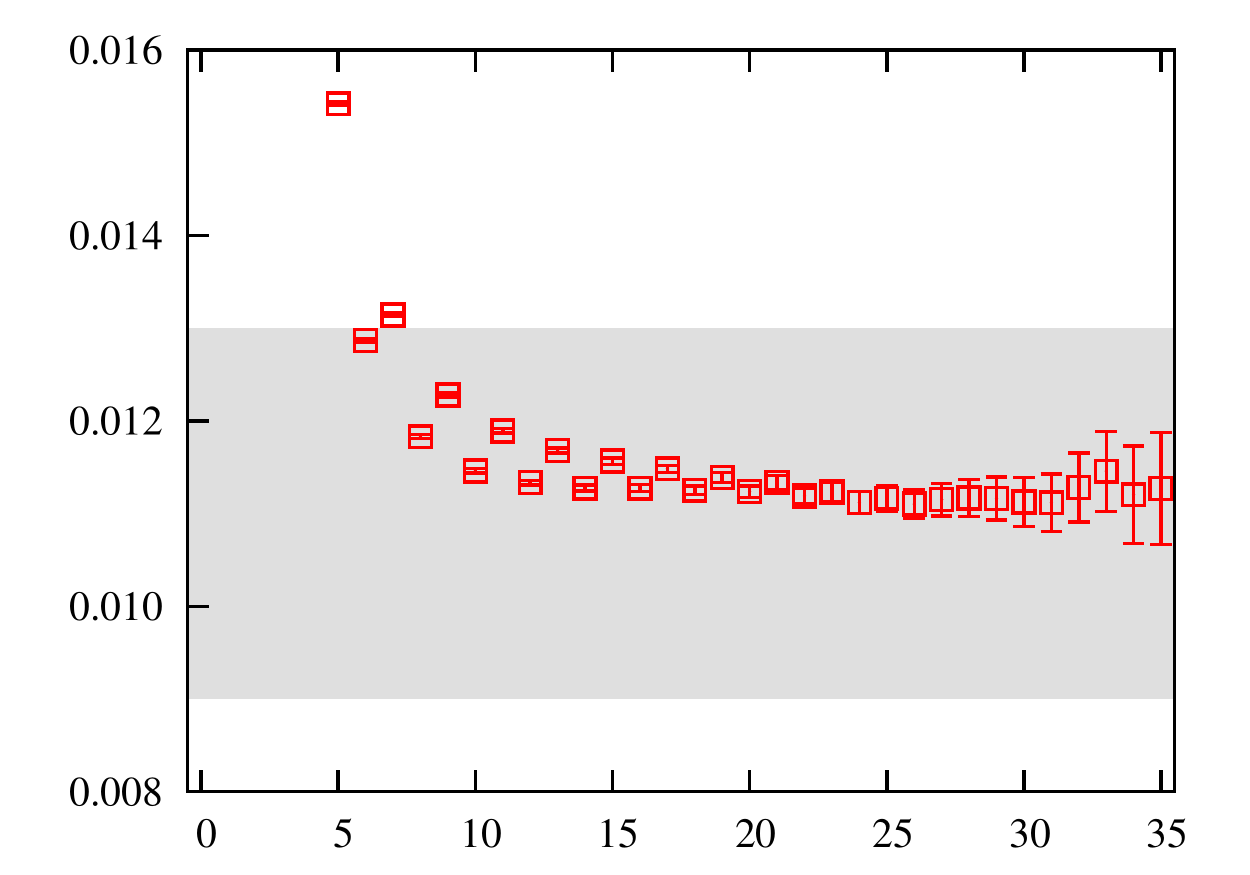}}}
\caption{Kaon, with momentum $\nicefrac{2\pi}{L}(1,1,0)$, effective mass ({\it top}) and amplitude ({\it bottom}) plots for ensemble F2.  Shaded bands indicate the choice of priors for $aM_K^{(0)}$ and $d_{110}^{(0)}$.}
\label{fig-KMZeff}
\end{figure}
Priors for the ground state energies and amplitudes are obtained from the long time behavior of the effective mass
\begin{eqnarray}
2M^{\rm eff}_{K({\bf p})}(t) &=&\cosh^{-1} \Bigg[ \frac{ C_K(t+2;{\bf p}) + C_K(t-2;{\bf p}) }{2 C_K(t;{\bf p}) } \Bigg], \nonumber \\
&& \\
{\rm prior}[aE^{(0)}_{K({\bf p})}] &\sim& M^{\rm eff}_{K({\bf p})}({\rm long}\ t),
\end{eqnarray}
and effective amplitude
\begin{eqnarray}
C_{K({\bf p})}^{\rm eff}(t) &=& \frac{ C_K(t;{\bf p}) }{ e^{-M^{\rm eff}_{K({\bf p})} t} + e^{-M^{\rm eff}_{K({\bf p})} (T-t)} }, \\
{\rm prior}[(d_{\bf p}^{(0)})^2] &\sim& C_{K({\bf p})}^{\rm eff}({\rm long}\ t).
\end{eqnarray}
Representative effective mass and amplitude plots are shown in Fig.~\ref{fig-KMZeff}.
Values for ground state priors and fit results are given in Table~\ref{tab-Kfits}.  Excited state mass and amplitude priors are based on the PDG~\cite{PDG:2012} and set according to
\begin{eqnarray}
d^{(n>1)}_{\bf p} &=& 0.01 \pm 0.5, \nonumber \\
E^{(1)}_{ K( {\bf p} ) } - E^{(0)}_{ K( {\bf p} ) } &=& (600\pm 300) \text{ MeV}, \nonumber \\
E^{(2)}_{ K( {\bf p} ) } - E^{(0)}_{ K( {\bf p} ) } &=& (1000\pm 500) \text{ MeV}, \nonumber \\
E^{(n+2)}_{K({\bf p})} - E^{(n>0)}_{K({\bf p})} &=& (600\pm 600) \text{ MeV}.
\end{eqnarray}
As with the $B$ meson fits, positive energy splittings are parametrically enforced.
\begin{table*}[t]
\begin{tabular}{clcccccccc}
\hline\hline
	& & & & & & & & & \\ [-3ex]
	\T ens   	&		&$aM_K^{(0)}$	& $d^{(0)}_{000}$	& $aE^{(0)}_{100}$		& $d^{(0)}_{100}$	& $aE^{(0)}_{110}$	& $d^{(0)}_{110}$	& $aE^{(0)}_{111}$		& $d^{(0)}_{111}$ 	\\ [0.5ex]
	\hline
	& & & & & & & & & \\ [-3ex]
	\T C1 	& prior	& 0.312(17)	& 0.2225(10)		& 0.41(11)			& 0.1936(52)		& 0.48(23)		& 0.178(13)		& 0.55(28)			& 0.173(23)	\\
			& fit		& 0.31211(15)	& 0.22283(25)		& 0.40657(58)			& 0.19455(70)		& 0.48461(76)		& 0.18005(81)		& 0.5511(16)			& 0.1693(15)	\\
			& & & & & & & & & \\
	\T C2	& prior	& 0.329(24)	& 0.2262(8)		& 0.45(15)			& 0.190(10)		& 0.55(15)		& 0.176(14)		& 0.61(31)			& 0.141(35)	\\
			& fit		& 0.32863(18)	& 0.22630(29)		& 0.45406(85)			& 0.19265(93)		& 0.5511(16)		& 0.1741(18)		& 0.6261(75)			& 0.1537(88)	\\
			& & & & & & & & & \\
	\T C3	& prior	& 0.356(25)	& 0.220(1)		& 0.475(75)			& 0.190(5)		& 0.58(20)		& 0.187(13)		& 0.65(30)			& 0.158(28)	\\
			& fit		& 0.35717(22)	& 0.22083(34)		& 0.47521(85)			& 0.19129(98)		& 0.5723(11)		& 0.1780(10)		& 0.6524(30)			& 0.1667(27)	\\
			& & & & & & & & & \\
	\T F1		& prior	& 0.229(60)	& 0.138(1)		& 0.32(24)			& 0.116(6)		& 0.39(34)		& 0.105(22)		& 0.43(40)			& 0.077(26) 	\\
			& fit		& 0.22865(11)	& 0.13786(13)		& 0.32024(66)			& 0.11618(65)		& 0.39229(86)		& 0.10636(71)		& 0.4515(25)			& 0.0987(21)	\\
			& & & & & & & & & \\
	\T F2		& prior	& 0.246(36)	& 0.137(1)		& 0.33(23)			& 0.117(6)		& 0.40(30)		& 0.105(10)		& 0.47(37)			& 0.105(24) 	\\
			& fit		& 0.24577(13)	& 0.13664(17)		& 0.33322(52)			& 0.11764(48)		& 0.40214(73)		& 0.10760(61)		& 0.4623(14)			& 0.10241(94)	\\
			& & & & & & & & & \\ [-3ex]
\hline\hline
\end{tabular}\caption{Kaon ground state priors and fit results.}
\label{tab-Kfits}
\end{table*}

\subsection{Three Point Correlators}
\label{app-3ptpriors}
%
\begin{figure}[b!]
{\scalebox{0.98}{\includegraphics[angle=-90,width=0.5\textwidth]{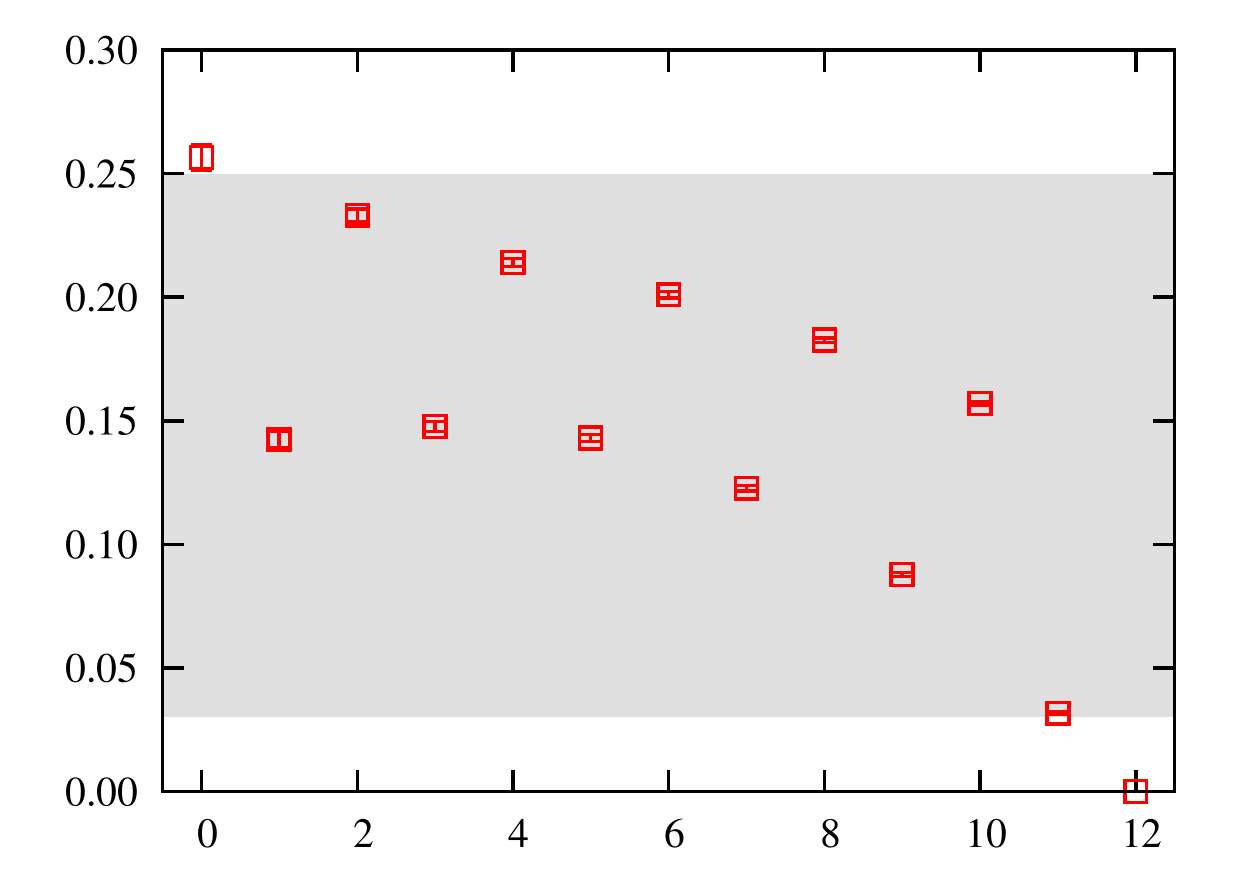}}}
\caption{Effective amplitude $A_{T(100)}^{\rm eff}(t)$ for the three point correlation function, with tensor current and momentum $\nicefrac{2\pi}{L}(1,0,0)$, plotted vs. $\nicefrac{t}{a}$ on ensemble C1.}
\label{fig-3ptZeff}
\end{figure}
Priors for ground state three point amplitudes are obtained from the long time ($t \approx \nicefrac{T}{2}$) behavior of the effective amplitude 
\begin{eqnarray}
A_{J(\bf p)}^{\rm eff}(t) &=& \frac{C_J^\alpha(t,T;{\bf p})}{\sqrt{C_{K({\bf p})}^{\rm eff}} \sqrt{C_B^{\alpha\alpha, \rm eff}}}  e^{M_K^{\rm eff}(T-t)}\ e^{M_B^{\rm eff}t}, \nonumber \\
&& \\
{\rm prior} [A_{J({\bf p})}^{(0,0)}] &\sim& A_{J({\bf p})}^{\rm eff}({\rm long}\ t).
\end{eqnarray}
A representative plot of the three point effective amplitude is shown in Fig.~\ref{fig-3ptZeff}.
Values for ground state amplitude priors and fit results are given in Table~\ref{tab-3ptfits}.  Excited state amplitude priors are set to
\begin{equation}
A_{J({\bf p})}^{(n,m)} = 0.01(1.0).
\end{equation}
\begin{turnpage}
\begin{table}[t]
\begin{tabular}{cc|cccc|ccc|ccc}
\hline\hline
	& & & & & & & & & & & \\ [-3ex]
	\T ens   	&		& $A^{(0,0)}_{V_t(000)}$	& $A^{(0,0)}_{V_t(100)}$	& $A^{(0,0)}_{V_t(110)}$	& $A^{(0,0)}_{V_t(111)}$	& $A^{(0,0)}_{V_k(100)}$	& $A^{(0,0)}_{V_k(110)}$	& $A^{(0,0)}_{V_k(111)}$	& $A^{(0,0)}_{T(100)}$	& $A^{(0,0)}_{T(110)}$	& $A^{(0,0)}_{T(111)}$ 	\\ [1ex]
	\hline\hline
	& & & & & & & & & & & \\ [-2ex]
	\T C1 	& prior	& 0.88(88)			& 0.71(71)			& 0.60(60)			& 0.54(54)			& 0.17(14)			& 0.14(11)			& 0.12(10)			& 0.14(11)			& 0.12(10)			& 0.12(10)	\\
	\hline
			& fit1		& 0.5261(46)			& 0.4051(39)			& 0.3431(37)			& 0.3034(50)			& 0.1464(26)			& 0.1095(17)			& 0.0863(13)			& 0.1263(18)			& 0.0964(13)			& 0.0776(15)	\\
			& fit2		& 0.5290(46)			& 0.4007(44)			& 0.3426(39)			& 0.3030(39)			& 0.1451(30)			& 0.1089(19)			& 0.0866(16)			& 0.1280(29)			& 0.0971(18)			& 0.0777(16)	\\
			& fit3		& 0.5276(40)			& 0.3982(50)			& 0.3448(44)			& 0.3044(41)			& 0.1439(39)			& 0.1062(26)			& 0.0853(20)			& 0.1282(33)			& 0.0954(21)			& 0.0771(16)	\\
			& & & & & & & & & & & \\
	\T C2	& prior	& 0.98(80)			& 0.75(67)			& 0.60(51)			& 0.45(40)			& 0.21(17)			& 0.18(14)			& 0.12(11)			& 0.19(15)			& 0.13(10)			& 0.12(10)	\\
	\hline
			& fit1		& 0.5202(55)			& 0.3738(38)			& 0.3062(25)			& 0.2660(32)			& 0.1526(27)			& 0.1068(13)			& 0.0835(18)			& 0.1337(24)			& 0.0933(11)			& 0.0741(16)	\\
			& fit2		& 0.5234(48)			& 0.3762(33)			& 0.3096(35)			& 0.2666(53)			& 0.1575(74)			& 0.1052(44)			& 0.0774(49)			& 0.1291(32)			& 0.0916(20)			& 0.0717(24)	\\	
			& fit3		& 0.5243(41)			& 0.3692(72)			& 0.3040(66)			& 0.2387(219)			& 0.1468(53)			& 0.1030(35)			& 0.0728(88)			& 0.1265(51)			& 0.0921(32)			& 0.0683(74)	\\
			& & & & & & & & & & & \\
	\T C3	& prior	& 0.93(79)			& 0.83(62)			& 0.67(53)			& 0.48(43)			& 0.18(14)			& 0.16(11)			& 0.13(11)			& 0.17(13)			& 0.13(10)			& 0.11(9)	\\
	\hline
			& fit1		& 0.4910(38)			& 0.3650(31)			& 0.3024(18)			& 0.2612(25)			& 0.1426(19)			& 0.1052(13)			& 0.0832(14)			& 0.1259(18)			& 0.0938(13)			& 0.0748(16)	\\
			& fit2		& 0.4926(36)			& 0.3604(48)			& 0.3017(38)			& 0.2585(70)			& 0.1460(34)			& 0.1063(13)			& 0.0847(16)			& 0.1273(26)			& 0.0941(12)			& 0.0752(13)	\\
			& fit3		& 0.4850(32)			& 0.3605(56)			& 0.2992(53)			& 0.2630(37)			& 0.1460(41)			& 0.1074(30)			& 0.0860(20)			& 0.1254(34)			& 0.0953(30)			& 0.0766(18)	\\		
			& & & & & & & & & & & \\
	\T F1		& prior	& 0.69(59)			& 0.59(47)			& 0.42(35)			& 0.31(25)			& 0.16(14)			& 0.13(10)			& 0.12(10)			& 0.16(14)			& 0.12(9)				& 0.09(8) 	\\
	\hline
			& fit1		& 0.5157(39)			& 0.3713(41)			& 0.3070(25)			& 0.2598(88)			& 0.1529(28)			& 0.1065(15)			& 0.0821(23)			& 0.1393(24)			& 0.0975(14)			& 0.0743(22)	\\
			& fit2		& 0.5188(25)			& 0.3705(42)			& 0.2999(96)			& 0.2565(100)			& 0.1495(38)			& 0.1043(22)			& 0.0796(25)			& 0.1371(34)			& 0.0967(19)			& 0.0743(23)	\\
			& fit3		& 0.5142(34)			& 0.3585(82)			& 0.2839(208)			& 0.2316(228)			& 0.1491(44)			& 0.1007(70)			& 0.0763(63)			& 0.1376(40)			& 0.0898(72)			& 0.0658(69)	\\		
			& & & & & & & & & & & \\
	\T F2		& prior	& 0.68(58)			& 0.57(45)			& 0.46(40)			& 0.46(39)			& 0.20(17)			& 0.17(14)			& 0.15(12)			& 0.17(15)			& 0.15(12)			& 0.13(10) 	\\
	\hline
			& fit1		& 0.4984(26)			& 0.3656(22)			& 0.2988(25)			& 0.2602(25)			& 0.1503(30)			& 0.1035(19)			& 0.0834(28)			& 0.1369(26)			& 0.0949(19)			& 0.0761(23)	\\
			& fit2		& 0.5046(23)			& 0.3725(33)			& 0.3055(56)			& 0.2586(72)			& 0.1442(112)			& 0.1003(45)			& 0.0848(116)			& 0.1491(82)			& 0.0949(57)			& 0.0800(77)	\\
			& fit3		& 0.5029(19)			& 0.3640(131)			& 0.3071(93)			& 0.2537(69)			& 0.1076(184)			& 0.0948(48)			& 0.0795(27)			& 0.0945(183)			& 0.0887(42)			& 0.0762(26)	\\ [1.7ex]
\hline\hline
\end{tabular}\caption{Three point ground state amplitude priors and fit results.  Fit1 results are from separate fits to each current and each momenta.  Fit2 results are from three separate fits, one to each current, but each including all momenta.  Fit3 results are from a single simultaneous fit to all three currents and including data at all momenta.}
\label{tab-3ptfits}
\end{table}
\end{turnpage}

\clearpage

\subsection{ Chiral/Continuum Extrapolation }
\label{app-ChPT}
We separate the priors into two groups.  Group I priors, shown with fit results in Table~\ref{tab-chiptfit_I}, allow us to incorporate uncertainty associated with input parameters.  
We set the prior for $r_1$ based on the value obtained in~\cite{Davies:2009}.  
For $F_\pi$ we tried values ranging from 131 MeV to 156 MeV (the value of $F_K$) and found negligible change in the fit results.  For the final result we use 144 MeV with an error of 10\%.
We set the $BB^*\pi$ coupling $g$ based on the recent works~\cite{gBBstarPi}.
The constant $B_0$ is set based on quark and meson masses from the PDG~\cite{PDG:2012}.
Values for $r_1/a$ are taken from~\cite{Bazavov:2010}.
Values of $aM_B$ are obtained using our best fit results for $aE_B^{{\rm sim}(0)}$ in Table~\ref{tab-Bfits} and Eq.~(\ref{eq-PhysMB}).  Priors for $aM_K$ are taken as our best fit results for $aM_K^{(0)}$ in Table~\ref{tab-Kfits}.
The priors for $\Delta^* = M_{B^*_s} - M_B$ are fixed based on the $B^*_s-B_s$ splitting from the PDG~\cite{PDG:2012} and ensemble dependent values for the $B_s-B$ splitting taken from~\cite{Na:2012}.
\begin{table}[t!]
\begin{tabular}{ccccccccc}
\hline\hline
	\T				&				&			\\
	\T Group I  		& prior			& fit    		\\ [0.5ex]
	\hline
	\T $r_1$ [fm]		& 0.3133(23)		& 0.3137(23)	\\
	\T $F_\pi$ [GeV]	& 0.144(14)		& 0.162(13)	\\
	\T $g$			& 0.51(20)		& 0.46(13)	\\
	\T $B_0$ [GeV]		& 2.4(1)			& 2.4(1)		\\
	\T $r_1/a$			& 2.647(3)		& 2.647(3)	\\
	\T $r_1/a$			& 2.618(3)		& 2.618(3)	\\
	\T $r_1/a$			& 2.644(3)		& 2.644(3)	\\
	\T $r_1/a$			& 3.699(3)		& 3.699(3)	\\
	\T $r_1/a$			& 3.712(4)		& 3.712(4)	\\
	\T $aM_B$		& 3.1891(18)		& 3.1891(18)	\\
	\T $aM_B$		& 3.2320(73)		& 3.2327(71)	\\
	\T $aM_B$		& 3.2095(77)		& 3.2091(77)	\\
	\T $aM_B$		& 2.2817(64)		& 2.2817(64)	\\
	\T $aM_B$		& 2.2799(87)		& 2.2802(87)	\\
	\T $aM_K$		& 0.31210(15)		& 0.31210(15)	\\
	\T $aM_K$		& 0.32864(17)		& 0.32864(17)	\\
	\T $aM_K$		& 0.35718(22)		& 0.35718(22)	\\
	\T $aM_K$		& 0.22865(11)		& 0.22865(11)	\\
	\T $aM_K$		& 0.24577(13)		& 0.24577(13)	\\
	\T $r_1\Delta^*$	& 0.1802(51)		& 0.1802(51)	\\
	\T $r_1\Delta^*$	& 0.1689(47)		& 0.1690(47)	\\
	\T $r_1\Delta^*$	& 0.1429(48)		& 0.1432(48)	\\
	\T $r_1\Delta^*$	& 0.1748(49)		& 0.1747(49)	\\
	\T $r_1\Delta^*$	& 0.1532(42)		& 0.1531(42)	\\  [0.5ex]
\hline\hline
\end{tabular}\caption{Group I priors and fit results for the chiral/continuum fit.  Quantities appearing in five consecutive rows have ensemble-dependent values corresponding to C1, C2, C3, F1, and F2.}
\label{tab-chiptfit_I}
\end{table}

Group II priors, shown with fit results in Table~\ref{tab-chiptfit_II}, are associated with fit parameters.  All group II priors are zero with widths chosen based on the typical size of the associated term in the extrapolation.
Prior widths for $\kappa$ and $\kappa_T$ are chosen based on values of the data for $f_\parallel$, $f_\perp$, and $f_T$ and the leading order terms in Eqs.~(\ref{eq-fpar}, \ref{eq-fperp}, \ref{eq-fT}).
The prior widths for $d_1$ include a factor of $\alpha_s$ based on the known $\mathcal{O}(\alpha_s a^2)$ discretization effects in the HISQ action.  For higher order discretization effects we take a prior width of one.
We take prior widths for the $f_i$ and $g_i$ to be one.
We express the coefficient $a_2^\perp$ in terms of other NLO valence quark analytic term coefficients using $a_2^\perp = a_1^\parallel + a_2^\parallel - a_1^\perp$~\cite{Aubin:2007}.

\begin{table}[t!]
\begin{tabular}{ccccccccc}
\hline\hline
	\T				&		& \multicolumn{3}{c}{fit}					\\
	\T Group II			& prior	& $f_\parallel$	& $f_\perp$	& $f_T$		\\ [0.5ex]
	\hline
	\T $r_1^{3/2}\kappa$& 0(3)	& \multicolumn{2}{c}{0.55(14)} & --			\\
	\T $r_1^2\,\kappa_T$	& 0(3)	& --			& --			& 0.84(57)	\\
	\T $d_1$			& 0(0.3)	& 0.18(20)	& -0.14(22)	& -0.27(22)	\\
	\T $d_2$			& 0(1)	& 0.32(91)	& -0.26(91)	& -0.40(95)	\\
	\T $f_1$			& 0(1)	& 0.34(0.92)	& 0.19(0.95)	& 0.60(0.91)	\\
	\T $f_2$			& 0(1)	& 0.08(1.00)	& -0.03(1.00)	& 0.17(0.99)	\\
	\T $f_3$			& 0(1)	& 0.08(1.00)	& 0.04(1.00)	& 0.12(1.00)	\\
	\T $f_4$			& 0(1)	& 0.03(1.00)	& 0.00(1.00)	& 0.04(1.00)	\\
	\T $g_1$			& 0(1)	& 0.00(1.00)	& 0.02(1.00)	& 0.00(1.00)	\\
	\T $g_2$			& 0(1)	& 0.00(1.00)	& 0.00(1.00)	& 0.00(1.00)	\\
	\T $g_3$			& 0(1)	& 0.00(1.00)	& 0.01(1.00)	& 0.00(1.00)	\\
	\T $g_4$ 			& 0(1)	& 0.00(1.00)	& 0.00(1.00)	& 0.00(1.00)	\\
	\T $h_1$ 			& 0(4)	& -3.11(50)	& 1.19(1.18)	& -2.04(2.42)	\\
	\T $h_2$			& 0(4)	& 0.42(28)	& -1.22(96)	& -1.82(1.93)	\\
	\T $h_3$			& 0(4)	& 0.182(79)	& 0.29(24)	& 0.40(45)	\\
	\T $a_1$			& 0(5)	& 5.21(1.38)	& 1.09(2.30)	& 1.88(3.09)	\\
	\T $a_2$			& 0(12)	& 11.76(3.91)	& --			& -6.24(7.12)	\\
	\T $a_3$			& 0(2)	& -1.34(61)	& -0.43(81)	& -0.25(1.16)	\\
	\T $a_4$			& 0(2)	& -0.06(2.00)	& -0.12(2.00)	& 0.07(2.00)	\\
	\T $a_5$			& 0(2)	& 0.13(2.00)	& 0.10(2.00)	& -0.01(2.00)	\\
	\T $a_6$			& 0(2)	& -0.59(1.92)	& -0.47(1.94)	& 0.19(1.95)	\\
	\T $a_7$			& 0(2)	& 0.01(2.00)	& -0.04(2.00)	& 0.02(2.00)	\\
	\T $a_8$			& 0(2)	& -0.14(1.98)	& -0.26(1.99)	& 0.15(1.99)	\\
	\T $a_9$			& 0(2)	& 0.00(2.00)	& 0.00(2.00)	& 0.00(2.00)	\\
	\T $a_{10}$		& 0(5)	& -4.12(81)	& -0.38(1.48)	& -0.19(2.04)	\\  [0.5ex]
\hline\hline
\end{tabular}\caption{Group II priors and fit results for the chiral/continuum fit.}
\label{tab-chiptfit_II}
\end{table}

\clearpage
\subsection{ Standard $z$ Expansion }
\label{app-stdzpriors}
\begin{table}[t!]
\begin{tabular}{ccc}
\hline\hline	
	\T Group I  						& prior			& fit    \\ [0.5ex]
	\hline
	\T $r_1$ [fm]						& 0.3133(23)		& 0.3133(23)	\\
	\T $\Delta^*_+$ [GeV]				& 0.04578(35)		& 0.04578(35)	\\
	\T $\Delta^*_T$	 [GeV]				& 0.046(35)		& 0.052(34)	\\
	\T $r_1M_B$						& 8.38197(27)		& 8.38197(27)	\\	
	\T $r_1M_K$						& 0.783821(21)	& 0.783821(21)\\ [0.5ex]
	\hline\hline
	\\
	\\
	\hline\hline
	\T Group II							& prior			& fit \\ [0.5ex]
	\hline
	\T $a_0^0$						& 0(2)			& 0.550(20)	\\
	\T $a_1^0$						& 0(2)			& -1.89(23)	\\
	\T $a_2^0$						& 0(2)			& 1.98(1.24)	\\
	\T $a_3^0$						& 0(2)			& -0.02(2.00)	\\
	\T $a_0^+$						& 0(2)			& 0.432(15)	\\
	\T $a_1^+$						& 0(2)			& -0.65(23)	\\
	\T $a_2^+$						& 0(2)			& -0.97(1.24)	\\
	\T $a_0^T$						& 0(2)			& 0.388(23)	\\
	\T $a_1^T$						& 0(2)			& -0.67(34)	\\
	\T $a_2^T$						& 0(2)			& -1.05(1.70)	\\ [0.5ex]
\hline\hline
\end{tabular}\caption{Priors and fit results for the simultaneous, standard $z$ expansion for $f_0$, $f_+$, and $f_T$. }
\label{tab-zpriors}
\end{table}
\setlength{\tabcolsep}{0.05in}
\begin{table*}[t!]
\begin{tabular}{c|cccccccccc}
	\T  			& $a^0_0$	& $a^0_1$	& $a^0_2$	& $a^0_3$	& $a^+_0$	& $a^+_1$	& $a^+_2$	& $a^T_0$	& $a^T_1$	& $a^T_2$ 	\\ [1ex]
	\hline
	& & & & & & & & & & \\ [-3ex]
	\T $a^0_0$	& 3.8400e-4&   1.9244e-3&   5.6298e-3&  -3.1229e-3&   2.1462e-4&   2.4187e-3&   9.6831e-3&   8.3579e-5&   8.0400e-4&   2.0606e-3 	\\ [1ex]
	\T $a^0_1$ 	&                         	&  5.3773e-2&   2.4109e-1&  -7.1455e-2&   2.4487e-3&   4.9308e-2&   2.3925e-1&   1.2560e-3&   2.1329e-2&   5.7572e-2 	\\ [1ex]
	\T $a^0_2$ 	&                         	&                        	& 1.5497&   1.0613e-1&   9.9620e-3&   2.4211e-1&   1.3622&   4.4186e-3&   9.6120e-2&   3.0591e-1	\\ [1ex]  
	\T $a^0_3$ 	&                         	&                       	&                         	&   3.9901&   7.6913e-5&   2.7607e-3&   1.3577e-2&  -2.6008e-4&  -7.0551e-3&  -2.8847e-2 	\\ [1ex]  
	\T $a^+_0$ 	&                         	&                   	&                         	&                         	&  2.3324e-4&   2.5548e-3&   8.3184e-3&   1.3602e-4&   1.4272e-3&   2.9191e-3 	\\ [1ex]  
	\T $a^+_1$ 	&                         	&                       	&                        	&                      	&                         	&   5.1033e-2&   2.2584e-1&   1.3456e-3&   2.3292e-2&   5.6690e-2 	\\ [1ex]  
	\T $a^+_2$ 	&                         	&                    	&                        	&                        	&                        	&                       	&  1.5378&   1.1423e-3&   5.16553e-2&   2.6917e-1 	\\ [1ex]  
	\T $a^T_0$ 	&                         	&                      	&                        	&                         	&                       	&                        	&                       	& 5.3232e-4&   5.5041e-3&   1.9645e-2 	\\ [1ex]  
	\T $a^T_1$ 	&                         	&                  	&                       	&                        	&                        	&                        	&                       	&                      	& 1.1248e-1&   4.2484e-1 	\\ [1ex]  
	\T $a^T_2$ 	&                         	&                      	&                         	&                        	&                        	&                  	&                        	&                       	&                    	&  2.9010 		\\ [1ex]  
\end{tabular}\caption{Covariance matrix for the physical extrapolated coefficients of the standard $z$ expansion.  
The covariance matrix only includes errors from the fit.  To construct form factors including the additional systematic errors of Sec.~\ref{sec-syserr}, a 4\% error must be added in quadrature to the error obtained from the covariance matrix.}
\label{tab-Zcovsimple}
\end{table*}

Table~\ref{tab-zpriors} lists Group I and II priors and fit results for the standard $z$ expansion.  
We set the prior for $r_1$ based on the value obtained in~\cite{Davies:2009}.  
Priors for the $M_B$ and $M_K$ are taken from the PDG~\cite{PDG:2012}.
For $\Delta^*_+$ we use the $B$ meson hyperfine splitting~\cite{PDG:2012} and for $\Delta_T^*$ we use the same central value but a width 100 times larger than that for $\Delta_+^*$.  For purposes of reconstructing the form factors, including the correlations, the covariance matrix associated with the coefficients of the $z$ expansion are given in Table~\ref{tab-Zcovsimple}.

\subsection{Modified $z$ Expansion}
\label{app-modzpriors}

Table~\ref{tab-modzpriorsI} lists Group I priors with widths that incorporate errors associated with various input parameters.  We take the scale $r_1$ from~\cite{Davies:2009}.    
For $F_\pi$ we tried values ranging from 131 MeV to 156 MeV (the value of $F_K$) and found negligible change in the fit results.  For the final result we use 144 MeV with an error of 10\%.  
We define the vector and tensor bound state masses of the Blaschke factors $P_{+,T}$ in terms of the splittings of Eqs.~(\ref{eq-f+polemass}) and~(\ref{eq-fTpolemass}) and use values based on the PDG~\cite{PDG:2012}.
We take the central value and width of $\Delta^*_+$ and the central value for $\Delta^*_T$ from the $M_{B^*}-M_B$ splitting.  For the $\Delta^*_T$ width we use $100\times$ the $M_{B^*}-M_B$ splitting error.
%
To absorb strange-quark mistuning effects, we use a target value $M_{\eta_s^{\rm phys}}$ from \cite{Davies:2009} and simulation values of $aM_{\eta_s}$ from ongoing analyses of $B_s \to K \ell \nu$ and $B_s \to \eta_s$.  
Values for $r_1/a$ are taken from~\cite{Bazavov:2010}.  Values of $aM_B$ are obtained using our best fit results for $aE_B^{{\rm sim}(0)}$ in Table~\ref{tab-Bfits} and Eq.~(\ref{eq-PhysMB}).  Priors for $aM_K$ are taken as our best fit results for $aM_K^{(0)}$ in Table~\ref{tab-Kfits}.

Table~\ref{tab-modzpriorsII} lists Group II priors associated with output fit parameters.  
We use common $b_i$ for all three form factors with priors for $b_1$ of zero with width 0.3 to reflect a factor $\alpha_s$ for the $(aE_K)^2$ term.  For $b_2$ we use a prior of zero with width one.
We express the analytic terms as functions of dimensionless quantities, {\it e.g. } $x_l$ and $\delta x_s$, which are ratios of meson masses and the chiral scale $\Lambda_\chi = 4\pi F_\pi$.  As a result, these terms are naturally $\mathcal{O}(1)$.  We therefore choose priors for $c_i$ to be zero with width one.  We know from previous works using the same ensembles that the sea quark contributions are smaller than the valence quark contributions.  We therefore take the coefficients $e^{(k)}$ to have prior zero with width 0.3.  The $am_b$ and $m_l$ dependent discretization effects are written in terms of quantities that are naturally $\mathcal{O}(1)$ so we choose priors of zero with width one for the $f^{(k)}$ and $g^{(k)}$.

\setlength{\tabcolsep}{0.15in}
\begin{table*}[t!]
\begin{tabular}{ccccccc}
\hline\hline	
	\T Group I  					& prior			& fit    			&& Group I						& prior			& fit		\\ [0.5ex]
	\hline
	\T $r_1$ [fm]					& 0.3133(23)		& 0.3132(23)		&& $aM_K^{\rm asqtad}$			& 0.36530(29)		& 0.36530(29)\\
	\T $F_\pi$ [GeV]				& 0.144(14)		& 0.135(14)		&& $aM_K^{\rm asqtad}$			& 0.38331(24)		& 0.38331(24)	\\
	\T $\Delta^*_+$ [GeV]			& 0.04578(35)		& 0.04578(35)		&& $aM_K^{\rm asqtad}$			& 0.40984(21)		& 0.40984(21)	\\
	\T $\Delta^*_T$	 [GeV]			& 0.046(35)		& 0.050(34)		&& $aM_K^{\rm asqtad}$			& 0.25318(19)		& 0.25318(19)	\\
	\T $M_{\eta_s^{\rm phys}}$ [GeV]	& 0.6858(40)		& 0.6858(40)		&& $aM_K^{\rm asqtad}$			& 0.27217(21)		& 0.27217(21)	\\	
	\T $r_1/a$						& 2.647(3)		& 2.647(3)		&& $aM_\pi^{\rm HISQ}$			& 0.15988(12)		& 0.15988(12)	\\
	\T $r_1/a$						& 2.618(3)		& 2.618(3)		&& $aM_\pi^{\rm HISQ}$			& 0.21097(16)		& 0.21096(16)	\\
	\T $r_1/a$						& 2.644(3)		& 2.644(3)		&& $aM_\pi^{\rm HISQ}$			& 0.29309(22)		& 0.29309(22)	\\
	\T $r_1/a$						& 3.699(3)		& 3.699(3)		&& $aM_\pi^{\rm HISQ}$			& 0.13453(11)		& 0.13453(11)	\\
	\T $r_1/a$						& 3.712(4)		& 3.712(4)		&& $aM_\pi^{\rm HISQ}$			& 0.18737(13)		& 0.18737(13)	\\
	\T $aM_B$					& 3.1891(18)		& 3.1891(18)		&& $aM_\pi^{\rm asqtad}$			& 0.15971(20)		& 0.15971(20)		\\
	\T $aM_B$					& 3.2322(73)		& 3.2318(62)		&& $aM_\pi^{\rm asqtad}$			& 0.22447(17)		& 0.22447(17)	\\
	\T $aM_B$					& 3.2096(77)		& 3.2110(72)		&& $aM_\pi^{\rm asqtad}$			& 0.31125(16)		& 0.31125(16)	\\
	\T $aM_B$					& 2.2818(64)		& 2.2824(55)		&& $aM_\pi^{\rm asqtad}$			& 0.14789(18)		& 0.14789(18)	\\
	\T $aM_B$					& 2.2796(88)		& 2.2790(60)		&& $aM_\pi^{\rm asqtad}$			& 0.20635(18)		& 0.20365(18)	\\
	\T $aM_K^{\rm HISQ}$			& 0.31210(15)		& 0.31210(15)		&& $aM_{\eta_s}^{\rm HISQ}$		& 0.41111(12)		& 0.41111(12)	\\
	\T $aM_K^{\rm HISQ}$			& 0.32864(17)		& 0.32864(17)		&& $aM_{\eta_s}^{\rm HISQ}$		& 0.41445(17)		& 0.41445(17)	\\
	\T $aM_K^{\rm HISQ}$			& 0.35718(22)		& 0.35717(22)		&& $aM_{\eta_s}^{\rm HISQ}$		& 0.41180(23)		& 0.41180(23)	\\
	\T $aM_K^{\rm HISQ}$			& 0.22865(11)		& 0.22865(11)		&& $aM_{\eta_s}^{\rm HISQ}$		& 0.294109(93)	& 0.294109(93)	\\
	\T $aM_K^{\rm HISQ}$			& 0.24577(13)		& 0.24576(13)		&& $aM_{\eta_s}^{\rm HISQ}$		& 0.29315(12)		& 0.29315(12)	\\ [0.5ex]
\hline\hline
\end{tabular}\caption{Group I priors for the simultaneous modified $z$ expansion for $f_0$, $f_+$, and $f_T$.  Quantities listed in five rows have ensemble-dependent values corresponding to C1, C2, C3, F1, and F2.}
\label{tab-modzpriorsI}
\end{table*}

\setlength{\tabcolsep}{0.05in}
\begin{table*}[t!]
\begin{tabular}{ccccccccccc}
\hline\hline
	\T				& 		& \multicolumn{3}{c}{fit} 					& &			&		& \multicolumn{3}{c}{fit} 						\\
	\T Group II   		& prior	& $f_0$		& $f_+$		& $f_T$		& \hspace{0.2in} & Group II 	& prior	& $f_0$		& $f_+$		& $f_T$     		\\ [0.5ex]
	\hline
	\\ [-3ex]
	\T $a_0$			& 0(2)	& 0.521(20)	& 0.422(19)	& 0.379(27)	& 	& $f_1^{(0)}$			& 0(1)	& 0.00(1.00)	& 0.01(1.00)	& 0.56(92)	\\
	\T $a_1$			& 0(2)	& -1.57(21)	& -0.57(22)	& -0.84(32)	& 	& $f_1^{(1)}$			& 0(1)	& 0.03(1.00)	& 0.00(1.00)	& 0.04(1.00)	\\
	\T $a_2$			& 0(2)	& 3.91(91)	& 1.52(1.14)	& -1.12(1.69)	& 	& $f_1^{(2)}$			& 0(1)	& 0.03(1.00)	& 0.00(1.00)	& 0.00(1.00)	\\
	\T $a_3$			& 0(2)	& -0.23(1.92)	& --			& --			& 	& $f_1^{(3)}$			& 0(1)	& 0.00(1.00)	& --			& --			\\
	\T $b_1$			& 0(0.3)	& -0.03(16)	& -0.03(16)	& -0.14(24)	& 	& $f_2^{(0)}$			& 0(1)	& 0.00(1.00)	& 0.01(1.00)	& 0.16(99)	\\
	\T $b_2$			& 0(1)	& 0.02(20)	& 0.02(19)	& 0.31(32) 	& 	& $f_2^{(1)}$			& 0(1)	& 0.00(1.00)	& 0.00(1.00)	& 0.01(1.00)	\\
	\T $c_1^{(0)}$		& 0(1)	& 0.06(65)	& -0.10(73)	& 0.50(84)	& 	& $f_2^{(2)}$			& 0(1)	& 0.00(1.00)	& 0.00(1.00)	& 0.00(1.00)	\\
	\T $c_1^{(1)}$		& 0(1)	& 0.34(85)	& 0.25(96)	& 0.17(98)	& 	& $f_2^{(3)}$			& 0(1)	& 0.00(1.00)	& --			& --			\\
	\T $c_1^{(2)}$		& 0(1)	& 0.81(0.95)	& -0.10(99)	& -0.02(1.00)	& 	& $f_3^{(0)}$			& 0(1)	& 0.00(1.00)	& 0.01(1.00)	& 0.11(1.00)	\\
	\T $c_1^{(3)}$		& 0(1)	& 0.00(1.00)	& --			& --			& 	& $f_3^{(1)}$			& 0(1)	& 0.01(1.00)	& 0.00(1.00)	& 0.01(1.00)	\\
	\T $c_2^{(0)}$		& 0(1)	& -0.13(35)	& -0.46(39)	& -0.42(48)	& 	& $f_3^{(2)}$			& 0(1)	& 0.00(1.00)	& 0.00(1.00)	& 0.00(1.00)	\\
	\T $c_2^{(1)}$		& 0(1)	& -0.46(51)	& -0.61(87)	& -0.34(90)	& 	& $f_3^{(3)}$			& 0(1)	& 0.00(1.00)	& --			& --			\\
	\T $c_2^{(2)}$		& 0(1)	& -1.68(82)	& 0.19(98)	& 0.03(1.00)	& 	& $f_4^{(0)}$			& 0(1)	& 0.00(1.00)	& 0.00(1.00)	& 0.04(1.00)	\\
	\T $c_2^{(3)}$		& 0(1)	& -0.01(1.00)	& --			& --			& 	& $f_4^{(1)}$			& 0(1)	& 0.00(1.00)	& 0.00(1.00)	& 0.00(1.00)	\\
	\T $c_3^{(0)}$		& 0(1)	& -0.02(98)	& -0.08(98)	& 0.08(1.00)	& 	& $f_4^{(2)}$			& 0(1)	& 0.00(1.00)	& 0.00(1.00)	& 0.00(1.00)	\\
	\T $c_3^{(1)}$		& 0(1)	& 0.03(1.00)	& -0.02(1.00)	& 0.01(1.00)	& 	& $f_4^{(3)}$			& 0(1)	& 0.00(1.00)	& --			& --			\\
	\T $c_3^{(2)}$		& 0(1)	& -0.01(1.00)	& 0.00(1.00)	& 0.00(1.00)	& 	& $g_1^{(0)}$			& 0(1)	& 0.00(1.00)	& 0.00(1.00)	& 0.04(1.00)	\\
	\T $c_3^{(3)}$		& 0(1)	& 0.00(1.00)	& --			& --			& 	& $g_1^{(1)}$			& 0(1)	& 0.00(1.00)	& 0.00(1.00)	& 0.00(1.00)	\\
	\T $d_1^{(0)}$		& 0(0.3)	& 0.01(24)	& 0.02(25)	& -0.26(23)	& 	& $g_1^{(2)}$			& 0(1)	& 0.00(1.00)	& 0.00(1.00)	& 0.00(1.00)	\\
	\T $d_1^{(1)}$		& 0(0.3)	& 0.05(27)	& 0.01(30)	& -0.06(30)	& 	& $g_1^{(3)}$			& 0(1)	& 0.00(1.00)	& --			& --			\\
	\T $d_1^{(2)}$		& 0(0.3)	& 0.01(29)	& 0.01(30)	& 0.01(30)	& 	& $g_2^{(0)}$			& 0(1)	& 0.00(1.00)	& 0.00(1.00)	& 0.00(1.00)	\\
	\T $d_1^{(3)}$		& 0(0.3)	& 0.00(30)	& --			& --			& 	& $g_2^{(1)}$			& 0(1)	& 0.00(1.00)	& 0.00(1.00)	& 0.00(1.00)	\\
	\T $d_2^{(0)}$		& 0(1)	& 0.02(90)	& 0.07(91)	& -0.39(96)	& 	& $g_2^{(2)}$			& 0(1)	& 0.00(1.00)	& 0.00(1.00)	& 0.00(1.00)	\\
	\T $d_2^{(1)}$		& 0(1)	& 0.10(96)	& 0.02(99)	& -0.13(1.00)	& 	& $g_2^{(3)}$			& 0(1)	& 0.00(1.00)	& --			& --			\\
	\T $d_2^{(2)}$		& 0(1)	& 0.01(99)	& 0.03(1.00)	& 0.02(1.00)	& 	& $g_3^{(0)}$			& 0(1)	& 0.00(1.00)	& 0.00(1.00)	& 0.01(1.00)	\\
	\T $d_2^{(3)}$		& 0(1)	& 0.00(1.00)	& --			& --			& 	& $g_3^{(1)}$			& 0(1)	& 0.00(1.00)	& 0.00(1.00)	& 0.00(1.00)	\\
	\T $e^{(0)}$		& 0(0.3)	& -0.02(29)	& 0.02(29)	& -0.02(30)	& 	& $g_3^{(2)}$			& 0(1)	& 0.00(1.00)	& 0.00(1.00)	& 0.00(1.00)	\\
	\T $e^{(1)}$		& 0(0.3)	& 0.01(30)	& 0.01(30)	& -0.01(30)	& 	& $g_3^{(3)}$			& 0(1)	& 0.00(1.00)	& --			& -- 			\\
	\T $e^{(2)}$		& 0(0.3)	& 0.01(30)	& 0.00(30)	& 0.00(30)	& 	& $g_4^{(0)}$			& 0(1)	& 0.00(1.00)	& 0.00(1.00)	& 0.00(1.00)	\\
	\T $e^{(3)}$		& 0(0.3)	& 0.00(30)	& --			& --			& 	& $g_4^{(1)}$			& 0(1)	& 0.00(1.00)	& 0.00(1.00)	& 0.00(1.00)	\\
	\T 				&		&			&			&			& 	& $g_4^{(2)}$			& 0(1)	& 0.00(1.00)	& 0.00(1.00)	& 0.00(1.00)	\\
	\T 				&		&			&			&			& 	& $g_4^{(3)}$			& 0(1)	& 0.00(1.00)	& --			& --			\\ [0.5ex]
\hline\hline
\end{tabular}\caption{Group II priors and fit results for the simultaneous modified $z$ expansion for $f_0$, $f_+$, and $f_T$.  The momentum-dependent discretization term coefficients $b_i$ are common to each form factor.  Note the fit ans\"atze for $f_+$ and $f_T$, Eq.~(\ref{eq-modzfpT}), has one fewer term in the sum than the fit ansatz for $f_0$, Eq.~(\ref{eq-modzf0}). }
\label{tab-modzpriorsII}
\end{table*}
\clearpage


\clearpage





\begin{thebibliography}{99}

\bibitem{Becirevic:2012}
   D.~Be\v{c}irevi\'{c}, N.~Ko\v{s}nik, F.~Mescia, and E.~Schneider,
   Phys. Rev. D {\bf 86}, 034034 (2012)
   [\href{http://arxiv.org/abs/1205.5811}{arXiv:1205.5811 [hep-ph]}].

\bibitem{Bobeth:2011}
   C.~Bobeth, G.~Hiller, D.~van~Dyk, and C.~Wacker,
   J. High Energy Phys. {\bf 01} (2012) 107
   [\href{http://arxiv.org/abs/1111.2558}{arXiv:1111.2558 [hep-ph]}].
   
\bibitem{Beaujean:2012}
   F.~Beaujean, C.~Bobeth, D.~van~Dyk, and C.~Wacker,
   J. High Energy Phys. 08 (2012) 030
   [\href{http://arxiv.org/abs/1205.1838}{arXiv:1205.1838 [hep-ph]}].

\bibitem{Altmannsofer:2012}
   W.~Altmannshofer and D.~M.~Straub,
   J. High Energy Phys. 08 (2012) 121
   [\href{http://arxiv.org/abs/1206.0273}{arXiv:1206.0273 [hep-ph]}].

\bibitem{Bobeth:2013}
   C.~Bobeth, G.~Hiller, and D.~van~Dyk,
   Phys. Rev. D {\bf 87}, 034016 (2013)
   [\href{http://arxiv.org/abs/1212.2321}{arXiv:1212.2321 [hep-ph]}].
   
\bibitem{Ball:2005}
    P.~Ball and R.~Zwicky,
    Phys. Rev. D {\bf 71}, 014029 (2005)
    [\href{http://xxx.lanl.gov/abs/hep-ph/0412079}{arXiv:hep-ph/0412079}].

\bibitem{Khodjamirian:2010}
   A.~Khodjamirian, Th.~Mannel, A.~A.~Pivovarov, and Y.-M.~Wang,
   J. High Energy Phys. 09 (2010) 089
   [\href{http://arxiv.org/abs/arXiv:1006.4945}{arXiv:1006.4945 [hep-ph]}].
   
\bibitem{Khodjamirian:2013}
   A.~Khodjamirian, Th.~Mannel, and Y.-M.~Wang,
   J. High Energy Phys. 02 (2013) 010
   [\href{http://arxiv.org/pdf/1211.0234v2.pdf}{arXiv:1211.0234 [hep-ph]}].

\bibitem{Becirevic:2000}
   D.~Be\v{c}irevi\'{c} and A.~B.~Kaidalov,
   Phys. Lett. B {\bf 478}, 417 (2000)
   [\href{http://arxiv.org/abs/hep-ph/9904490}{arXiv:hep-ph/9904490}].
   
\bibitem{Al-Haydari:2009}
   A.~Al-Haydari, A.~Ali~Khan, V.~M.~Braun, S.~Collins, M.~G\"{o}ckeler, G.~N.~Lacagnina, M.~Panero, A.~Sch\"{a}fer, G.~Schierholz
   (QCDSF),
   Eur. Phys. J. A {\bf 43}, 107-120 (2010)
   [\href{http://xxx.lanl.gov/abs/0903.1664}{arXiv:0903.1664 [hep-lat]}].   

\bibitem{Liu:2011}
   Z.~Liu, S.~Meinel, A.~Hart, R.~Horgan, E.~M\"{u}ller, M.~Wingate
   [\href{http://arxiv.org/pdf/1101.2726v1.pdf}{arXiv:1101.2726 [hep-ph]}].

\bibitem{Zhou:2012}
   R.~Zhou, S.~Gottlieb, J.~A.~Bailey, D.~Du, A.~X.~El-Khadra, R.~D.~Jain, A.~S.~Kronfeld, R.~S.~Van~de~Water, Y.~Liu, and Y.~Meurice
   (Fermilab Lattice and MILC),
   [\href{http://arxiv.org/abs/1211.1390}{arXiv:1211.1390 [hep-lat]}].
   
\bibitem{Lees:2012}
   J.~P.~Lees {\it et al.}
   (BABAR),
   Phys. Rev. D {\bf 86}, 032012 (2012)
   [\href{http://arxiv.org/abs/arXiv:1204.3933}{arXiv:1204.3933 [hep-ex]}].

\bibitem{Wei:2009}
   J.-~T.~Wei {\it et al.}
   (Belle),
   Phys. Rev. Lett. {\bf 103}, 171801 (2009)
   [\href{http://arxiv.org/abs/arXiv:0904.0770}{arXiv:0904.0770 [hep-ex]}].

\bibitem{Aaltonen:2011}
   T.~Aaltonen {\it et al.}
   (CDF),
   Phys. Rev. Lett. {\bf 107}, 201802 (2011)
   [\href{http://arxiv.org/abs/arXiv:1107.3753}{arXiv:1107.3753 [hep-ex]}].

\bibitem{Aaij:2012}
   R.~Aaij {\it et al.}
   (LHCb),
   J. High Energy Phys. 07 (2012) 133
   [\href{http://arxiv.org/abs/1205.3422}{arXiv:1205.3422 [hep-ex]}].
   
\bibitem{Aaij:2012b}
   R.~Aaij {\it et al.}
   (LHCb),
   J. High Energy Phys. 02 (2013) 105
   [\href{http://arxiv.org/abs/1209.4284}{arXiv:1209.4284 [hep-ex]}].

\bibitem{Bouchard:PRL}
   C.~M.~Bouchard, G.~P.~Lepage, C.~Monahan, H.~Na, and J.~Shigemitsu
   (HPQCD),
   [\href{http://arxiv.org/pdf/1306.0434.pdf}{arXiv:1306.0434 [hep-lat]}],
   to appear in Phys. Rev. Lett.
   
\bibitem{Bazavov:2010}
   A.~Bazavov, C.~Bernard, C.~DeTar, S.~Gottlieb, U.~M.~Heller, J.~E.~Hetrick, J.~Laiho, L.~Levkova, P.~B.~Mackenzie, M.~B.~Oktay, R.~Sugar, D.~Toussaint, and R.~S.~Van~de~Water
   (MILC),
   Rev. Mod. Phys. {\bf 82}, 1349 (2010)
   [\href{http://arxiv.org/abs/0903.3598}{arXiv:0903.3598 [hep-lat]}].

\bibitem{Lepage:1992}
   G.~P.~Lepage, L.~Magnea, C.~Nakhleh, U.~Magnea, and K.~Hornbostel
   (HPQCD),
   Phys. Rev. D {\bf 46}, 4052 (1992)
   [\href{http://arxiv.org/abs/hep-lat/9205007}{arXiv:hep-lat/9205007}].

\bibitem{Na:2012}
   H.~Na, C.~J.~Monahan, C.~T.~H.~Davies, R.~Horgan, G.~P.~Lepage and J.~Shigemitsu
   (HPQCD),
   Phys. Rev. D {\bf 86}, 034506 (2012)
   [\href{http://arxiv.org/abs/1202.4914}{arXiv:1202.4914 [hep-lat]}].

\bibitem{Follana:2007}
   E.~Follana, Q.~Mason, C.~Davies, K.~Hornbostel, G.~P.~Lepage, J.~Shigemitsu, H.~Trottier, and K.~Wong
   (HPQCD),
   Phys. Rev. D {\bf 75}, 054502 (2007)
   [\href{http://arxiv.org/abs/hep-lat/0610092}{arXiv:hep-lat/0610092}].

\bibitem{Na:2010}
   H.~Na, C.~T.~H.~Davies, E.~Follana, G.~P.~Lepage, and J.~Shigemitsu
   (HPQCD),
   Phys. Rev. D {\bf 82}, 114506 (2010)
   [\href{http://arxiv.org/abs/1008.4562}{arXiv:1008.4562 [hep-lat]}].

\bibitem{Na:2011}
   H.~Na, C.~T.~H.~Davies, E.~Follana, J.~Koponen, G.~P.~Lepage, and J.~Shigemitsu
   (HPQCD),
   Phys. Rev. D {\bf 84}, 114505 (2011)
   [\href{http://arxiv.org/abs/1109.1501}{arXiv:1109.1501 [hep-lat]}].

\bibitem{Monahan:2013}
   C.~Monahan, J.~Shigemitsu, and R.~Horgan
   (HPQCD),
   Phys. Rev. D {\bf 87}, 034017 (2013)
   [\href{http://arxiv.org/abs/arXiv:1211.6966}{arXiv:1211.6966 [hep-lat]}].

\bibitem{Gulez:2007}
   E.~Gulez, A.~Gray, M.~Wingate, C.~T.~H.~Davies, G.~P.~Lepage, and J.~Shigemitsu,
   Phys. Rev. D {\bf 73}, 074502 (2006); {\bf 75}, 119906(E) (2007)
   [\href{http://arxiv.org/abs/hep-lat/0601021}{arXiv:hep-lat/0601021}].

\bibitem{Lepage:2002}
   G.~P.~Lepage, B.~Clark, C.~T.~H.~Davies, K.~Hornbostel, P.~B.~Mackenzie, C.~Morningstar and H.~Trottier,
   Nucl. Phys. Proc. Suppl.  {\bf 106}, 12 (2002)
   [\href{http://arxiv.org/abs/hep-lat/0110175}{arXiv:hep-lat/0110175}].
   
\bibitem{Gregory:2010}
   E.~B.~Gregory, C.~T.~H.~Davies, I.~D.~Kendall, J.~Koponen, K.~Wong, E.~Follana, E.~G\'{a}miz, G.~P.~Lepage, E.~H.~M\"{u}ller, H.~Na, and J.~Shigemitsu
   (HPQCD),
   Phys. Rev. D {\bf 83}, 014506 (2011)
   [\href{http://arxiv.org/abs/1010.3848}{arXiv:1010.3848 [hep-lat]}].
   
\bibitem{Aubin:2007}
   C.~Aubin and C.~Bernard,
   Phys. Rev. D {\bf 76}, 014002 (2007)
   [\href{http://arxiv.org/abs/0704.0795}{arXiv:0704.0795 [hep-lat]}].
   
 \bibitem{MILC:2010}
   A.~Bazavov, C.~Bernard, C.~DeTar, W.~Freeman, Steven~Gottlieb, U.~M.~Heller, J.~E.~Hetrick, J.~Laiho, L.~Levkova, M.~Oktay, J.~Osborn, R.L.~Sugar, D.~Toussaint, and R.S.~Van~de~Water
   (MILC),
   Phys. Rev. D {\bf 82}, 074501 (2010)
   [\href{http://arxiv.org/abs/1004.0342}{arXiv:1004.0342 [hep-lat]}].
   
\bibitem{Bailey:2009}
   J.~Bailey, C.~Bernard, C.~DeTar, M.~Di~Pierro, A.~X.~El-Khadra, R.~T.~Evans, E.~D.~Freeland, E.~G\'{a}miz, S.~Gottlieb, U.~M.~Heller, J.~E.~Hetrick, A.~S.~Kronfeld, J.~Laiho, L.~Levkova, P.~B.~Mackenzie, M.~Okamoto, J.~N.~Simone, R.~Sugar, D.~Toussaint, and R.~S.~Van~de~Water
   (Fermilab Lattice and MILC),
   Phys. Rev. D {\bf 79}, 054507 (2009)
   [\href{http://arxiv.org/abs/0811.3640}{arXiv:0811.3640 [hep-lat]}].   
   
\bibitem{Boyd:1996}
   C.~G.~Boyd, B.~Grinstein, and R.~F.~Lebed,
   Nucl. Phys. {\bf B461}, 493 (1996)
   [\href{http://arxiv.org/abs/hep-ph/9508211}{arXiv:hep-ph/9508211}].

\bibitem{Arnesen:2005}
   M.~C.~Arnesen, B.~Grinstein, I.~Z.~Rothstein, and I.~W.~Stewart,
   Phys. Rev. Lett. {\bf 95}, 071802 (2005)
   [\href{http://arxiv.org/abs/hep-ph/0504209}{arXiv:hep-ph/0504209}].
   
\bibitem{Bourrely:2010}
   C.~Bourrely, L.~Lellouch, and I.~Caprini,
   Phys. Rev. D {\bf 79}, 013008 (2009); {\bf 82}, 099902(E) (2010)
   [\href{http://arxiv.org/abs/0807.2722}{arXiv:0807.2722 [hep-ph]}].

\bibitem{Becher:2006}
   T.~Becher and R.~J.~Hill,
   Phys. Lett. B {\bf 633}, 61 (2006)
   [\href{http://arxiv.org/abs/hep-ph/0509090}{arXiv:hep-ph/0509090}].
   
\bibitem{Beneke:2002}
  M.~Beneke, A.~P.~Chapovsky, M.~Diehl and T.~Feldmann,
  Nucl.\ Phys.\ {\bf B643}, 431 (2002)
  [\href{http://arxiv.org/pdf/hep-ph/0206152.pdf}{arXiv:hep-ph/0206152}].
   
\bibitem{Hill:2006}
   R.~J.~Hill,
   Phys. Rev. D {\bf 73}, 014012 (2006)
   [\href{http://arxiv.org/abs/hep-ph/0505129}{arXiv:hep-ph/0505129}].
   
\bibitem{Bijnens:2010}
   J.~Bijnens and I.~Jemos,
   Nucl. Phys. {\bf B840}, 54 (2010); {\bf B844}, 182(E) (2011)
   [\href{http://arxiv.org/abs/1006.1197}{arXiv:1006.1197 [hep-ph]}]. 
   
\bibitem{Davies:2008}
   C.~T.~H.~Davies,  K.~Hornbostel, I.~D.~Kendall, G.~P.~Lepage, C.~McNeile, J.~Shigemitsu, and H.~Trottier
   (HPQCD),
   Phys. Rev. D {\bf 78} 114507 (2008)
   [\href{http://arxiv.org/abs/0807.1687}{arXiv:0807.1687 [hep-lat]}].
   
\bibitem{Davies:2010}
   C.~T.~H.~Davies, C.~McNeile, E.~Follana, G.~P.~Lepage, H.~Na, and J.~Shigemitsu
   (HPQCD),
   Phys. Rev. D {\bf 82} 114504 (2010)
   [\href{http://arxiv.org/abs/1008.4018}{arXiv:1008.4018 [hep-lat]}].
   
\bibitem{Bobeth:2007}
   C.~Bobeth, G.~Hiller, and G.~Piranishvili,
   J. High Energy Phys. 12 (2007) 040
   [\href{http://arxiv.org/abs/0709.4174}{arXiv:0709.4174 [hep-ph]}].
   
\bibitem{PDG:2012}
    J.~Beringer {\it et al.}
    (Particle Data Group),
    Phys. Rev. D {\bf 86}, 010001 (2012)
    [\href{http://pdg.lbl.gov/}{pdg.lbl.gov}].
    
\bibitem{Jegerlehner:2008}
   F.~Jegerlehner,
   Nucl. Phys. Proc. Suppl. {\bf 181-182}, 135-140 (2008)
   [\href{http://arxiv.org/abs/arXiv:0807.4206}{arXiv:0807.4206 [hep-ph]}].
    
 \bibitem{CKM:2013}
    J.~Charles {\it et al.}
    (CKMfitter Group), 
    Eur. Phys. J. {\bf C41}, 1-131 (2005) 
    [\href{http://arxiv.org/abs/hep-ph/0406184}{hep-ph/0406184}], 
    updated results and plots available at: \href{http://ckmfitter.in2p3.fr}{ckmfitter.in2p3.fr}.   

\bibitem{Altmannshofer:2008}
   W.~Altmannshofer, P.~Ball, A.~Bharucha, A.~J.~Buras, D.~M.~Straub, and M.~Wick,
    J. High Energy Phys. 01 (2009) 019
   [\href{http://arxiv.org/abs/0811.1214}{arXiv:0811.1214 [hep-ph]}].
   
\bibitem{ASe}
   W.~Altmannshofer and D.~M.~Straub (private communication).
   
\bibitem{gdev}
   This is accomplished with the use of {\tt gvar} data types in the {\tt lsqfit} Python package, available at \href{http://www.physics.gla.ac.uk/HPQCD/}{www.physics.gla.ac.uk/HPQCD}.

\bibitem{Demir:2000}
   D.~A.~Demir, K.~A.~Olive, and M.~B.~Voloshin,
   Phys. Rev. D {\bf 66}, 034015 (2002)
   [\href{http://arxiv.org/abs/hep-ph/0204119}{arXiv:hep-ph/0204119}].

\bibitem{Koponen:2013}
   J.~Koponen, C.~T.~H.~Davies, G.~C.~Donald, E.~Follana, G.~P.~Lepage, H.~Na, and J.~Shigemitsu
   (HPQCD),
   [\href{http://arxiv.org/pdf/1305.1462v1.pdf}{arXiv:1305.1462 [hep-lat]}].   
   
\bibitem{Bernard:2002}
   C.~Bernard
   (MILC),
   Phys. Rev. D {\bf 65}, 054031 (2002)
   [\href{http://arxiv.org/abs/hep-lat/0111051}{arXiv:hep-lat/0111051}].

\bibitem{Davies:2009}
   C.~T.~H.~Davies, E.~Follana, I.~D.~Kendall, G.~P.~Lepage, and C.~McNeile
   (HPQCD),
   Phys. Rev. D {\bf 81}, 034506 (2010)
   [\href{http://arxiv.org/abs/0910.1229}{arXiv:0910.1229 [hep-lat]}].
    
\bibitem{gBBstarPi}
   H.~Ohki, H.~Matsufuru, and T.~Onogi,
   Phys. Rev. D {\bf 77}, 094509 (2008)
   [\href{http://arxiv.org/pdf/0802.1563v1.pdf}{arXiv:0802.1563 [hep-lat]}];
   J.~Bulava, M.~A.~Donnellan, and R.~Sommer
   (ALPHA),
   PoS {\bf Lattice2010} 303, (2010)
   [\href{http://arxiv.org/pdf/1011.4393v1.pdf}{arXiv:1011.4393 [hep-lat]}];
   W.~Detmold, C.-J.~D.~Lin, and S.~Meinel
   Phys. Rev. D {\bf 85}, 114508 (2012)
   [\href{http://arxiv.org/pdf/1203.3378v2.pdf}{arXiv:1203.3378 [hep-lat]}].


%
%


\end{thebibliography}
\end{document}